\documentclass[prd,onecolumn,showpacs,superscriptaddress,nofootinbib,floatfix,11pt]{revtex4-2}

\usepackage[utf8]{inputenc}
\usepackage{amsmath,amssymb}
\usepackage{slashed}
\usepackage[colorlinks=true, 
linkcolor=blue,
breaklinks=true,
urlcolor=magenta,
citecolor=blue]{hyperref}
\usepackage[usenames,dvipsnames]{color}
\usepackage{appendix}
\usepackage{braket,bm}
\usepackage{multirow}
\usepackage{enumitem}
\usepackage{color}
\usepackage{cancel}
\usepackage{orcidlink}
\usepackage{mathrsfs}
\usepackage{graphicx}
\usepackage{tikz}
\usepackage[normalem]{ulem}
\usepackage{booktabs}
\usepackage{array}
\usepackage{overpic}
\usepackage{float}
\usepackage{subfigure} 
\usepackage{threeparttable}
\allowdisplaybreaks[4]
\graphicspath{{figs/}}

\usetikzlibrary{calc}
\usetikzlibrary{intersections}
\usetikzlibrary{trees}
\usetikzlibrary{decorations.pathmorphing}
\usetikzlibrary{decorations.markings}
\usetikzlibrary{arrows.meta}
\usetikzlibrary{patterns}
\tikzset{
   global scale/.style={
      scale=#1,
      every node/.append style={scale=#1}},
   photon/.style={
   decorate,decoration={
      snake,
      amplitude=0.5mm,
      segment length=2.5mm,post length=0,pre length=0},draw=red},
   nucleon/.style={draw=black, postaction={decorate},
      decoration={markings,mark=at position .65 with{\arrow[draw=black]{latex}}}},
   pion/.style={draw=blue, postaction={decorate},
      decoration={markings,mark=at position .55 with{\arrow[draw=blue]{}}}},
    nucleonstar/.style={draw=black, postaction={decorate},
      decoration={markings, mark=at position 0.7 with {\arrow[draw=black]{latex}}}},
    }

\newcommand{\itp}{\affiliation{ Institute of Theoretical Physics, Chinese Academy of Sciences, Beijing 100190, China}}

\newcommand{\ucas}{\affiliation{School of Physical Sciences, University of Chinese Academy of Sciences, Beijing 100049, China}}

\newcommand{\huu}{\affiliation{School for Theoretical Physics, School of Physics and Electronics, Hunan University, Changsha 410082, China}}

\newcommand{\huhep}{\affiliation{Hunan Provincial Key Laboratory of High-Energy Scale Physics and Applications,\\
Hunan University, Changsha 410082, China}}

\newcommand{\md}{\mathrm{d}}

\newcommand{\mA}{\mathcal{A}}

\begin{document}
\title{Nucleon polarizabilities revisited:\\ insights from the Bernab\'eu--Ericson--Ferro Fontan--Tarrach sum rule}

\author{Xiong-Hui Cao\orcidlink{0000-0003-1365-7178}}
\email{xhcao@itp.ac.cn}
\itp

\author{Ling-Yun Dai\orcidlink{0000-0002-4070-4729}}
\email{dailingyun@hnu.edu.cn}
\huu\huhep

\author{Feng-Kun Guo\orcidlink{0000-0002-2919-2064}}\email{fkguo@itp.ac.cn}
\itp\ucas


\begin{abstract}
We present a complete dispersive evaluation of the Bernab\'eu--Ericson--Ferro Fontan--Tarrach (BEFT) sum rule for the nucleon dipole polarizability difference $\alpha_{E1}-\beta_{M1}$.  
The analysis is based on hyperbolic dispersion relations, which separate the $s$- and $t$-channel contributions without double counting and are particularly suited to backward real Compton scattering. 
This framework makes it possible to precisely determine the polarizability difference from independent dispersive input rather than from fits to low-energy Compton scattering data. 
The $s$-channel part is updated with modern single-pion photoproduction multipoles and model estimates of non-$\pi N$ inelastic channels. 
For the $t$-channel, we use contemporary dispersive input for $\gamma\gamma\to\pi\pi/K\bar K_{I_t=0}$ together with Roy--Steiner equation solutions for $\pi\pi/K\bar K_{I_t=0}\to N\bar N$, and we extend the calculation to the isovector $\pi\eta/K\bar K_{I_t=1}$ channel associated with the $a_0(980)$. 
The resulting treatment resolves the long-standing puzzle of earlier BEFT evaluations driven by the scalar-isoscalar two-pion channel and gives a controlled estimate of the subleading isovector contribution.
Combining the resulting BEFT values with the Baldin sum rule, we obtain, in units of $10^{-4}\,\mathrm{fm}^3$,
\[
\alpha_{E1}^p=11.17(52),\quad \beta_{M1}^p=2.84(52),\quad
\alpha_{E1}^n=12.18(59),\quad \beta_{M1}^n=3.03(59).
\]
For the proton, these values provide a benchmark with a precision comparable to state-of-the-art dispersive and chiral effective field theory extractions.
For the neutron, whose polarizabilities are much less directly constrained experimentally, the result provides an independent dispersive benchmark with an uncertainty comparable to that of the proton. 
\end{abstract}

\maketitle
\newpage
\tableofcontents

\section{Introduction}\label{sec:Intro}

Low-energy Compton scattering (CS) off the nucleon is a standard tool for investigating nucleon electromagnetic structure. 
At threshold and sufficiently low energies, gauge and Lorentz invariance force the process to reduce to Thomson scattering, which depends only on the charge and mass of the target. 
Measurable deviations appear at higher energies. 
Some are governed by the anomalous magnetic moment and by mechanisms such as neutral-pion exchange, while the remainder reflects the internal hadronic structure and excitations of the target.
The leading effects at low energy are the nucleon electric and magnetic dipole polarizabilities, $\alpha_{E1}$ and $\beta_{M1}$. 
These polarizabilities quantify the deformation of the nucleon charge and current distributions under an external electromagnetic field, reflecting the ease of photon-induced transitions.
Consequently, $\alpha_{E1}$ and $\beta_{M1}$ serve as direct probes of the symmetries and low-energy dynamics of the underlying constituents; see Refs.~\cite{Bernard:1995dp,Drechsel:2002ar,Schumacher:2005an,Griesshammer:2012we,Holstein:2013kia,Hagelstein:2015egb,Pascalutsa:2024hvy} for reviews.

These two polarizabilities are not equally well determined: only the sums $\alpha_{E1}+\beta_{M1}=14.0(2)\times 10^{-4}\,\mathrm{fm}^3$~\cite{Gryniuk:2015eza} for the proton and $15.2(4)\times 10^{-4}\,\mathrm{fm}^3$~\cite{Levchuk:1999zy} for the neutron are known to better than about $3\%$, whereas the individual values $\alpha_{E1}$ and $\beta_{M1}$, or equivalently their difference $\alpha_{E1}-\beta_{M1}=(6.8\text{--}10.9) \times 10^{-4}\,\mathrm{fm}^3$ for the proton~\cite{Krupina:2017pgr}, cannot currently be extracted with comparable precision.
The current  Particle Data Group (PDG) averages for the individual values are~\cite{ParticleDataGroup:2026mpi}
\begin{align}
    \begin{aligned}
        \alpha_{E1}^{p}&=11.5(4),\quad &\beta_{M1}^{p}&=2.31(29), \\ 
        \alpha_{E1}^{n}&=11.8(1.1), & \beta_{M1}^{n}&=3.7(1.2).
    \end{aligned}
    \label{eq:polvalues}
\end{align}
Above and hereafter, all polarizabilities are quoted in units of $10^{-4}\,\mathrm{fm}^3$.

In the 1970s, Bernab\'eu, Ericson, Ferro Fontan and Tarrach (BEFT)~\cite{Bernabeu:1974zu,Bernabeu:1977hp} derived a sum rule for the dipole polarizability difference $\alpha_{E1}-\beta_{M1}$ of the nucleon, which relates this quantity to the $s$-channel photoabsorption cross section and to the dynamics from the $t$-channel intermediate states such as $\pi\pi$. 
Unlike the well-established forward Baldin (or Baldin--Lapidus) sum rule~\cite{baldin1960polarizability,lapidus1963scattering} for $\alpha_{E1}+\beta_{M1}$, which has long served as a stringent data-driven constraint, the BEFT sum rule has received far less attention; its application was summarized in Ref.~\cite{Schumacher:2005an}. 
Holstein and Nathan~\cite{Holstein:1994tw} performed the first systematic implementation of the BEFT sum rule, incorporating $\pi\pi$ final-state-interaction (FSI) in $\gamma\gamma\to\pi\pi$ by fitting the available experimental data. 
However, Drechsel et al.~\cite{Drechsel:2002ar} later reported a significantly different value of $\alpha_{E1}-\beta_{M1}$ for nucleons.
The discrepancy is almost entirely attributable to the poorly constrained scalar-isoscalar $\pi\pi$ sector, in particular the contribution from the $\sigma$ resonance, also known as the $f_0(500)$. 
If this spread were assigned as a systematic error, the resulting uncertainties in the polarizabilities would far exceed the precision required to discriminate between different theoretical models.
To reconcile this tension, Schumacher~\cite{Schumacher:2006cy,Schumacher:2007ez,Schumacher:2011gi,Schumacher:2019ikn} modeled the $t$-channel contribution, taking narrow-width approximations for the $\sigma$ and $f_0(980)$, thereby obtaining an intermediate value between the two. 
However, such an approach fails to capture the full nonperturbative $\pi\pi$ continuum and rescattering dynamics, which are crucial for an accurate description of the scalar-isoscalar channel; see~\cite{Pelaez:2015qba,Yao:2020bxx} for reviews.
These issues motivate a comprehensive reevaluation of the BEFT sum rule using state-of-the-art experimental and theoretical inputs.

Moreover, because the BEFT and Baldin sum rules are complementary, a combined data-driven treatment can determine $\alpha_{E1}$ and $\beta_{M1}$ separately. 
This is particularly significant in light of the current tension between the proton polarizabilities extracted from the MAMI~\cite{A2:2021wmm} ($\alpha_{E1}^p=10.99 \pm 0.16 \pm 0.47 \pm 0.17 \pm 0.34$, $\beta_{M1}^p=3.14 \pm 0.21 \pm 0.24 \pm 0.20 \pm 0.35$) and $\mathrm{HI}\gamma\mathrm{S}$~\cite{Li:2022vnz} ($\alpha_{E1}^p=13.8 \pm 1.2 \pm 0.1 \pm 0.3$, $\beta_{M1}^p=0.2 \mp 1.2 \pm 0.1 \mp 0.3$) CS measurements.
The same approach also provides a reliable extraction of the nucleon isovector polarizability combination $(\alpha_{E1}-\beta_{M1})^{p-n}$, which has long been the source of one of the largest uncertainties in dispersive estimates of the electromagnetic contribution to the proton-neutron mass difference~\cite{Walker-Loud:2012ift,Thomas:2014dxa,Erben:2014hza,Tomalak:2018dho,Gasser:2015dwa,Gasser:2020hzn,Gasser:2020mzy}, and our result is presented in the companion paper~\cite{Cao:2026bdc}.

This article is organized as follows. 
In Sec.~\ref{sec:Kinematics}, we outline the kinematics and analyticity properties of real Compton scattering (RCS) off nucleons. 
Sec.~\ref{sec:Amps} introduces the invariant amplitudes along with the definitions of nucleon dipole polarizabilities, while Sec.~\ref{sec:DRs} reviews the dispersive formalism for RCS, covering both fixed-$t$ and hyperbolic dispersion relations (DRs). 
In Sec.~\ref{sec:Derivation}, we present a detailed derivation of the BEFT sum rule, including its generalizations to incorporate $t$-channel $K\bar{K}$ and isovector contributions. 
The saturation of the sum rule in the $s$ and $t$ channels is investigated in Secs.~\ref{sec:BEFT_s} and~\ref{sec:BEFT_t}, respectively. 
Sec.~\ref{sec:Results} is devoted to our numerical results, and our main findings are summarized in Sec.~\ref{sec:Summary}. 
Finally, technical details regarding the backward optical theorem, isospin conventions, chiral matching scheme, and dispersively improved Breit--Wigner parameterization are relegated to Appendices~\ref{app:backward_optical},~\ref{app:isospin},~\ref{app:ChPT}, and~\ref{app:BW}.

\section{Kinematics and the Mandelstam plane}\label{sec:Kinematics}

We start with a non-forward RCS off nucleons, $\gamma(q)+N(p)\to \gamma(q')+N(p')$, the familiar Mandelstam variables are 
\begin{align}
    s=(q+p)^2, \quad t=\left(q-q^{\prime}\right)^2, \quad u=\left(q-p^{\prime}\right)^2,
\end{align}
with the constraint $s+t+u=2m_N^2$. The variable $\nu$ is defined by 
\begin{align}
    \nu=\frac{s-u}{4m_N}=\frac{s-m_N^2+t / 2}{2 m_N}=\frac{-u+m_N^2-t / 2}{2 m_N}=\omega+\frac{t}{4m_N}.
\end{align}
In these equations, $\omega$ is the incident photon energy in the laboratory (lab) frame and $m_N$ is the nucleon mass. 
In the lab frame, the kinematic invariant $\nu$ reads $\nu=(\omega+\omega')/2$ with
\begin{align}\label{eq:Energy_lab}
    \omega^{\prime}=\omega+\frac{t}{2 m_N}=\omega\left[1+\frac{\omega}{m_N}(1-z)\right]^{-1},
\end{align}
where $z=\cos\theta_\text{lab}$ is the photon scattering angle, and we also have
\begin{align}\label{eq:stu_lab}
    s=m_N^2+2 m_N \omega, \quad u=m_N^2-2 m_N \omega^{\prime},\quad t=-4\omega\omega' \sin^2\frac{\theta_\text{lab}}{2}.
\end{align}

For the $t$-channel reaction $\gamma\gamma\to N\bar N$ with the center of mass (c.m.) system momenta $q_t$ for photons and $p_t$ for nucleons and scattering angle $z_t=\cos\theta_t$, the c.m. kinematics for $t\geq 4m_N^2$ read
\begin{align}
    q_t=\frac{\sqrt{t}}{2},\quad p_t=\sqrt{\frac{t}{4}-m_N^2},\quad z_t=\frac{m_N\nu}{p_t q_t}.
\end{align}
Note that the relations $q_t^2=t/4$ and $ p_t^2=t/4-m_N^2$ hold regardless of whether $t$ lies above or below the threshold $4m_N^2$.

\begin{figure}[t]
    \centering
    \includegraphics[width=1\linewidth]{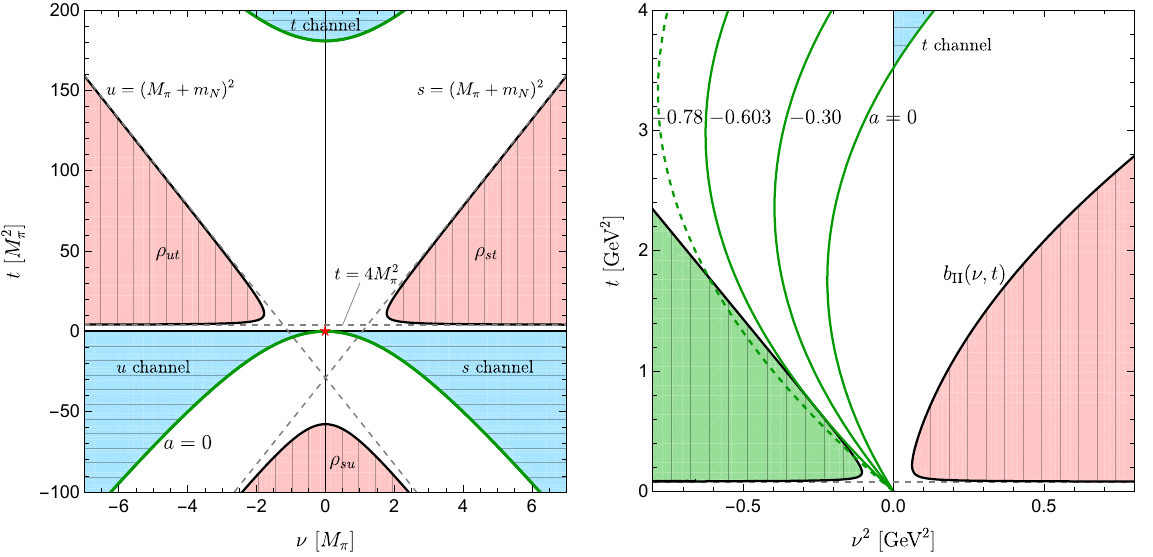}
    \caption{Left panel: The Mandelstam plane for RCS. The physical regions are blue horizontally hatched. The spectral regions are red vertically hatched.
    Right panel: Hyperbolic paths for RCS in the $\nu^2$--$t$ plane. In the right panel, the green vertically hatched region is derived from the semi-major axis of the convergent Lehmann--Martin ellipse, cf. Eq.~\eqref{eq:Nst}.}
    \label{fig:MandelstamPlane}
\end{figure}
The physical regions of various channels in the Mandelstam plane are shown in the left panel of Fig.~\ref{fig:MandelstamPlane} by the blue horizontally hatched areas. 
The boundaries of the physical regions in the $s, u$ and $t$ channels are determined by the zeros of the Kibble function~\cite{Kibble:1960zz}
\begin{align}
    \Phi(s, t, u)=t\left(su-m_N^4\right)=0.
\end{align}
In particular, the RCS experiment takes place in the $s$-channel region, limited by the line $t=0$ (forward scattering) and the lower right part of the hyperbola $su=m_N^4$ or $16 m_N^2 \nu^2=t(t-4m_N^2)$ (backward scattering). 
The $u$-channel physical region is obtained by crossing ($\nu\to -\nu$), and the $t$-channel physical region in the upper part of Fig.~\ref{fig:MandelstamPlane} corresponds to the process $\gamma+\gamma \to N+\bar{N}$ and requires a value of $t \geq 4 m_N^2$.

\begin{figure}[t]
\centering

\begin{minipage}{0.3\textwidth}
\centering
\begin{tikzpicture}[scale=0.7]
    \draw[photon, very thick] (-2,3) to (-1,2);
    \draw[pion, dashed, very thick] (-1,2) to (1,2);
    \draw[photon, very thick] (1,2) to (2,3);
    \draw[nucleon, very thick] (-2,-1) to (-1,0);
    \draw[nucleon, very thick] (-1,0) to (1,0);
    \draw[nucleon, very thick] (1,0) to (2,-1);
    \draw[pion, dashed, very thick] (-1,2) to (-1,0);
    \draw[pion, dashed, very thick] (1,2) to (1,0);
    \node[above left] at (-1.5, 1.9) {$\gamma$};
    \node[left] at (-0.2, 1) {$\pi$};
    \node[below left] at (-1.5, 0.1) {$N$};
\end{tikzpicture}
\\(a) $\rho_{st}$
\label{fig:uni_dia_a}
\end{minipage}
\hfill
\begin{minipage}{0.3\textwidth}
\centering
\begin{tikzpicture}[scale=0.7]
    \draw[photon, very thick] (-2,3) to (-1,2);
    \draw[pion, dashed, very thick] (-1,2) to (1,2);
    \draw[nucleon, very thick] (1,2) to (2,3);
    \draw[nucleon, very thick] (-2,-1) to (-1,0);
    \draw[nucleon, very thick] (-1,0) to (1,0);
    \draw[photon, very thick] (1,0) to (2,-1);
    \draw[pion, dashed, very thick] (-1,2) to (-1,0);
    \draw[nucleon, very thick] (1,0) to (1,2);
\end{tikzpicture}
\\(b) $\rho_{su}$
\label{fig:uni_dia_b}
\end{minipage}
\hfill
\begin{minipage}{0.3\textwidth}
\centering
\begin{tikzpicture}[scale=0.7]
    \draw[photon, very thick] (-2,3) to (-1,2);
    \draw[nucleon, very thick] (-1,2) to (1,2);
    \draw[nucleon, very thick] (1,2) to (2,3);
    \draw[nucleon, very thick] (-2,-1) to (-1,0);
    \draw[pion, dashed, very thick] (-1,0) to (1,0);
    \draw[photon, very thick] (1,0) to (2,-1);
    \draw[nucleon, very thick] (-1,0) to (-1,2);
    \draw[pion, dashed, very thick] (1,2) to (1,0);
\end{tikzpicture}
\\(c) $\rho_{su}$
\label{fig:uni_dia_c}
\end{minipage}

\caption{Unitarity box diagrams that are used to calculate the double spectral regions of $\gamma N \to \gamma N$ scattering.}
\label{fig:uni_dia}
\end{figure}
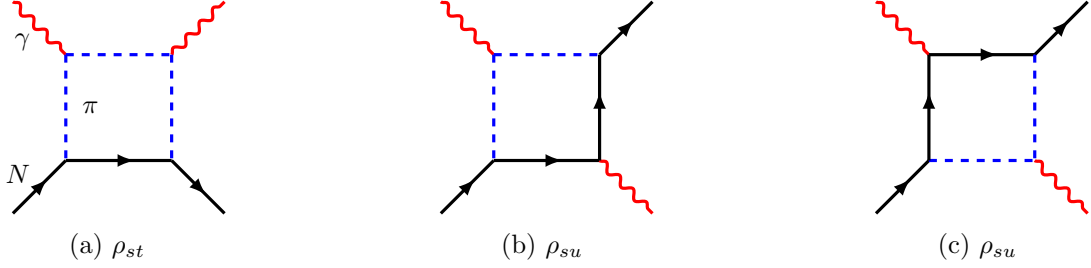

In Fig.~\ref{fig:MandelstamPlane} the red vertically hatched areas are the regions of nonvanishing Mandelstam's double spectral functions~\cite{Drechsel:1999rf}. 
These spectral regions are defined within the Mandelstam plane where two of the three Mandelstam variables $s,t$, and $u$ take on values corresponding to a physical, i.e., on-shell intermediate state. 
The boundaries of these regions are only a consequence of unitarity. 
Therefore, it is sufficient to consider the lowest intermediate states in the unitarity diagrams shown in Fig.~\ref{fig:uni_dia} in all channels. 
For a pedagogical discussion of these aspects, we refer to Sec.~6.3 of the textbook~\cite{Itzykson:1980rh}. 
Regarding the spectral function $\rho_{st}$, we obtain the boundary~\cite{Drechsel:1999rf}
\begin{align}\label{eq:Tst}
    t=T_{s t}=\frac{4 M_\pi^2\left(s-m_N^2\right)^2}{\left(s-\left(m_N-M_\pi\right)^2\right)\left(s-\left(m_N+M_\pi\right)^2\right)},
\end{align}
and for the spectral function $\rho_{su}$ we find
\begin{align}
    t=T_{s u}=\frac{\left(s-m_N^2\right)^2}{\left(m_N+M_\pi\right)^2-s}.
\end{align}
The boundary of the spectral function $\rho_{ut}$ follows from crossing symmetry. 

\section{Invariant amplitudes and nucleon polarizabilities}\label{sec:Amps}

The invariant amplitudes for RCS can be constructed as\footnote{The differential cross section in terms of the amplitude $T_{fi}$ is given by $\frac{\mathrm{d} \sigma}{\mathrm{d} \Omega}=\frac{1}{64 \pi^2 s} \frac{1}{4}\sum_{\text{spins}}\left|T_{f i}\right|^2$.}
\begin{align}\label{eq:Tfi}
    T_{f i}=\epsilon_\mu^{\prime *}(q') \bar{u}(p') T^{\mu \nu} u(p)\epsilon_\nu(q),
\end{align}
where $\epsilon$ and $\epsilon'$ are the polarization vectors of the incoming and outgoing photons, respectively, and $u$ and $\bar u$ are the nucleon spinors. 
The RCS tensor can be expressed on an orthogonal basis suggested by Prange~\cite{PhysRev.110.240} by means of six invariant amplitudes $T_i(\nu,t), i=1\cdots6$,
\begin{align}\label{eq:Prange}
    T^{\mu\nu}=&-\frac{P^{\prime \mu} P^{\prime \nu}}{P^{\prime 2}}\left(T_1+\slashed{K} T_2\right)-\frac{N^\mu N^\nu}{N^2}\left(T_3+\slashed{K} T_4\right) \nonumber\\
    &+i \frac{P^{\prime \mu} N^\nu-P^{\prime \nu} N^\mu}{P^{\prime 2} K^2} \gamma_5 T_5+i \frac{P^{\prime \mu} N^\nu+P^{\prime \nu} N^\mu}{P^{\prime 2} K^2} \gamma_5 \slashed{K} T_6,
\end{align}
where
\begin{align}
\begin{aligned}
    P^\mu &=\frac{1}{2}\left(p+p^{\prime}\right)^\mu,\quad K^\mu=\frac{1}{2}\left(q+q^{\prime}\right)^\mu,\quad Q^\mu=\frac{1}{2}\left(q^{\prime}-q\right)^\mu,\\
    P^{\prime \mu} &=P^\mu-\frac{(P \cdot K)}{K^2} K^\mu,\quad N^\mu=\epsilon^{\mu \nu \alpha \beta} P_\nu^{\prime} Q_\alpha K_\beta,
\end{aligned}
\end{align}
with $\epsilon^{\mu \nu \alpha \beta}$ being the antisymmetric Levi-Civita tensor with the sign convention $\epsilon^{0123}=-1$. 
The amplitudes $T_i(\nu,t)$ do not have kinematic singularities. 
However, the Prange amplitudes suffer from kinematic constraints leading to linear dependencies at certain values of $\nu$ and $t$, which arise from the vanishing of $P^{\prime 2}, N^2$ and $P^{\prime 2} K^2$ in the denominators of the decomposition~\eqref{eq:Prange}. 
L'vov~\cite{Lvov:1980wp} therefore proposed a different tensor basis, leading to the linear combinations of $T_i$,
\begin{align}
    & A_1=\frac{1}{t}\left[T_1+T_3+\nu\left(T_2+T_4\right)\right], \quad A_2=\frac{1}{t}\left[2 T_5+\nu\left(T_2+T_4\right)\right], \nonumber\\
    & A_3=\frac{m_N^2}{m_N^4-s u}\left[T_1-T_3-\frac{t}{4 \nu}\left(T_2-T_4\right)\right], \quad A_4=\frac{m_N^2}{m_N^4-s u}\left[2 m_N T_6-\frac{t}{4 \nu}\left(T_2-T_4\right)\right], \nonumber\\
    & A_5=\frac{1}{4 \nu}\left[T_2+T_4\right], \quad A_6=\frac{1}{4 \nu}\left[T_2-T_4\right] .
\end{align}
The L'vov amplitudes $A_i(\nu, t)$ are even functions of $\nu$ and have no kinematic singularities or kinematic constraints.
They have poles at $s=m_N^2$ and $u=m_N^2$ due to the elastic nucleon intermediate state, i.e., the Born terms $A_i^\text{B} (\nu,t)$~\cite{Lvov:1996rmi}. 
The electric and magnetic static dipole polarizabilities $\alpha_{E1}, \beta_{M1}$ are defined by subtracting the Born terms $A_i^\text{B}$ from the L'vov amplitudes $A_i$ at the origin of the Mandelstam plane, $\nu=t=0$ (see the red star in Fig.~\ref{fig:MandelstamPlane})~\cite{Lvov:1996rmi}
\begin{align}\label{eq:EMpola_def}
    \alpha_{E1}+\beta_{M1}=-\frac{1}{2 \pi}\left[A_3^{\mathrm{nB}}(0,0)+A_6^{\mathrm{nB}}(0,0)\right], \quad \alpha_{E1}-\beta_{M1}=-\frac{1}{2 \pi}\left[A_1^{\mathrm{nB}}(0,0)\right],
\end{align}
where $A_i^{\text{nB}}(\nu, t)=A_i(\nu, t)-A_i^{\text{B}}(\nu, t)$ are the non-Born parts of the invariant amplitudes. 
The L'vov amplitudes can be reduced~\cite{Babusci:1998ww,Schumacher:2005an} to the spin-independent Gell-Mann--Goldberger--Thirring (GGT) amplitudes $f(\omega)$~\cite{Gell-Mann:1954ttj},
\begin{align}
    \label{eq:f0} f_0(\omega) &=-\frac{\omega^2}{2 \pi}\left[A_3(\omega, 0)+A_6(\omega, 0)\right],\quad \\
    \label{eq:fpi} f_\pi(\omega) &=-\frac{\omega \omega^{\prime}}{2 \pi}\left(1+\frac{\omega \omega^{\prime}}{m_N^2}\right)^{1 / 2}\left[A_1\left(\omega\frac{1+\frac{\omega}{m_N}}{1+2 \frac{\omega}{m_N}}, -4 \omega \omega'\right)-\frac{t}{4 m_N^2} A_5\left(\omega\frac{1+\frac{\omega}{m_N}}{1+2 \frac{\omega}{m_N}}, -4 \omega \omega'\right)\right],
\end{align}
where the subscripts ``$0,\pi$" correspond to the forward ($\theta_\text{lab}=0$) and backward ($\theta_\text{lab}=\pi$) direction, respectively. 
This reflects the fact that the combinations in~\eqref{eq:EMpola_def} correspond to forward and backward Compton scatterings, respectively.

\section{Fixed-$t$ and hyperbolic dispersion relations}\label{sec:DRs}

The invariant amplitudes $A_i$ are free of kinematical singularities and zeros, and obey the crossing symmetry $A_i(-\nu,t)=A_i(\nu,t)$. 
Considering further analyticity, these amplitudes fulfill the fixed-$t$ DR, provided the appropriate high-energy behaviour is ensured. 
To evaluate the high-energy asymptotic behaviour for RCS, we invoke the predictions of Regge theory for $\nu\to\infty$ and fixed-$t$~\cite{Lvov:1996rmi,Drechsel:2002ar}:
\begin{align}\label{eq:Regge}
    \begin{aligned}
    & A_1, A_2 \sim \nu^{\alpha_{\mathrm{M}}(t)}, \quad\left(A_3+A_6\right) \sim \nu^{\alpha_{\mathrm{P}}(t)-2}, \\
    & A_3, A_5 \sim \nu^{\alpha_{\mathrm{M}}(t)-2}, \quad A_4 \sim \nu^{\alpha_{\mathrm{M}}(t)-3},
    \end{aligned}
\end{align}
where $\alpha_{\mathrm{M}}(t) \sim 0.5$ (for $t \leqslant 0$ ) is a meson Regge trajectory, and where $\alpha_{\mathrm{P}}(t)$ is the Pomeron trajectory which has an intercept $\alpha_{\mathrm{P}}(0) \simeq 1$. 
Note that the Pomeron dominates the high-energy behaviour of the combination of $A_3+A_6$. 
We expect that the invariant amplitudes $A_{3,4,5,6}$ are given by the unsubtracted fixed-$t$ dispersion integrals, \begin{align}\label{eq:unsub_DR}
    A_i^{\mathrm{nB}}(\nu,t)=\frac{1}{\pi} \int_{\nu_\text{thr}}^{\infty} \mathrm{d} \xi\left[\frac{1}{\xi-\nu-i 0^{+}}+ \frac{1}{\xi+\nu-i 0^{+}}\right] \operatorname{Im}_s A_i(\xi,t)=\frac{2}{\pi} \int_{\nu_\text{thr}}^{\infty} \mathrm{d}\xi~\frac{\xi \operatorname{Im}_s A_i(\xi,t)}{\xi^2-\nu^2-i 0^{+}},
\end{align} 
where $\operatorname{Im}_s$ corresponds to the $s$-channel discontinuities. 
A precise estimate can be obtained in the particular case of forward RCS by using the optical theorem, which reads in terms of the amplitudes $A_i$ as
\begin{align}
    \operatorname{Im}_s \left[A_3(\nu, 0)+A_6(\nu, 0)\right]=-\frac{\sigma_\text{tot}(\nu)}{2\nu},
\end{align}
using Eq.~\eqref{eq:f0} and $\operatorname{Im}_s f_0(\nu)=\frac{\nu}{4\pi}\sigma_\text{tot}(\nu)$~\cite{Gell-Mann:1954ttj}. One can immediately obtain the Baldin sum rule~\cite{baldin1960polarizability,lapidus1963scattering} using Eq.~\eqref{eq:EMpola_def},
\begin{align}\label{eq:BL_SR}
    \alpha_{E1}+\beta_{M1}=-\frac{1}{\pi^2} \int_{\nu_\text{thr}}^{\infty} \mathrm{d}\nu~\frac{\nu \operatorname{Im}_s \left[A_3(\nu,0)+A_6(\nu,0)\right]}{\nu^2}=\frac{1}{2 \pi^2} \int_{\nu_{\text{thr}}}^{\infty} \md\nu~\frac{\sigma_{\text{tot}}(\nu)}{\nu^2}.
\end{align}
It provides the most stringent empirical constraint on the sum of nucleon polarizabilities.

From the asymptotic behaviour of Eq.~\eqref{eq:Regge}, one must introduce subtractions for the DRs of the amplitudes $A_1$ and $A_2$. 
The divergence of the unsubtracted dispersive integrals is essentially caused by $t$-channel $\pi^0$ exchange for $A_2$ and by the $\sigma$ enhancement in the $\pi\pi$ $S$ wave for $A_1$~\cite{Lvov:1996rmi}, which encodes the correlated two-pion continuum with quantum numbers $I=J=0$.
The convergence can be ensured by considering once-subtracted fixed-$t$ DRs at $\nu=0$ as proposed in Ref.~\cite{Drechsel:1999rf}.\footnote{Just recently, this approach based on the subtracted fixed-$t$ DR has been successfully extended to virtual Compton scattering off the proton~\cite{Danilkin:2026wkl}.} 
The price to pay for this approach is the appearance of the subtraction functions (combinations of polarizabilities) that have to be fitted from the data. 
Unfortunately, the predictive power of the subtracted DRs for the polarizabilities is greatly weakened by the lack of full consistency among the existing low-energy RCS data~\cite{Krupina:2017pgr}. 

From a theoretical perspective, fixed-$t$ DRs have the shortcomings that the dispersion integrals get contributions from the unphysical region between the boundaries of the physical $s$- and $u$-channel regions, see Fig.~\ref{fig:MandelstamPlane}. 
Moreover, in fixed-$t$ DRs the crossed-channel contributions are always ambiguous, and properly including them without double counting is  challenging~\cite{Lvov:1996rmi,Drechsel:1999rf}. 
In order to improve the convergence for larger values of $s$ and to take into account the contribution from the $t$-channel consistently without double counting, hyperbolic DRs~\cite{Hite:1973pm,Hohler:1983} have been applied to RCS~\cite{Drechsel:2002ar}. 
A natural choice is the family of $s$--$u$ crossing-symmetric hyperbolas 
\begin{align}
    (s-a)(u-a)=b,
\end{align}
which do not enter the double spectral function regions (see Fig.~\ref{fig:MandelstamPlane}). 
Two special cases of hyperbolic DRs are particularly useful. The first is fixed-$b$ DRs, used in partial-wave dispersive analyses,\footnote{The parameter $a$ is usually chosen to maximize the domain of validity of the hyperbolic dispersive representation in a specified region on the complex energy plane.} e.g., Roy--Steiner equation analyses~\cite{Hohler:1983,Buettiker:2003pp,Ditsche:2012fv,Hoferichter:2015hva,Pelaez:2020gnd,Cao:2022zhn,Cao:2025hqm}. 
The second is known as interior DRs~\cite{Hohler:1983}, because all hyperbolae pass through the interior of the physical regions of the three channels. 
In the latter case, the relation $b=\lambda(0,m_N^2,a)$ holds, i.e.,
\begin{align}\label{eq:hyperbola}
    (s-a)(u-a)=\lambda(0,m_N^2,a),
\end{align}
where $\lambda(x,y,z)=x^2+y^2+z^2-2(xy+xz+yz)$ is the standard K\"all\'en triangle function. 
It is easy to verify that all hyperbolae in Eq.~\eqref{eq:hyperbola} pass through the $s$- and $u$-channel thresholds; the hyperbola corresponding to $a=0$ also passes through the $t$-channel threshold, see Fig.~\ref{fig:MandelstamPlane}. From the energy shift formula~\eqref{eq:Energy_lab}, one obtains
\begin{align}
    (s-a)(u-a)=\lambda(0,m_N^2,a)+\left[a+m_N^2 \frac{1+\cos \theta_{\rm{lab}}}{1-\cos \theta_{\rm{lab}}}\right] t.
\end{align}
Therefore, along each of the hyperbolae, the photon scattering angle $\theta_{\rm{lab}}$ in the lab frame is fixed. The parameter $a$ appearing in Eq.~\eqref{eq:hyperbola} is then given by 
\begin{align}
    a=-m_N^2 \frac{1+\cos \theta_{\rm{lab}}}{1-\cos \theta_{\rm{lab}}},
\end{align}
establishing a one-to-one correspondence between $a$ and the lab scattering angle.
Consequently, DRs based on Eq.~\eqref{eq:hyperbola} are also called fixed-angle DRs, as first discussed in nucleon CS by Hearn and Leader~\cite{Hearn:1962zz} and later investigated in detail by K\"oberle~\cite{Koberle:1968zz}. 
Along a path at fixed $a$, one can write down a dispersion integral as~\cite{Hite:1973pm}
\begin{align}\label{eq:HDR}
    A_i(s, t, a)= {}& A_i^{\rm{B}}(s, t, a)+A_i^{t\text {-pole}}(s, t, a) \nonumber\\
    & +\frac{1}{\pi} \int_{\left(m_N+M_\pi\right)^2}^{\infty} \mathrm{d}s^{\prime} \operatorname{Im}_s A_i\left(s^{\prime}, t^\prime, a\right)\left[\frac{1}{s^{\prime}-s}+\frac{1}{s^{\prime}-u}-\frac{1}{s^{\prime}-a}\right] \nonumber\\
    & +\frac{1}{\pi} \int_{4 M_\pi^2}^{\infty} \mathrm{d} t^{\prime} \frac{\operatorname{Im}_t A_i\left(s^\prime, t^{\prime}, a\right)}{t^{\prime}-t},
\end{align}
where the external kinematics $(s,t,u)$ and the internal kinematics $(s',t',u')$ are related by 
\begin{align}
    (s-a)(u-a)=(s'-a)(u'-a)=\lambda(0,m_N^2,a),\quad s+t+u=s'+t'+u'=2m_N^2.
\end{align}
The amplitude $A_i^{t\text {-pole}}(s, t, a)$ entering into Eq.~\eqref{eq:HDR} describes the contribution of pseudoscalar mesons, $\pi,\eta,\eta'$, see Refs.~\cite{Lvov:1998jza,Schumacher:2007ez} for further discussion. 
Some curves corresponding to fixed values of $a$ are shown in the right panel of Fig.~\ref{fig:MandelstamPlane}. 
Not all values of the parameter $a$ yield physically acceptable hyperbolae. 
The most restrictive requirement in nucleon CS is that the partial-wave expansion of the imaginary part $\operatorname{Im}_t A_i$ entering the $t$-channel dispersive integral in Eq.~\eqref{eq:HDR} must converge. Hence the representation is valid only inside the Lehmann--Martin ellipse~\cite{Lehmann:1958ita,Martin:1965jj,Martin:1966zsy}. 
For $t\geq 4M_\pi^2$, the constraint is given by~\cite{Ditsche:2012fv,Pelaez:2020gnd}
\begin{align}
    \frac{p_t^2 q_t^2}{m_N^2}-N_{st}^2(t)\leq \nu^2\leq N_{st}^2(t),
\end{align}
where the upper bound of $\nu$ can be obtained by the explicit expression $T_{st}$ [cf.~Eq.~\eqref{eq:Tst}],
\begin{align}\label{eq:Nst}
    N_{st}(t)=\frac{t(t-2M_\pi^2)+4M_\pi\sqrt{t}\sqrt{(t-4M_\pi^2)m_N^2+M_\pi^4}}{4m_N(t-4M_\pi^2)}.
\end{align}
Therefore, the $t$-channel partial-wave expansion of $\operatorname{Im}_t A_i$ converges if and only if $-0.603~\text{GeV}^2 \leq a\leq 0$,\footnote{Our lower bound $-0.603~\text{GeV}^2$ differs slightly from the value $-0.594~\text{GeV}^2$ quoted in Ref.~\cite{Drechsel:2002ar}, because we take the charged-state masses $m_N=m_p=938.27$~MeV and $M_\pi=M_{\pi^{\pm}}=139.57$~MeV as reference unless otherwise specified.} corresponding to $101^\circ\leq \theta_\text{lab}\leq 180^\circ$~\cite{Drechsel:2002ar}. 
This shows that hyperbolic (fixed-angle) DRs cannot be applied at all angles with equal reliability. 
In particular, hyperbolic DRs are ideally suited to backward angles ($\theta_\text{lab}= 180^\circ$), corresponding to $a=0$~\cite{Drechsel:2002ar,Schumacher:2005an}. 
Most importantly, hyperbolic dispersion theory naturally avoids double counting of physical contributions between the $s$- ($u$-) and $t$-channels in a crossing symmetric way, whereas this issue is ambiguous for fixed-$t$ dispersion theory.

\section{Derivation of the BEFT sum rule}\label{sec:Derivation}

In this section, we present the derivation of the BEFT sum rule considering the decomposition into parts of the $s$- and $t$-channels. 
In particular, for the $t$-channel part, we provide a comprehensive extension beyond the $\pi\pi$ intermediate state, including the isoscalar ($I=0$) $K\bar{K}$ channel and the isovector ($I=1$) $\pi\eta, K\bar{K}$ channels.

Our goal is to construct a Baldin-like sum rule for the difference $\alpha_{E1}-\beta_{M1}$. 
From Eq.~\eqref{eq:EMpola_def}, one may write a DR for the amplitude $A_1$ directly, while a preferable choice is the combination $A_1-\frac{t}{4m_N^2}A_5$ corresponding to the backward GGT amplitude $f_\pi$, cf.~\eqref{eq:fpi}. 
We define the invariant amplitude combination 
\begin{align}\label{eq:A1new}
    \tilde{A}_1(\nu,t)=A_1(\nu,t)-\frac{t}{4m_N^2}A_5(\nu,t),
\end{align}
and apply the hyperbolic DR for $a=0$. 
Then we arrive at
\begin{align}\label{eq:HDR_A1}
    \tilde{A}_1^\mathrm{nB}(\nu, t)= {}& \frac{1}{\pi} \int_{\left(m_N+M_\pi\right)^2}^{\infty}\md s'~\left(\frac{1}{s^{\prime}-s}+\frac{1}{s^{\prime}-u}-\frac{1}{s^{\prime}}\right) \operatorname{Im}_s \tilde{A}_1\left(\nu', t^{\prime}\right) \nonumber\\
    & +\frac{1}{\pi} \int_{4M_\pi^2}^{\infty}\frac{\md t'}{t^{\prime}-t} \operatorname{Im}_t \tilde{A}_1\left(\nu', t^{\prime}\right),
\end{align}
where $s' u'=m_N^4$. 
Assuming for the GGT amplitude $f_\pi$ a standard Regge behavior $\sim s^{\alpha_\Delta(u)}$ for $s\to +\infty, t\to -\infty, u\to a$~\cite{Ditsche:2012fv}, we have [cf.~Eq.~\eqref{eq:stu_lab} and~\eqref{eq:fpi}] 
\begin{align}
    \tilde{A}_1 \sim s^{\alpha_\Delta(0)-3 / 2}.
\end{align}
Here $\alpha_\Delta(0)\simeq 0.03$ for the leading $\Delta_\delta$ (isospin $I=3/2$, parity $P=+1$ and signature $(-1)^{J-1/2}=-1$) Regge trajectory~\cite{Huang:2010aro}. 
Such a high-energy behavior of $\tilde{A}_1$ guarantees convergence of the $s$-channel part of the hyperbolic DR~\eqref{eq:HDR_A1}. 
The discussion of the $t$‐channel integral in Eq.~\eqref{eq:HDR_A1} is analogous.
Using Eqs.~\eqref{eq:EMpola_def} and~\eqref{eq:HDR_A1}, one can write down an unambiguous decomposition of the difference $\alpha_{E1}-\beta_{M1}$,
\begin{align}
\label{eq:HDR_diff_s} & (\alpha_{E1}-\beta_{M1})^s=-\frac{1}{2 \pi^2} \int_{\left(m_N+M_\pi\right)^2}^{\infty}\frac{\mathrm{d} s^{\prime}}{s^{\prime}}~ \frac{s^{\prime}+m_N^2}{s^{\prime}-m_N^2} \operatorname{Im}_s \tilde{A}_1\left(\nu', t^{\prime}\right), \\
\label{eq:HDR_diff_t} & (\alpha_{E1}-\beta_{M1})^t=-\frac{1}{2 \pi^2} \int_{4M_\pi^2}^{\infty}\frac{\mathrm{d} t^{\prime}}{t^{\prime}}~ \operatorname{Im}_t \tilde{A}_1\left(\nu', t^{\prime}\right).
\end{align}

\subsection{The $s$-channel part of the BEFT sum rule}

Note that Eq.~\eqref{eq:fpi} relates the invariant amplitude $\tilde{A}_1$ in backward kinematics to the backward spin-independent amplitude $f_\pi(\omega)$. 
Consequently, the $s$-channel imaginary part $\operatorname{Im}_s \tilde{A}_1$ can be expressed in terms of the absorptive part of $f_\pi(\omega)$. 
To derive the backward unitarity relation, it is convenient to work in the c.m. frame with total energy $W=\sqrt{s}$ and photon momentum $k=m_N\omega/W$, where the lab [$f_\pi(\omega)$] and c.m. [$\hat{f}_\pi(W)$] frame spin-independent amplitudes are related by $f_\pi(\omega)=(W/m_N)\hat{f}_\pi(W)$. 
Labeling the two-body $\gamma N$ helicity state by $a=(\lambda,h)$, with photon helicity $\lambda=\pm 1$, nucleon helicity $h=\pm 1/2$, and net helicity $\mu_a=\lambda-h$, angular-momentum conservation in the backward direction ($\theta=\pi$) requires both helicities to flip, leading to the final state $\bar{a}=(-\lambda,-h)$ with $\mu_{\bar{a}}=-\mu_a$. 
Projecting the c.m. helicity amplitude $\hat{F}_{\bar{a}a}(\theta)$ onto the spin-independent structure and expanding in partial waves of total angular momentum $J$, we obtain
\begin{align}
    \hat{f}_\pi(W) = \frac{1}{4}\sum_{a} B_a \hat{F}_{\bar{a}a}(\pi) = \frac{1}{4}\sum_{J, a} (2J+1) B_a d^J_{\mu_a,-\mu_a}(\pi) \hat{F}^J_{\bar{a}a}(W)\,,
\end{align}
where $d^J_{\mu_a,-\mu_a}(\pi)=(-1)^{J+\mu_a}$ is the Wigner $d$-function at $\theta=\pi$, and $B_a = -(-1)^{1/2-h}$ is the spin-projection factor connecting the canonical amplitude to the helicity basis.

Since the two-body helicity states $|JM;a\rangle$ are not parity eigenstates, we relate them to states of definite parity by coupling the electric ($E_L$) and magnetic ($M_L$) photon multipoles,
\begin{align}
    |L m; E/M\rangle = \frac{1}{\sqrt{2}}\left(|L m; \lambda=+1\rangle \pm |L m; \lambda=-1\rangle\right)\,,
\end{align}
which carry parities $(-1)^L$ and $(-1)^{L+1}$, respectively, to the positive-parity nucleon spin to form total angular momentum $J=L\pm 1/2$. 
The reflection property of the resulting Clebsch--Gordan coefficients implies that under parity $\mathcal{P}$, the partial-wave helicity states transform as $\mathcal{P}|JM;a\rangle = \xi_J |JM;\bar{a}\rangle$ with $\xi_J=(-1)^{J+1/2}$. 
Decomposing the $s$-channel partial-wave unitarity relation into intermediate states of definite total parity $P=\pm 1$, parity conservation dictates that the helicity-flip absorptive part is related to the diagonal one by $(\operatorname{Im}\hat{F}^J)^P_{\bar{a}a} = P \xi_J (\operatorname{Im}\hat{F}^J)^P_{aa}$. 
Remarkably, the product of the spin-projection factor, backward Wigner $d$-function, and parity phase is independent of $J$, $\lambda$, and $h$:
\begin{align}
    B_a d^J_{\mu_a,-\mu_a}(\pi) \xi_J = -(-1)^{1/2-h}(-1)^{J+\lambda-h}(-1)^{J+1/2} = -1\,.
\end{align}
Summing over all partial waves and applying the optical theorem to the diagonal absorptive parts in each parity sector, the c.m. amplitude becomes $\operatorname{Im}\hat{f}_\pi(W) = \frac{k}{4\pi}[\sigma(W, P=-1) - \sigma(W, P=+1)]$ (see Appendix~\ref{app:backward_optical} for a more detailed derivation). 
Converting back to the lab frame via $f_\pi(\omega)=(W/m_N)\hat{f}_\pi(W)$ and $(W/m_N)k=\omega$, we arrive at
\begin{align}\label{eq:fpi_Im}
    \operatorname{Im} f_\pi(\omega)=\frac{\omega}{4 \pi}\left[\sigma(\omega, \Delta P=\text{yes})-\sigma(\omega, \Delta P=\text{no})\right]\,,
\end{align}
where $\Delta P=\text{yes}$ ($P=-1$ for $E1, M2, E3, \cdots$) and $\Delta P=\text{no}$ ($P=+1$ for $M1, E2, M3, \cdots$) denote photoexcitation processes with and without a parity change relative to the ground-state nucleon, respectively. 
In other words, photoexcitations preserving the nucleon parity ($\Delta P=\text{no}$) contribute to the polarizability difference with a sign opposite to those that flip the parity ($\Delta P=\text{yes}$).
Inserting the optical theorem for $f_\pi$~\eqref{eq:fpi_Im} and the relation~\eqref{eq:fpi} into Eq.~\eqref{eq:HDR_diff_s}, we obtain
\begin{align}\label{eq:BEFT_s}
    (\alpha_{E1}-\beta_{M1})^s=\frac{1}{2 \pi^2} \int_{M_\pi+\frac{M_\pi^2}{2 m_N}}^{\infty}\frac{\mathrm{d} \omega}{\omega^2}~\sqrt{1+\frac{2 \omega}{m_N}}[\sigma(\Delta P=\mathrm{yes})-\sigma(\Delta P=\mathrm{no})].
\end{align}
This is the $s$-channel part of the BEFT sum rule, first obtained by Bernab\'eu et al.~\cite{Bernabeu:1974zu}, although no proof was given in that paper.\footnote{We remark that the first term on the right-hand side of Eq.~(7) in Ref.~\cite{Bernabeu:1974zu} should not be present and the corrected expression is provided in Ref.~\cite{Bernabeu:1977hp}.} 
Unlike the total photoabsorption cross section, this integrand entering the $s$-channel part of the BEFT sum rule is written in terms of the cross section difference $\sigma(\Delta P=\mathrm{yes})-\sigma(\Delta P=\mathrm{no})$, which is not directly experimentally observable and is inconvenient for practical calculations. 
The main contribution to $(\alpha_{E1}-\beta_{M1})^s$ comes from the photoproduction of $\pi N$ states~\cite{Lvov:1996rmi,Drechsel:1999rf}.
For these states, the cross section difference can be expressed in terms of the standard Chew--Goldberger--Low--Nambu (CGLN) amplitudes~\cite{Chew:1957tf} via\footnote{Here, our conventions for the CGLN amplitudes follow those of Ref.~\cite{Drechsel:1992pn}.}
\begin{align}\label{eq:sigma_CGLN}
{[\sigma(\omega, \Delta P=\text{yes})} & -\sigma(\omega, \Delta P=\text{no})]^{\pi N}=2 \pi \frac{q_\pi}{k} \sum_{\ell=0}^{\infty}(-1)^\ell(\ell+1)^2 \nonumber\\
& \times\left[(\ell+2)\left(\left|E_{\ell+}\right|^2-\left|M_{(\ell+1)-}\right|^2\right)+\ell\left(\left|M_{\ell+}\right|^2-\left|E_{(\ell+1)-}\right|^2\right)\right]^{\pi N},
\end{align}
where
\begin{align}
    q_\pi=|\mathbf{q}_\pi|=\frac{\sqrt{\lambda(s,M_\pi^2,m_N^2)}}{2\sqrt{s}}\,,\quad k=|\mathbf{k}_\gamma|=\frac{\sqrt{\lambda(s,0,m_N^2)}}{2\sqrt{s}}\,,
\end{align}
are the pion and photon momenta in the c.m. system, respectively. 
The final state $\pi N$ of photoproduction is described by an orbital angular momentum $\ell$ of the pion relative to the recoiling nucleon, with parity $(-1)^{\ell+1}$ after considering the pion intrinsic parity. 
The total angular momentum of the final state, $J$, must be equal to the total angular momentum of the initial state, $J=|\ell\pm 1/2|=|L\pm 1/2|$. 
Using parity conservation, we find for the electric multipole $(-1)^L=(-1)^{\ell+1} \rightarrow L=\ell \pm 1$ and for the magnetic multipole $(-1)^{L+1}=(-1)^{\ell+1} \rightarrow L=\ell$, respectively. 
The allowed multipoles and the corresponding quantum numbers are summarized in Table~\ref{tab:multipole}.
\begin{table}[t]
    \centering
    \caption{Allowed multipoles for photoproduction of pions from nucleons.}
    \label{tab:multipole}
    \begin{tabular}{cccccc}
    \hline\hline 
    $J$ & $L$ & Parity & Multipole & Notation & Minimum $\ell$ \\
    \hline
    $\ell+1/2$ & $\ell+1$ & $(-1)^L$ & Electric $2^{\ell+1}$ & $E_{\ell +}$ & 0 \\
    $\ell-1/2$ & $\ell-1$ & $(-1)^L$ & Electric $2^{\ell-1}$ & $E_{\ell -}$ & 2 \\
    $\ell+1/2$ & $\ell$ & $(-1)^{L+1}$ & Magnetic $2^\ell$ & $M_{\ell +}$ & 1 \\
    $\ell-1/2$ & $\ell$ & $(-1)^{L+1}$ & Magnetic $2^{\ell}$ & $M_{\ell -}$ & 1 \\
    \hline\hline
    \end{tabular}
\end{table}
The discontinuity of the $\pi N$ intermediate state can also be derived directly from the standard relativistic helicity-amplitude formalism for nucleon CS listed in Appendix~B of Ref.~\cite{Lvov:1996rmi}. 
As a consistency check, we also reproduce Eqs.~\eqref{eq:HDR_diff_s} and~\eqref{eq:sigma_CGLN}.

\subsection{The $t$-channel part of the BEFT sum rule}

The $t$-channel part~\eqref{eq:HDR_diff_t} cannot be recast directly through cross sections because a large part of the integration region is unphysical, that is, $4M_\pi^2\leq t< 4m_N^2$. 
Nevertheless, the key point is that the imaginary part in Eq.~\eqref{eq:HDR_diff_t} is governed by $t$-channel intermediate states with even total angular momentum $J$. 
This selection rule follows from the Bose symmetry of the two-photon state (including the Landau--Yang theorem), charge conjugation $C=+1$, natural parity $P=(-1)^J=+1$ associated with $\tilde{A}_1$, and $t$-channel isospin $I_t\leq 1$ restricted by the $N\bar{N}$ state, i.e.,~\cite{Schumacher:2005an}
\begin{align}\label{eq:JPC}
    J^{PC}=0^{++},\ 2^{++},\cdots,\quad I_t=0,1.
\end{align}
Among the lightest hadronic states carrying the quantum numbers~\eqref{eq:JPC} that contribute to $\operatorname{Im}_t \tilde{A}_1$ are the two-pseudoscalar-meson continua: $\pi\pi$ and $K\bar K$ for $I_t=0$, and $\pi\eta$ and $K\bar K$ for $I_t=1$. 
Other two-body continua, such as $\eta\eta$ and $\pi\eta'$, are negligible here, since their couplings to the lower channels are rather weak~\cite{Dudek:2016cru,Briceno:2017qmb}.

We next evaluate the $\pi\pi$ intermediate state, which dominates the integral in Eq.~\eqref{eq:HDR_diff_t} below $K\bar K$ threshold. 
To calculate this contribution with the unitarity relation, we have to construct the partial-wave amplitudes $\gamma \gamma \rightarrow \pi \pi$ and $\pi \pi \rightarrow \bar{N} N$. Following the derivation in Appendix~B of Ref.~\cite{Drechsel:1999rf} (see also Appendix~A of Ref.~\cite{Drechsel:2002ar}), we start by decomposing the $t$-channel helicity amplitude for $\gamma\gamma\to \bar{N} N$ into a partial wave series,
\begin{align}
    T_{\lambda_{\bar{N}} \lambda_N, \lambda_\gamma \lambda_\gamma^{\prime}}^t(\nu, t)=\sum_{J} \frac{2 J+1}{2} T_{\lambda_{\bar{N}} \lambda_N, \lambda_\gamma \lambda_\gamma^{\prime}}^{J(\gamma \gamma \rightarrow \bar{N}N)}(t) d_{\Lambda_\gamma \Lambda_N}^J\left(\theta_t\right),
\end{align}
where $\Lambda_\gamma=\lambda_\gamma-\lambda_\gamma',\, \Lambda_N=\lambda_{\bar{N}}-\lambda_N$ and $\theta_t$ is the scattering angle in the $t$-channel, which is related to $\nu$ and $t$ by $\cos\theta_t=\frac{4 m_N \nu}{\sqrt{t(t-4 m_N^2)}}$. 
The unitarity relation from the Cutkosky cutting rule~\cite{Cutkosky:1960sp} (or the discontinuity calculus~\cite{Jing:2025qmi}) reads
\begin{align}\label{eq:unitarity_v1}
    2 \operatorname{Im} T^{\gamma \gamma \rightarrow \bar{N}N}=\frac{1}{(4 \pi)^2} \frac{p_\pi}{\sqrt{t}} \int \md \Omega_\pi\left[T^{\gamma \gamma \rightarrow \pi \pi}\right] \left[T^{\pi \pi \rightarrow \bar{N}N}\right]^*,
\end{align}
where $p_\pi=\left|\mathbf{p}_\pi\right|=\sqrt{t/4-M_\pi^2}$ is the c.m. momentum of the pion. The partial-wave expansion for $\gamma\gamma \rightarrow \pi \pi$ reads
\begin{align}\label{eq:PW_gg->pipi}
    T_{\Lambda_\gamma}^{\gamma \gamma \rightarrow \pi \pi}\left(t, \theta_{\pi \pi}\right)= \sum_{J\,\text{even}} \frac{2 J+1}{2} T_{\Lambda_\gamma}^{J\left(\gamma \gamma \rightarrow \pi \pi\right)}(t)\sqrt{\frac{\left(J-\Lambda_\gamma\right)!}{\left(J+\Lambda_\gamma\right)!}} P_J^{\Lambda_\gamma}\left(\cos \theta_{\pi \pi}\right).
\end{align}
Due to parity arguments, only two helicity modes $\lambda_\gamma=+1, \lambda_\gamma'=\pm 1$ ($\Lambda_\gamma=0,2$) are independent. These two-photon-related helicity amplitudes have partial-wave expansions with only even $J$. Analogously, the partial-wave expansion for $\pi\pi\to N \bar N$ reads
\begin{align}\label{eq:PW_pipi->NN}
    T_{\Lambda_N}^{\pi \pi \rightarrow \bar{N}N}(t, \Theta)=\sum_J \frac{2 J+1}{2} T_{\Lambda_N}^{J(\pi \pi \rightarrow \bar{N}N)}(t)\sqrt{\frac{\left(J-\Lambda_N\right)!}{\left(J+\Lambda_N\right)!}} P_J^{\Lambda_N}(\cos \Theta),
\end{align}
with $\Lambda_N=0,1$. The $J$-even and $J$-odd cases correspond to $t$-channel isospin $I_t=0$ and 1, respectively. Only the $J$-even component ($I_t=0$) contributes to the unitarity relation~\eqref{eq:unitarity_v1}. 
Combining Eqs.~\eqref{eq:PW_gg->pipi} and~\eqref{eq:PW_pipi->NN}, we can now construct the imaginary parts of the Compton $t$-channel partial waves
\begin{align}\label{eq:unitarity_v2}
    2 \operatorname{Im} T_{\lambda_{\bar{N}} \lambda_N, \lambda_\gamma \lambda_\gamma^{\prime}}^{J\left(\gamma \gamma \rightarrow \bar{N}N\right)}(t)=\frac{1}{8 \pi} \frac{p_\pi}{\sqrt{t}}\left[T_{\Lambda_\gamma}^{J\left(\gamma \gamma \rightarrow \pi \pi\right)}(t)\right]\left[T_{\Lambda_N}^{J(\pi \pi \rightarrow \bar{N}N)}(t)\right]^*\,.
\end{align}
The partial-wave amplitudes $T_{\Lambda_N=0}^{J(\pi \pi \rightarrow \bar{N}N)}$ of Eq.~\eqref{eq:PW_pipi->NN} are related to the Frazer--Fulco amplitudes $f_{+}^J(t)$~\cite{Frazer:1960zza,Frazer:1960zzb} by
\begin{align}
    T_{\Lambda_N=0}^{J(\pi \pi \rightarrow \bar{N}N)}(t) & =\frac{16 \pi}{p_t}\left(p_t p_\pi\right)^J f_{+}^J(t), \\
    T_{\Lambda_N=1}^{J(\pi \pi \rightarrow \bar{N}N)}(t) & =8 \pi \frac{\sqrt{t}}{p_t}\left(p_t p_\pi\right)^J f_{-}^J(t),
\end{align}
where the subscript $+/-$ refers to parallel/antiparallel antinucleon-nucleon helicities. For the reaction $\gamma \gamma \rightarrow \pi \pi$, we use the partial-wave amplitudes $F_{J \Lambda_\gamma}(t)$, which are related to the partial-wave helicity amplitudes $T_{\Lambda}^{J(\gamma \gamma \rightarrow \pi \pi)}$ in Eq.~\eqref{eq:PW_gg->pipi} by\footnote{We adopt the partial-wave amplitude convention of Dai--Pennington~\cite{Dai:2014zta}, which is related to those used by Drechsel et al.~\cite{Drechsel:1999rf} and Gourdin--Martin~\cite{gourdin1960pion} by $2e^2F_{J \Lambda_\gamma}^\text{\cite{Dai:2014zta}}=F_{J \Lambda_\gamma}^\text{\cite{Drechsel:1999rf}}=\sqrt{2J+1}F_{J \Lambda_\gamma}^\text{\cite{gourdin1960pion}}$.}
\begin{align}
    T_{\Lambda_\gamma}^{J(\gamma \gamma \rightarrow \pi \pi)}(t)=\frac{4 e^2}{\sqrt{2 J+1}} F_{J \Lambda_\gamma}(t).
\end{align}
For $\gamma\gamma\to \bar{N}N$, the relation between the $t$-channel helicity amplitudes $T_{\lambda_{\bar{N}} \lambda_N, \lambda_\gamma \lambda_\gamma^{\prime}}^{t}$ and the L'vov amplitudes $A_i$ can be found in Appendix~B of Ref.~\cite{Drechsel:1999rf}. We thus find the expression of $\tilde{A}_1$ in Eq.~\eqref{eq:A1new},
\begin{align}\label{eq:A&T_relation}
    \tilde{A}_1=\frac{1}{t\sqrt{t-4m_N^2}}\left(T^t_{\frac{1}{2}\,\frac{1}{2},1\,1}+T^t_{\frac{1}{2}\,\frac{1}{2},-1\,-1}\right),
\end{align}
where backward kinematics $\nu^2=\frac{tp_t^2}{4m_N^2}$ is used. 
Inserting the partial-wave expansion of Eq.~\eqref{eq:unitarity_v2} into Eqs.~\eqref{eq:unitarity_v1} and~\eqref{eq:A&T_relation}, we can finally express the $t$-channel $2 \pi$ component $\operatorname{Im}_t \tilde{A}_1(\nu, t)^{2 \pi}$ in the backward direction in terms of the partial-wave amplitudes for $\gamma \gamma \rightarrow \pi \pi$ and $\pi \pi \rightarrow \bar{N}N$,\footnote{Following the conventions of Ref.~\cite{Dai:2014zta}, the two-photon amplitudes decompose as $F^C=-\sqrt{2/3}F^{I=0}-\sqrt{1/3}F^{I=2}$ and $F^N=-\sqrt{1/3}F^{I=0}+\sqrt{2/3}F^{I=2}$, where the identical-particle factor $1/\sqrt{2}$ is absorbed into the definition of $F^N$ such that the physical neutral matrix element is $\sqrt{2}F^N$. 
In the $t$-channel unitarity relation, the neutral mode $\pi^0\pi^0$ carries a symmetry phase-space factor $1/2$. 
Contracting with the isoscalar $\bar{N}N$ amplitude $f_+^J$, the intermediate charge sum yields $F^C + \frac{1}{2}(\sqrt{2}F^N) = F^C + F^N/\sqrt{2} = -\sqrt{3/2}F^{I=0}$.}
\begin{align}
    \operatorname{Im}_t \tilde{A}_1\left(\pm\frac{\sqrt{t}p_t}{2m_N}, t\right)^{2 \pi}=\frac{2 e^2 p_\pi}{t\sqrt{t} p_t^2}\sum_{J\,\text{even}}\sqrt{2J+1}(p_\pi p_t)^J \left(-\sqrt{\frac{3}{2}} F^{I=0}_{J\,\Lambda_\gamma=0}(t)\right)f^{J\,*}_{+}(t),
\end{align}
where only the $I=0$ component contributes as mentioned above. 

If we restrict ourselves in the $t$-channel imaginary part to intermediate states with $2\pi$ up to the $D$-wave ($J \leq 2$), the DR~\eqref{eq:HDR_diff_t} takes the convenient form
\begin{align}\label{eq:BEFT_pipi_t}
    (\alpha_{E1}-\beta_{M1})^t_{\pi\pi}= {}&- \frac{8\sqrt{3}\alpha}{\sqrt{2}\pi} \int_{4M_\pi^2}^{\infty}\frac{\mathrm{d} t}{t^2}\frac{1}{4m_N^2-t}\sqrt{\frac{t-4M_\pi^2}{t}} \nonumber\\
    & \times\left[f_+^0(t)F^{0\,*}_{0\,0}(t)-\sqrt{5}\left(m_N^2-\frac{t}{4}\right)\left(\frac{t}{4}-M_\pi^2\right) f_{+}^2(t) F_{2\,0}^{0\,*}(t)\right].
\end{align}
The $s$- and $t$-channel sum rules~\eqref{eq:BEFT_s} and~\eqref{eq:BEFT_pipi_t} are commonly called the BEFT sum rule~\cite{Bernabeu:1974zu,Bernabeu:1977hp}, where only the $\pi\pi$ intermediate state is involved. 
Because the isoscalar $S$-wave involves the well-known $f_0(980)$ resonance near 1~GeV, which strongly couples to the $K\bar{K}$ channel, the BEFT sum rule~\eqref{eq:BEFT_pipi_t} can be readily extended through a coupled-channel formulation~\cite{Hoferichter:2012wf, Cao:2024zlf} to cover the $K\bar{K}$ channel analogous to previous dispersive analyses on nucleon form factors,
\begin{align}\label{eq:BEFT_KK_I=0}
    (\alpha_{E1}-\beta_{M1})^t_{K \bar K_{I_t=0}}= {}&- \left(\frac{2}{\sqrt{3}}\right)\frac{8\sqrt{3}\alpha}{\sqrt{2}\pi} \int_{4 M_K^2}^{\infty}\frac{\mathrm{d} t}{t^2}\frac{1}{4 m_N^2-t}\sqrt{\frac{t-4 M_K^2}{t}} h_+^{0,0}(t)F^{0\,*}_{K,0\,0}(t)\nonumber\\
    = {}& - \frac{8\sqrt{2}\alpha}{\pi} \int_{4 M_K^2}^{\infty}\frac{\mathrm{d} t}{t^2}\frac{1}{4 m_N^2-t}\sqrt{\frac{t-4 M_K^2}{t}} h_+^{0,0}(t)F^{0\,*}_{K,0\,0}(t),
\end{align}
where $h_{+}^{J=0,I=0}$ is the $S$-wave isoscalar $K \bar{K} \rightarrow N \bar{N}$ scattering amplitude,\footnote{Note that, in contrast to $\pi N$ scattering, the combinations $I_t=0$ with odd $J$ and $I_t=1$ with even $J$ are also allowed due to the lack of Bose symmetry in the $\bar{K} K$ system.} $2/\sqrt{3}$ is the corresponding isospin factor.

Let us now consider the contribution of a scalar ``$\mathbf{S}$" in the narrow-width approximation. 
In this case, the $S$-wave contribution to the BEFT sum rule~\eqref{eq:BEFT_pipi_t} can be rewritten in a simplified pole form.
Starting from the $\pi\pi\to N \bar N$ amplitude $f^0_+(t)$ on the first Riemann sheet in the vicinity of the pole, one has~\cite{Hoferichter:2023mgy}\footnote{We adopt the convention of the couplings, $\mathcal{L}_{\mathbf{S} N N}=-g_{\mathbf{S}  N N} \bar{\psi}_N \psi_N \mathbf{S} $.}
\begin{align}\label{eq:f0+_I}
    \frac{g_{\mathbf{S} N N}}{g_{\mathbf{S} \pi \pi}}=i \sqrt{6} \frac{\sigma_\pi\left(M_\mathbf{S}^2\right)}{4 m_N^2-M_\mathbf{S}^2} f_{+}^{0,\rm{I}}\left(M_\mathbf{S}^2\right),
\end{align}
where we use the $g_{\mathbf{S}\pi\pi}$ convention as in Ref.~\cite{Garcia-Martin:2011nna}.\footnote{Note that $g_{\mathbf{S} \pi\pi}^\text{\cite{Moussallam:2011zg}}=\sqrt{2} g_{\mathbf{S} \pi\pi}^\text{\cite{Garcia-Martin:2011nna}}$.} 
Following Refs.~\cite{Pennington:2006dg,Moussallam:2011zg}, one can identify the residues of the amplitudes $F^0_{0\, 0}(t)$ in terms of the corresponding coupling constants on the second Riemann sheet,
\begin{align}\label{eq:FS_II}
    \left.F_{0\,0}^{0, \rm{II}}(t)\right|_\text{pole} & =\frac{M_\mathbf{S}^2 F_{\mathbf{S} \gamma \gamma}}{8\pi \alpha}\frac{g_{\mathbf{S} \pi \pi}}{M_\mathbf{S}^2-t-i\epsilon} \nonumber\\
    & =i \frac{M_\mathbf{S}^2 g_{\mathbf{S} \pi \pi} F_{\mathbf{S} \gamma \gamma}}{8 \alpha}\delta(M_\mathbf{S}^2-t)+\text{regular part},
\end{align}
with the two-photon decay width
\begin{align}
    \Gamma(\mathbf{S}\to \gamma \gamma)=\frac{M_\mathbf{S}^3}{64 \pi}|F_{\mathbf{S} \to \gamma \gamma}|^2 .
\end{align}
Substituting Eqs.~\eqref{eq:f0+_I} and~\eqref{eq:FS_II} into Eq.~\eqref{eq:BEFT_pipi_t} yields
\begin{align}\label{eq:pole_form}
    (\alpha_{E1}-\beta_{M1})^t_{\mathbf{S}}=\frac{g_{\mathbf{S} N N} F_{\mathbf{S} \gamma \gamma}}{2 \pi M_\mathbf{S}^2}.
\end{align}
In Refs.~\cite{Schumacher:2006cy,Schumacher:2007ez,Schumacher:2011gi,Schumacher:2019ikn} (see also reviews~\cite{Schumacher:2013hu,Holstein:2013kia}), Eq.~\eqref{eq:pole_form} is commonly used to estimate the contributions of $\sigma, f_0(980)$ and even $a_0(980)$ resonances to the BEFT sum rule. 
However, as follows from the above derivation, Eq.~\eqref{eq:pole_form} is valid only under the narrow-width approximation and hence fails for the broad $\sigma$ resonance. 

The contribution of the isovector $S$-wave channels ($\pi\eta$ and $K\bar{K}_{I_t=1}$) follows from a coupled-channel dispersion relation analogous to the isoscalar case. 
Taking into account that the intermediate $K^+K^-$ and $K^0\bar{K}^0$ states consist of distinguishable particles and summing over the physical charge modes, the isovector $K\bar{K}$ channel enters with an overall sign opposite to the isoscalar case.\footnote{Adopting the kaon isospin phase convention in Appendix~\ref{app:isospin}, the two-photon amplitudes decompose into $s$-channel isospin components as $F_K^C = -(F_K^0 + F_K^1)/\sqrt{2}$ for $K^+K^-$ and $F_K^N = -(F_K^0 - F_K^1)/\sqrt{2}$ for $K^0\bar{K}^0$. 
Meanwhile, the corresponding $t$-channel $\bar{N}N$ strong amplitudes decompose as $h_+^{0,0} - h_+^{0,1}$ for $K^+K^-\to\bar{p}p$ and $h_+^{0,0} + h_+^{0,1}$ for $K^0\bar{K}^0\to\bar{p}p$ [cf.~\eqref{eq:iso_KK->NN}].  
Contracting the two-photon and strong amplitudes yields the physical charge sum $F_K^C [h_+^{0,0} - h_+^{0,1}]^* + F_K^N [h_+^{0,0} + h_+^{0,1}]^* = -\sqrt{2} F_K^0 h_+^{0,0*} + \sqrt{2} F_K^1 h_+^{0,1*}$, which directly establishes the opposite signs for the isoscalar and isovector kaon terms, respectively.} 
Incorporating the appropriate two-body phase spaces and matching to the two-channel unitarity relation, the isovector $t$-channel spectral contribution to the polarizability difference is given by
\begin{align}\label{eq:BEFT_t_I=1}
    (\alpha_{E1}-\beta_{M1})^t_{\pi\eta/K \bar K_{I_t=1}}= {}&\frac{8\alpha}{\pi} \int_{(M_\pi+M_\eta)^2}^{\infty}\frac{\mathrm{d} t}{t^2}\frac{1}{4 m_N^2-t}\frac{\sqrt{\lambda(t,M_\pi^2,M_\eta^2)}}{t} g_+^{0}(t)F^{*}_{\pi\eta,0\,0}(t) \nonumber\\
    & +\frac{8\sqrt{2}\alpha}{\pi} \int_{4 M_K^2}^{\infty}\frac{\mathrm{d} t}{t^2}\frac{1}{4 m_N^2-t}\sqrt{\frac{t-4 M_K^2}{t}} h_+^{0,1}(t)F^{1\,*}_{K,0\,0}(t),
\end{align}
where $g_{+}^{J=0}$ is the $S$-wave isovector $\pi\eta \rightarrow N \bar{N}$ scattering amplitude and $\sqrt{2}$ is the corresponding isospin factor. 
Unfortunately, apart from recent progress in dispersive analyses of the $I=1$ two-photon amplitudes $F^1$~\cite{Lu:2020qeo,Deineka:2024mzt}, the $\pi\eta/K\bar{K}_{I_t=1} \to N\bar{N}$ amplitudes in the pseudophysical subthreshold region remain poorly known. 
Noting that the $I=1$ contribution is dominated by the $a_0(980)$ resonance in the low-energy regime, we can employ the narrow-width approximation~\eqref{eq:pole_form} to Eq.~\eqref{eq:BEFT_t_I=1}, which simply yields\cite{Schumacher:2007ez}
\begin{align}\label{eq:BEFT_a0_pole}
    \left(\alpha_{E 1}-\beta_{M 1}\right)_{a_0}^t=\frac{g_{a_0 N N} F_{a_0 \gamma \gamma}}{2 \pi M_{a_0}^2}.
\end{align}
A complete and more rigorous dispersive analysis is presented in Sec.~\ref{sec:BEFT_t}.

\section{Saturation of the $s$-channel BEFT sum rule}\label{sec:BEFT_s}

In this section, we describe the calculation of the $s$-channel contributions entering the BEFT sum rule. 
Due to the energy denominator $1/\omega^2$ in the dispersion integral~\eqref{eq:BEFT_s}, the most important contribution is from the $\pi N$ intermediate states~\eqref{eq:sigma_CGLN}, while mechanisms involving more pions or heavier excited states in the intermediate states are largely suppressed. 
Partial-wave analyses of single-pion photoproduction have advanced substantially in past decades; see Refs.~\cite{Doring:2025sgb,Burkert:2025coj} for recent reviews. 
Previous studies of the BEFT sum rule exhibit reasonable consistency with respect to the $\pi N$ contribution in the proton channel~\cite{Schumacher:2005an}. 
However, for the neutron channel, the lack of high-precision data forced earlier partial-wave analyses to rely on theoretical constraints rather than direct fits to experimental data (for example, the widely used MAID results~\cite{Drechsel:2007if}). 
In recent years, new phenomenological analyses~\cite{Kamano:2016bgm,Hunt:2018wqz,Kamano:2019gtm,Osmanovic:2021rck,Ma:2020hpe,Cao:2021kvs,Ronchen:2022hqk,Briscoe:2023gmb,Guo:2022hud,Zhuge:2024iuw,Schneider:2025ndq}, lattice QCD (LQCD)~\cite{Gao:2025loz} and LQCD-based~\cite{Zhuge:2026dxl} results have appeared, providing more reliable inputs for applications such as the BEFT sum rule.

\begin{figure}[t]
    \centering
    \includegraphics[width=1\linewidth]{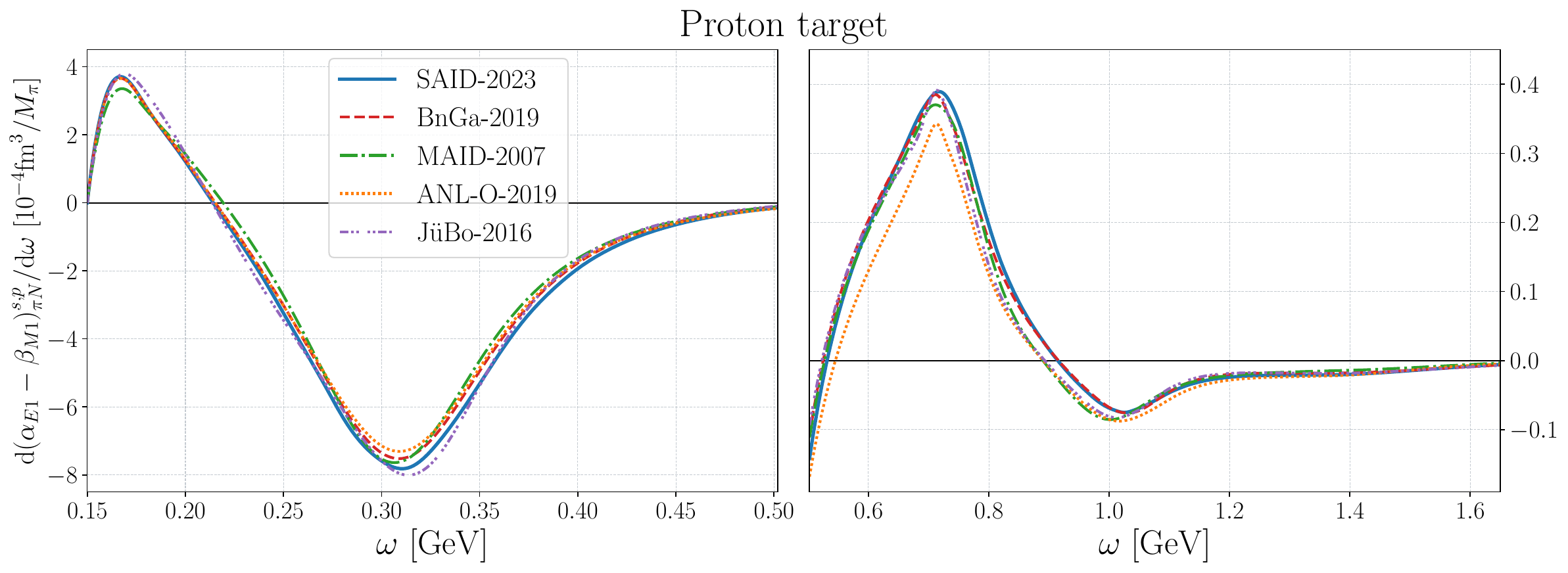}
    \caption{$s$-channel integrand for the proton target defined via $(\alpha_{E1}-\beta_{M1})^{s}_{\pi N}=\int^\infty_{M_\pi+\frac{M_\pi^2}{2 m_N}} \md\omega\frac{\mathrm{d}(\alpha_{E1}-\beta_{M1})^{s,p}_{\pi N}}{\mathrm{d}\omega}$ evaluated using three different multipole analyses: SAID-2023~\cite{GWU}, BnGa-2019~\cite{BG}, and MAID-2007~\cite{Mainz}. 
    Predictions from the dynamical coupled-channel models of ANL-O-2019~\cite{ANL-O} and J\"uBo-2016~\cite{JuBo} are also shown for comparison.}
    \label{fig:s_proton}
\end{figure}

\begin{figure}[t]
    \centering
    \includegraphics[width=1\linewidth]{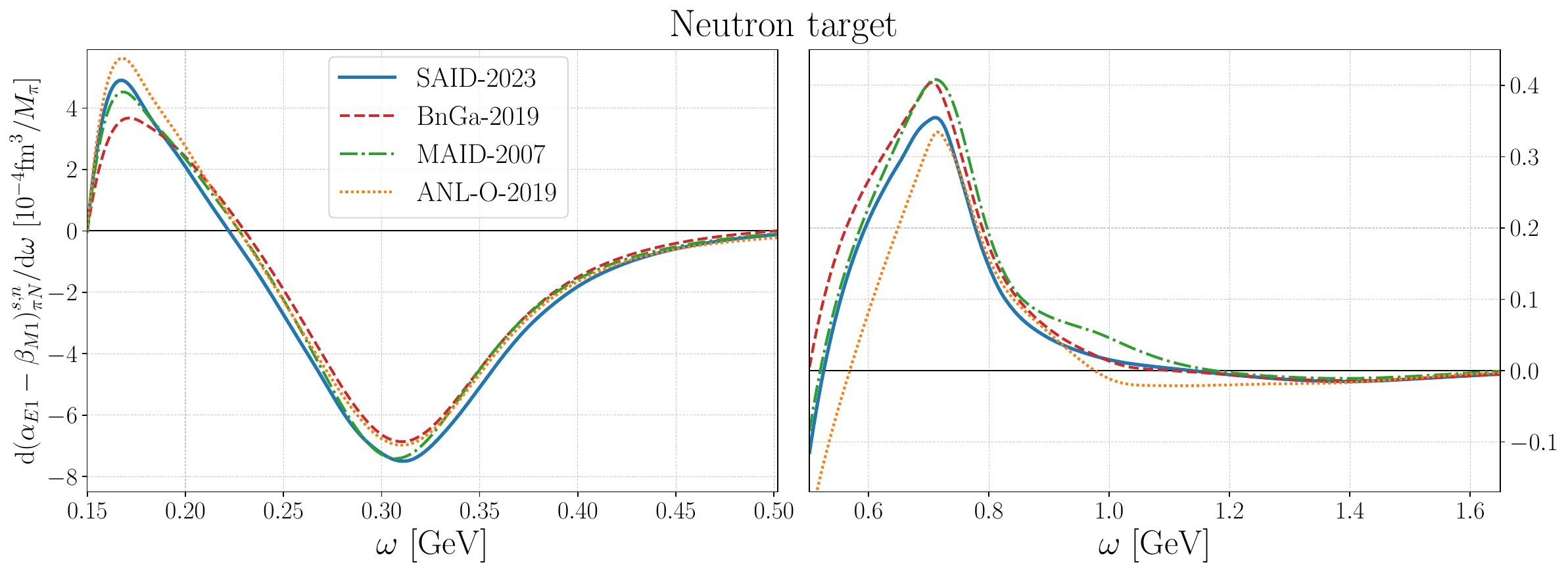}
    \caption{Same as Fig.~\ref{fig:s_proton} but for neutron target.}
    \label{fig:s_neutron}
\end{figure}

In our calculation, we evaluate the $\pi N$ contribution using the single-pion photoproduction multipole amplitudes from the latest analysis by the SAID group~\cite{Briscoe:2023gmb,GWU}. The expansion in Eq.~\eqref{eq:sigma_CGLN} into multipoles is truncated at a maximum angular momentum $\ell\leq 4$, within the energy range $\sqrt{s}<2~$GeV ($\omega<1.66$~GeV). 
We also examined the contributions of higher partial waves $\ell>4$. 
In Eq.~\eqref{eq:BEFT_s} these contributions are below the percent level and therefore can be safely neglected.
In Figs.~\ref{fig:s_proton} and~\ref{fig:s_neutron}, we plot the integrand of Eq.~\eqref{eq:BEFT_s}, $\mathrm{d}(\alpha_{E1}-\beta_{M1})^{s,p/n}_{\pi N}/\mathrm{d}\omega$, as obtained from the SAID multipoles, where $p/n$ corresponds to the proton/neutron target, respectively.
The integrand is clearly peaked at low energies and practically vanishes above $0.5$~GeV, showing a good convergence of the $s$-channel BEFT sum rule. 
In addition, to estimate possible uncertainties in evaluating the $\pi N$ contributions of the BEFT sum rule, older MAID~\cite{Drechsel:2007if,Mainz} and more recent Bonn-Gatchina (BnGa)~\cite{Anisovich:2016vzt,BG} partial-wave results are also plotted for comparison. 
As expected, the results for the proton target are in good agreement with each other, whereas those for the neutron target exhibit some moderate discrepancies. 
We adopt the average of the three multipole sets as the central value and take the largest deviation between the central value and each set as a conservative estimate of the uncertainties associated with the single-pion contribution to the $s$-channel BEFT sum rule, and obtain
\begin{align}\label{eq:s_piN_values}
    \begin{aligned}
    (\alpha_{E1}-\beta_{M1})^{s,p}_{\pi N}=-4.96(33),\\
    (\alpha_{E1}-\beta_{M1})^{s,n}_{\pi N}=-3.69(52).
    \end{aligned}
\end{align}
As a cross-check, we also employ the multipoles from the dynamical coupled-channel models of the Argonne-Osaka (ANL-O-2019)~\cite{ANL-O} and J\"ulich-Bonn (J\"uBo-2016)~\cite{JuBo} collaborations to re-evaluate the single-pion contribution. 
This yields $-4.97$ for the proton and $-3.54$ for the neutron using ANL-O-2019, and $-5.24$ for the proton using J\"uBo-2016. 
These values are fully consistent with the results obtained from the three multipole datasets within uncertainties.

The $\pi N$ channel consistently reproduces the measured photoabsorption cross section in the energy range $\omega\lesssim 500~$MeV~\cite{Lvov:1996rmi}, while at higher energies inelastic resonant and nonresonant mechanisms should be included to fully describe the non-$\pi N$ inelastic channels.
However, for the non-$\pi N$ inelastic contribution, a precise calculation analogous to the single-pion case is not possible, since an available partial-wave analysis has not yet been performed. 
Accordingly, a model-based treatment is required to assess the above inelastic contributions.

\begin{figure}[t]
    \centering
    \includegraphics[width=.6\linewidth]{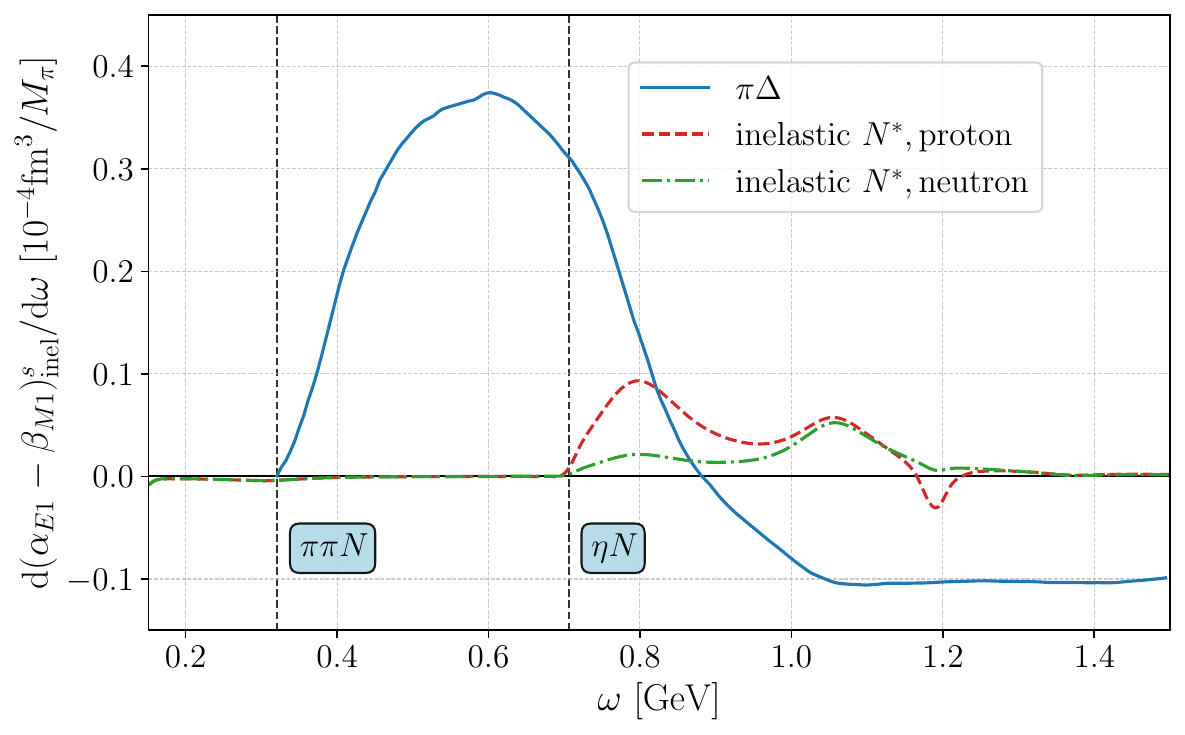}
    \caption{Integrand of the $s$-channel non-$\pi N$ inelastic contributions for the nucleon, including resonant contributions from inelastic $N^*$ states and nonresonant contributions modeled by the $\pi \Delta$ mechanism. }
    \label{fig:s_pipiN}
\end{figure}
First, the resonant non-$\pi N$ $(\pi\pi N$ and $\eta N$) inelastic intermediate states are approximated by the inelastic decay channels of the $\pi N$ resonances.
Following Appendix~D of Ref.~\cite{Lvov:1996rmi}, we take the inelastic contribution to share the helicity structure of the single pion photoproduction amplitudes. 
In this approximation, we calculate the resonant part of the pion photoproduction multipoles using the Breit--Wigner (BW) parameterization of Refs.~\cite{Walker:1968xu,Lvov:1996rmi}, which is then scaled by a corresponding factor to include the inelastic decays of the resonances. 
The BW resonance masses, widths, and the corresponding resonant photo-decay amplitudes are taken from Refs.~\cite{Workman:2011vb,Workman:2012jf}.
We show in Fig.~\ref{fig:s_pipiN} that the dominant inelastic resonant contribution originates from the $N^*(1535) \to \eta N$ and $\Delta(1700)\to \pi\pi N$ decay modes.
A common way to include nonresonant contributions in two-pion photoproduction is to approximate the $\gamma N\to \pi\Delta$ channel via the one-pion exchange (OPE) diagram~\cite{Holstein:1994tw}; see Ref.~\cite{Lvov:1996rmi} for more details.
However, models containing only resonant mechanisms plus this OPE term still deviate from experimental photoabsorption cross-section data. 
This discrepancy can be resolved by introducing a nonresonant phenomenological $S$-wave correction for $\gamma N\to \pi\Delta$ and adjusting the $S$-wave amplitude to fit the cross section.\footnote{Other $\rho N$ and $\sigma N$ contributions have been absorbed into the adjusted effective $S$-wave $\gamma N\to \pi\Delta$ amplitude.} The nonresonant contribution is shown in Fig.~\ref{fig:s_pipiN}. 
It should be stressed that this approach produces identical results for proton and neutron targets, since the resulting amplitude is isoscalar.
Finally, we obtain the estimates
\begin{align}\label{eq:s_pipiN_values}
    \begin{aligned}
    (\alpha_{E1}-\beta_{M1})^{s,p}_\mathrm{inel}=(0.16+0.56)\,,\\
    (\alpha_{E1}-\beta_{M1})^{s,n}_\mathrm{inel}=(0.09+0.56)\,,
    \end{aligned}
\end{align}
where the first and second terms in each parentheses denote the effective resonant and nonresonant contributions, respectively.
The resulting non-$\pi N$ inelastic contributions are about one order of magnitude smaller than the corresponding single-pion contributions in Eq.~\eqref{eq:s_piN_values}, as expected.
We do not assign explicit uncertainties to these subleading estimates. 
A rigorous error budget for non-$\pi N$ intermediate states is currently out of reach. 
Moreover, because the sum of the $\pi N$ and inelastic channels is constrained by empirical total photoabsorption cross sections, quoting independent uncertainties would incur somewhat double counting. 
A self-consistent separation with full theoretical control would necessitate a comprehensive coupled-channel partial-wave analysis, which lies beyond the scope of this work.

\section{Saturation of the $t$-channel BEFT sum rule}\label{sec:BEFT_t}

\subsection{Isoscalar channel}

Since the dispersive integrals are ultraviolet convergent, the most important imaginary part in the $t$-channel BEFT sum rule from $4 M_\pi^2 \rightarrow \infty$ is given by the lowest isoscalar $\pi\pi/K \bar K$ intermediate states, cf.~Eqs.~\eqref{eq:BEFT_pipi_t} and~\eqref{eq:BEFT_KK_I=0}. 
As a consequence, the $t$-channel integrals can be calculated by including the experimental information on the $t$-channel process through $\pi \pi/K \bar K$ intermediate states as $\gamma \gamma \rightarrow \pi \pi/K\bar{K} \rightarrow N \bar{N}$. 
For the $\pi\pi/K\bar{K} \rightarrow N \bar{N}$ amplitudes, significant progress has been made in the past decade. 
By solving Roy--Steiner equations~\cite{Hite:1973pm} for the $\pi N$ system, which are built upon unitarity, analyticity and crossing symmetry, model-independent lowest partial-wave amplitudes for $\pi\pi/K\bar{K} \rightarrow N \bar{N}$ have been obtained from the pseudophysical threshold $t=4 M_\pi^2$ to the nucleon-antinucleon threshold $t=4 m_N^2$~\cite{Hoferichter:2015hva,Cao:2022zhn}. 
In practice, we utilize the numerical results given in Ref.~\cite{Cao:2024zlf}, from which we take the amplitudes $f_+^{0},f_+^{2}$ and $h_+^{0,0}$ for use in Eqs.~\eqref{eq:BEFT_pipi_t} and~\eqref{eq:BEFT_KK_I=0}. 

The amplitudes for $\gamma \gamma \rightarrow \pi\pi/K\bar K$ have traditionally been considered less reliable, especially for the $S$-wave component. 
Fortunately, over the last two decades, renewed experimental and theoretical efforts have significantly improved the situation.
This progress has been strongly motivated by the role of two-photon processes in providing one of the most important constraints on precise data-driven determination of the hadronic light-by-light contribution to the muon anomalous magnetic moment $(g-2)_\mu$; see Refs.~\cite{Danilkin:2019mhd,Aoyama:2020ynm} for recent reviews.
In particular, modern dispersive analyses lead to cross sections and differential cross sections that are in excellent agreement with the experimental data for both $\pi^+ \pi^-$ and $\pi^0 \pi^0$ channels~\cite{Pennington:2006dg,Oller:2008kf,Mao:2009cc,Garcia-Martin:2010kyn,Hoferichter:2011wk,Dai:2014zta,Danilkin:2018qfn}.\footnote{The correlation between the meson polarizability uncertainties and those in the photon-photon collision cross sections is revealed in Ref.~\cite{Dai:2016ytz}.} 
To match the results of the Roy--Steiner equation, we require an $S$-wave coupled-channel analysis as input. 
The corresponding coupled-channel Muskhelishvili--Omn\`es (MO) dispersive representation was first considered in Ref.~\cite{Mao:2009cc}.
For the $S$-wave, the $n$-th subtracted DR can be written for the matrix function $\left[\mathbf{\Omega}^0_0(t)\right]^{-1}\left(\mathbf{F}_{0\,0}^0(t)-\mathbf{F}_{0\,0}^{0,\rm{Born}}(t)\right)/t$, resulting in a modified MO representation~\cite{Garcia-Martin:2010kyn}\footnote{The MO representations admit two formally equivalent formulations: either by subtracting only the Born term from the partial waves and treating the residual LHC contributions via their discontinuities (modified representation), or by incorporating the full LHC input explicitly into the inhomogeneous term (standard representation).}
\begin{align}\label{eq:MO_rep}
    \mathbf{F}_{0\,0}^0(t)=\mathbf{F}_{0\,0}^{0,\rm{Born}}(t)+\mathbf{\Omega}^0_0(t)\Biggl[ & \mathbf{P}_n(t)+\frac{t^{n+1}}{\pi}\int_{\mathcal{C}_L}\md t^\prime~\frac{\left[\mathbf{\Omega}^0_0(t')\right]^{-1} \operatorname{Im}\mathbf{F}_{0\,0}^{0,\rm{Res}}(t')}{(t')^{n+1}(t'-t)}\nonumber\\
    & -\frac{t^{n+1}}{\pi}\int_{4 M_\pi^2}^{\infty}\md t^\prime~\frac{ \operatorname{Im}\left[\mathbf{\Omega}^0_0(t')\right]^{-1}\mathbf{F}_{0\,0}^{0,\rm{Born}}(t')}{(t')^{n+1}(t'-t)}\Biggr],
\end{align}
where $\mathbf{F}_{0\,0}^0\equiv \left(F^0_{\pi,0\,0}, F^0_{K,0\,0}\right)^T$;\footnote{Our isoscalar partial-wave amplitude convention is related to those used by Garc\'ia-Mart\'in--Moussallam~\cite{Garcia-Martin:2010kyn} and Danilkin--Vanderhaeghen~\cite{Danilkin:2018qfn} by $2\sqrt{2/(2J+1)}\,e^2\mathbf{F}_{J, \Lambda_\gamma=0}^{I=0} = e^2 \bigl(h_{J,++}^{I=0},\, k_{J,++}^{I=0}\bigr)^{T,\,\text{\cite{Garcia-Martin:2010kyn}}} = \bigl(h_{I=0,++}^{(J)},\, k_{I=0,++}^{(J)}\bigr)^{T,\,\text{\cite{Danilkin:2018qfn}}}$.} $\mathbf{P}_n(t)$ is an $n$-th order vector-valued subtraction polynomial with vanishing subtraction constant to comply with Low's theorem in the soft-photon limit~\cite{Low:1954kd}; $\mathcal{C}_L$ denotes the integration path along the resonance-exchange left-hand cuts (LHCs); $\mathbf{\Omega}^0_0$ denotes the isoscalar $S$-wave Omn\`es matrix and $\mathbf{F}_{0\,0}^{0,\rm{Res}}$ comprises the LHC contributions excluding the Born-term contribution $\mathbf{F}_{0\,0}^{0,\rm{Born}}$.

\begin{figure}[t]
    \centering
    \includegraphics[width=.65\linewidth]{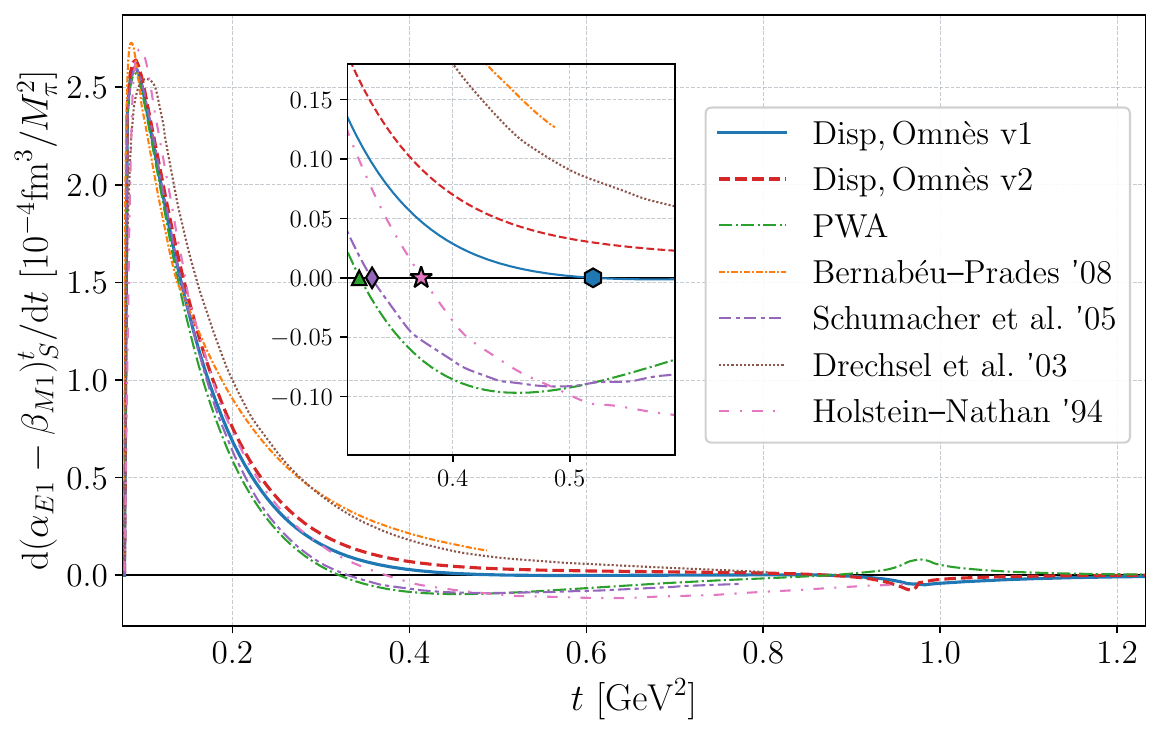}
    \caption{The integrand for the $t$-channel $S$-wave contribution to the polarizability difference $\alpha_{E1}-\beta_{M1}$.}
    \label{fig:t_Swave}
\end{figure}

The integrand for the $S$-wave piece is given in Fig.~\ref{fig:t_Swave}, in which three different curves are shown, corresponding to three different representations of the $\gamma \gamma \to \pi\pi/K\bar K$ amplitudes: a modified MO representation with 2 subtractions~\cite{Garcia-Martin:2010kyn} (``Disp, Omn\`es v1"), a modified MO representation with 1 subtraction~\cite{Danilkin:2018qfn} (``Disp, Omn\`es v2"),\footnote{The isoscalar $S$-wave Omn\`es matrix is matched to the latest $N/D$ results of Ref.~\cite{Danilkin:2020pak}.} and a partial-wave amplitude analysis supplemented by a dispersive Omn\`es constraint in the low-energy region $\sqrt{t}<0.6~$GeV in a single-channel manner~\cite{Dai:2014lza,Dai:2014zta} (``PWA"). 
The phases of these amplitudes were slightly adjusted to match the Roy--Steiner equation results.
In addition, we also show several previous results that account for $\pi\pi$ FSI for comparison~\cite{Holstein:1994tw,Drechsel:2002ar,Schumacher:2005an,Bernabeu:2008wt}. 

Near the $\pi\pi$ threshold ($\sqrt{t}<0.45$~GeV), all calculations agree well because the Born term dominates, as shown in Fig.~\ref{fig:t_Swave}. 
In the region $0.45~\text{GeV}<\sqrt{t}<0.75$~GeV, significant discrepancies among previous phenomenological predictions can be traced to whether the integrand contains a zero in the low energy region.\footnote{Note that the integrand in Eq.~\eqref{eq:BEFT_pipi_t} also acquires at least one additional zero near $\sqrt{t}=0.94$~GeV, derived by the $K\bar K$ channel (see Ref.~\cite{Dong:2020hxe} for a discussion about the universality of such a zero in the presence of strong $S$-wave attraction in an intermediate coupled channel) and contained in the $S$-wave $\pi\pi\to N \bar N$ amplitude $f^0_+$. The zero discussed here refers specifically to a possible zero of the $S$-wave $\gamma\gamma\to \pi\pi$ amplitude $F^0_{0\,0}$, whose location lies roughly between 0.3 and 0.8~GeV.}
This feature was first noted by Holstein and Nathan~\cite{Holstein:1994tw}. 
The presence of this zero reduces the $t$-channel $S$-wave BEFT sum rule contribution from 16.1 (pure Born contribution) to 10.3, with the consequence that the total sum of the $s$- and $t$-channel contributions is considerably smaller than the accepted experimental value. 
Drechsel et al.~\cite{Drechsel:2002ar} repeated this analysis and found no such zero for $\sqrt{t}<0.88$~GeV and obtained a value of 17.6, significantly larger than the value of Holstein and Nathan. 

Our two results based on the MO dispersive representation lie between the Holstein--Nathan and Drechsel et al. curves and show similar trends.
In particular, the ``Disp, Omn\`es v1" solution exhibits a relatively high zero at $\sqrt{t}=0.72$~GeV (with an additional zero near $\sqrt{t}=0.81$~GeV), while the ``Disp, Omn\`es v2" solution shows no such zero. 
Both solutions show a visible cusp at the $K\bar K$ threshold, corresponding to a small $f_0(980)$ contribution, which enters the BEFT sum rule with a negative sign.
Interestingly, the curve corresponding to ``PWA" lies significantly below both MO dispersive solutions and even yields an opposite sign for the $f_0(980)$ contribution. 
The corresponding value of 10.1 is even smaller than that of Holstein and Nathan. 
This issue can be traced back to the fact that the partial-wave helicity amplitude $F_{2\,2}^0$ is much larger than $F_{2\,0}^0$ because the $f_2(1270)$ resonance couples predominantly to the helicity-different two-photon state ($\lambda_\gamma=-\lambda_\gamma^\prime$).\footnote{Considering the gauge-invariant coupling term, $f_{\mu \nu} F^{\mu \rho} F_{\ \rho}^\nu$, where $f_{\mu\nu}$ is a symmetric, traceless tensor for $f_2(1270)$ and $F^{\mu\rho}$ is the electromagnetic field tensor, the amplitudes with equal photon helicities ($\lambda_\gamma=\lambda_\gamma^\prime$) vanish~\cite{Drechsel:1999rf}.}
Above $\sqrt{t}\simeq 0.4$~GeV the $F_{2\,2}^0$ contribution already exceeds the $S$-wave $F_{0\,0}^0$ and also $F_{2\,0}^0$;\footnote{Owing to the proximity of the pion pole in the OPE Born term to the physical region, the $D$-wave projection receives substantial non-polynomial contributions even at low energies. 
This contrasts with purely hadronic $\pi\pi$ scattering, where the $D$-wave is strongly suppressed below $1~\mathrm{GeV}$ by the $(p/\Lambda)^{2L}$ centrifugal barrier (with $\Lambda \sim m_\rho$) and because the lowest tensor resonance, $f_2(1270)$, lies well above $1~\mathrm{GeV}$.} see Fig.~1 in Ref.~\cite{Dai:2014lza}. 
That hierarchy makes a partial-wave analysis particularly challenging for reliably determining the comparatively small $F_{0\,0}^0$ and $F_{2\,0}^0$ components, and the lack of precise experimental constraints on the two-photon cross section below $\sqrt{t}= 0.8$~GeV further exacerbates this difficulty.
Therefore, we conservatively adopt the average of the two results obtained from the modified MO dispersive representation as our central value,
\begin{align}\label{eq:Swave_value}
    (\alpha_{E1}-\beta_{M1})^{t}_{S}=12.82(87)_{\rm MO}\,,
\end{align}
with the error estimated as one half of the difference between the two results, while other uncertainties are deferred to a later discussion.

\begin{figure}[t]
    \centering
    \includegraphics[width=.65\linewidth]{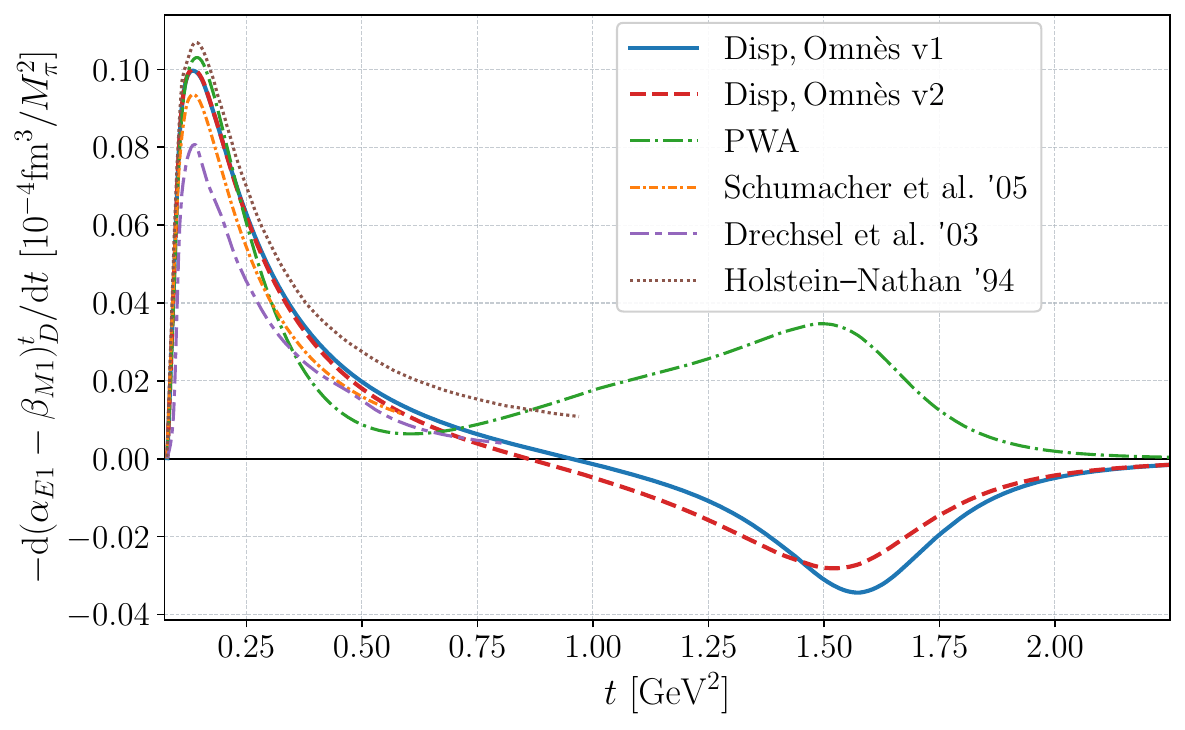}
    \caption{Same as Fig.~\ref{fig:t_Swave} but for $D$-wave.}
    \label{fig:t_Dwave}
\end{figure}

For the $D$-wave piece, we use the dispersive results~\cite{Garcia-Martin:2010kyn,Danilkin:2018qfn} for $\gamma\gamma\to \pi\pi$. 
The integrand is shown in Fig.~\ref{fig:t_Dwave}, and previous phenomenological results are also shown for comparison. 
A few points are worth noting in order.
First, below $\sqrt{t}\simeq 0.8$~GeV the Born term dominates, and all results agree well. 
Second, the ``PWA" and the ``Disp, Omn\`es" solutions differ markedly in their description of the $f_2(1270)$ because of the discrepancies in the two-photon helicity amplitude $F^0_{2\,0}$, which was mentioned above. 
Finally, in the ``Disp, Omn\`es" result, the $f_2(1270)$ contribution largely cancels against the low-energy part, yielding a minor and still negative contribution,
\begin{align}\label{eq:Dwave_value}
    (\alpha_{E1}-\beta_{M1})^{t}_{D}=-0.44(2)_{\rm MO} \,,
\end{align}
to be compared with the previous value of $-1.7$~\cite{Holstein:1994tw} and the ``PWA" result of $-2.39$.
It is evident that the value of Eq.~\eqref{eq:Dwave_value} is significantly smaller than the $S$-wave counterpart~\eqref{eq:Swave_value}, thereby providing a numerical justification for neglecting higher partial waves.

\subsection{Isovector channel}

The role of $t$-channel $I=1$ intermediate states dominated by the $a_0(980)$ resonance in the BEFT sum rule was first discussed by Schumacher~\cite{Schumacher:2007ez} in the sense of the narrow-width approximation, cf.~Eq.~\eqref{eq:BEFT_a0_pole}.
While there are good reasons to assume that the narrow-width or pole approximation captures some significant contributions to the BEFT sum rule in the isovector channel, we will assess scalar rescattering contributions explicitly by calculating the isovector intermediate states for $\gamma \gamma \rightarrow \pi \eta/K\bar{K}_{I_t=1} \rightarrow N \bar{N}$ because $a_0(980)$ lies close to the $K\bar K$ threshold.

\begin{figure}[t]
    \centering
    \includegraphics[width=1\linewidth]{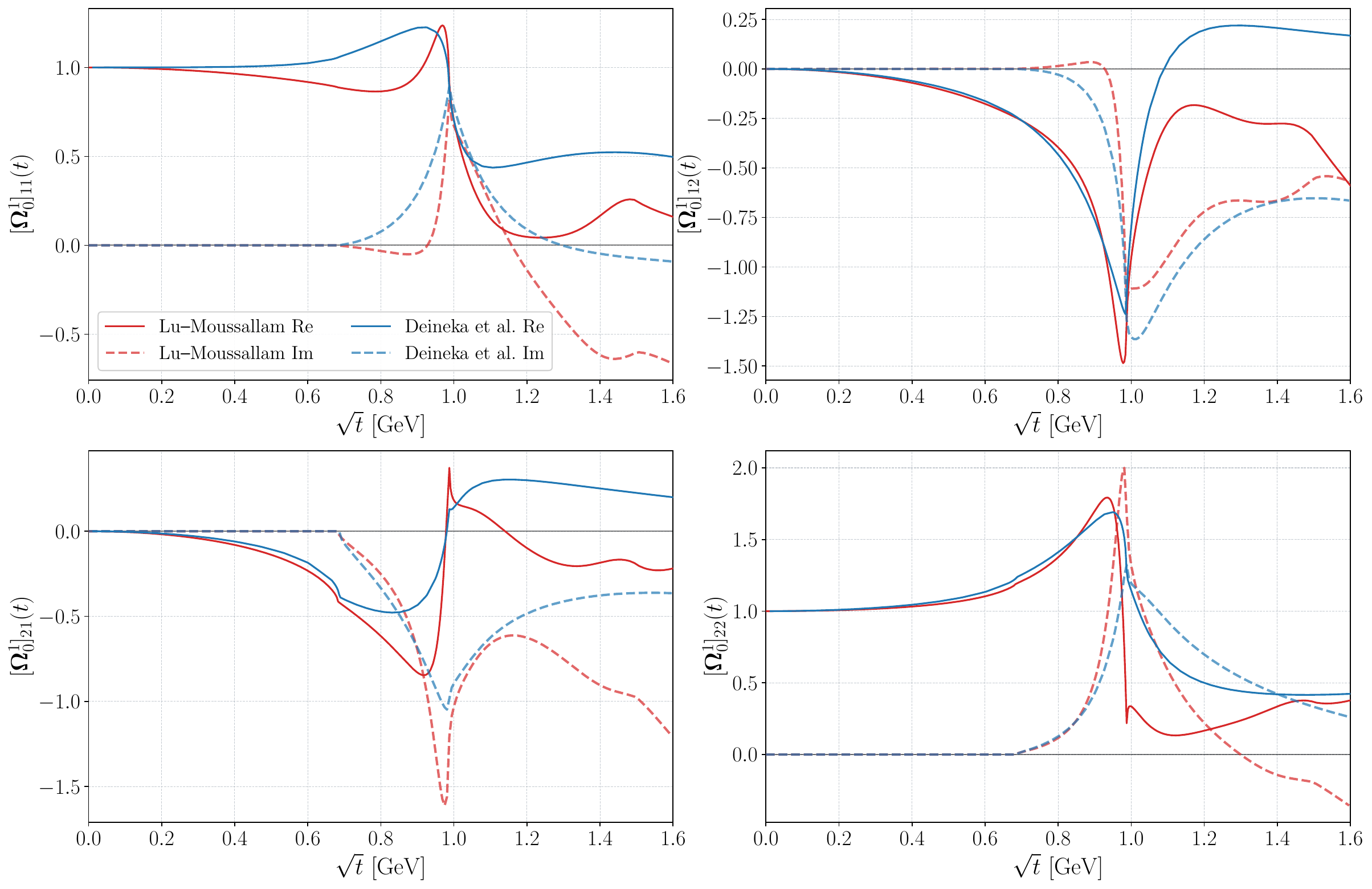}
    \caption{The real (solid) and imaginary (dashed) part of the matrix elements of the $S$-wave $I=1$ Omn\`es matrix taken from Lu--Moussallam~\cite{Lu:2020qeo} (red) and Deineka et al.~\cite{Deineka:2024mzt} (blue). Note that in Lu--Moussallam~\cite{Lu:2020qeo} we adopt their fit~II result, whereas we employ the solution corresponding to Eq.~(26) in Deineka et al.~\cite{Deineka:2024mzt}.}
    \label{fig:Isovector_Omnes}
\end{figure}

For the $S$-wave two photon amplitude entering Eq.~\eqref{eq:BEFT_t_I=1}, we employ two recent modified MO representations~\cite{Lu:2020qeo,Deineka:2024mzt} [cf.~Eq.~\eqref{eq:MO_rep}], both of which provide a reliable description of the existing $\gamma\gamma\to \pi\eta/K\bar{K}_{I_t=1}$ data.
The discrepancies between the two solutions are subtle and trace back to two sources. 
First, the dispersive relations are implemented with a different number of subtractions: Ref.~\cite{Lu:2020qeo} uses a once-subtracted dispersion relation, whereas Ref.~\cite{Deineka:2024mzt} favors the unsubtracted version. 
Second, the coupled-channel $T$-matrix for $\pi\eta/K\bar{K}$ scattering, which is crucial for the calculation of the $I=1$ Omn\`es matrix, is constructed in two different ways: Ref.~\cite{Lu:2020qeo} employs a chiral $K$-matrix type parametrization, while Ref.~\cite{Deineka:2024mzt} uses a more data-driven $N/D$ representation supplemented by conformal mapping. 
As displayed in Fig.~\ref{fig:Isovector_Omnes}, the resulting Omn\`es matrices exhibit similar trends. 
The difference in the cusp strength near the $K\bar K$ threshold reflects somewhat the distinct $a_0(980)$ pole nature in the two solutions.

\begin{figure}[t]
    \centering
    \includegraphics[width=1\linewidth]{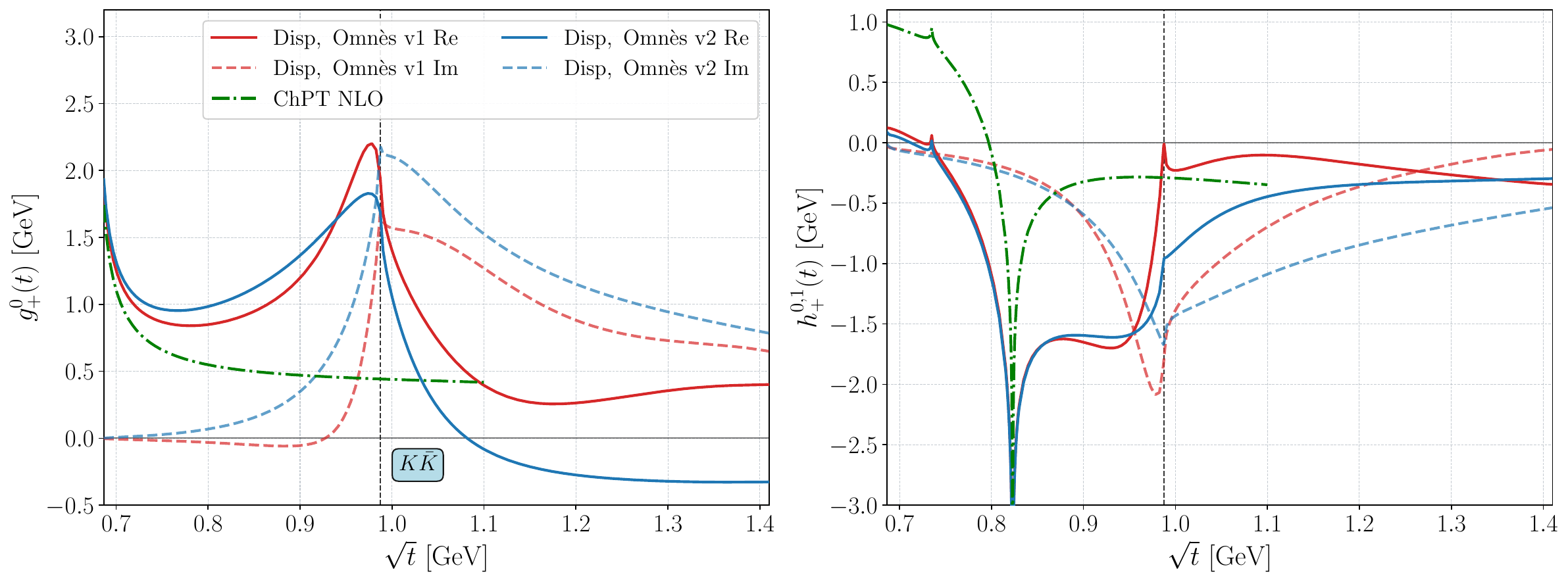}
    \caption{Results for real (solid) and imaginary (dashed) part of $g^0_+(t)$ and $h^{0,1}_+(t)$. We also show the NLO ChPT results for comparison.}
    \label{fig:Amp_isovector_NNbar}
\end{figure}

The $\pi\eta/K\bar{K}_{I_t=1} \to N\bar{N}$ amplitudes are not experimentally accessible in the pseudophysical subthreshold region ($(M_\pi+M_\eta)^2\leq t\leq 4m_N^2$). 
Nevertheless, the $\pi \eta \rightarrow \bar{N} N$ and $\bar{K} K \rightarrow \bar{N} N$ $S$-waves $g_{+}^0(t)$ and $h_{+}^{0,1}(t)$ fulfill a unitarity relation [cf.~\cite{Ditsche:2012fv} and Appendix~\ref{app:isospin}]
\begin{align}\label{eq:unitarity_v3}
    \operatorname{Im} \mathbf{g}(t)=\mathbf{T}^*(t) \mathbf{\Sigma}(t) \mathbf{g}(t)\,, \quad \mathbf{g}(t)=\begin{pmatrix}g_{+}^0(t) \\ \frac{c_0^\eta}{c_0^{KN}} h_{+}^{0,1}(t)\end{pmatrix}\,,\quad \frac{c_0^\eta}{c_0^{KN}}=\sqrt{2}\,,
\end{align}
where
\begin{align}
    \mathbf{\Sigma}(t)=\operatorname{diag}\left(\frac{\lambda^{1/2}(t,M_\pi^2,M_\eta^2)}{t} \theta\left(t-\left(M_\pi+M_\eta\right)^2\right), \sqrt{1-\frac{4M_K^2}{t}} \theta\left(t-4M_K^2\right)\right)\,,
\end{align}
and the convention of the $T$-matrix $\mathbf{T}$ is described in Ref.~\cite{Lu:2020qeo}.
Following the Roy--Steiner equation analysis for the $\pi\pi/N\bar{N}$ system~\cite{Hoferichter:2012wf}, one can apply this approach to the problem at hand. 
The dispersive MO representation of $\mathbf{g}(t)$ is then given by
\begin{align}\label{eq:g_DR}
    \mathbf{g}(t)= \boldsymbol{\Delta}(t)+\left(t-4m_N^2\right)\mathbf{\Omega}^1_0(t)\left[\mathbf{P}_1(t)-\frac{t^2}{\pi} \int_{(M_\pi+M_\eta)^2}^{\infty} \mathrm{d} t^{\prime} \frac{\operatorname{Im} \left[\mathbf{\Omega}^1_0\left(t^{\prime}\right)\right]^{-1} \mathbf{\Delta}\left(t^{\prime}\right)}{t^{\prime 2}\left(t^{\prime}-4m_N^2\right)\left(t^{\prime}-t\right)}\right]\,.
\end{align}
We consider here only the twice-subtracted version of the dispersion relation. 
At low energies, i.e., close to or even below thresholds, the corresponding inhomogeneity $\mathbf{\Delta}$ for the $\pi N\to \eta N$ and $K(\bar{K})N\to K(\bar{K})N$ systems and the first-order subtraction polynomial $\mathbf{P}_1$ can be matched to the chiral representation from next-to-leading-order (NLO) $\mathrm{SU}(3)$ chiral perturbation theory (ChPT).\footnote{The physical threshold $t \ge 4m_N^2$ for $\pi\eta/K\bar{K} \to N\bar{N}$ lies well beyond the validity range of baryon ChPT, while in the low-$t$ subthreshold region of interest the nucleons are far off-shell. The amplitude is therefore obtained via crossing symmetry and analytic continuation from the crossed processes $\pi N\to \eta N$ and $K(\bar{K})N\to K(\bar{K})N$, where both mesons are soft and standard single-baryon ChPT applies directly.}
We perform the matching in the limit in which $\pi\eta/K\bar K$ rescattering is switched off, as done in~\cite{Hoferichter:2012wf} (see also Refs.~\cite{Chen:2015jgl,Granados:2017cib,Wu:2023uva}), i.e., $\mathbf{\Omega}^1_0=\mathbf{1}$, see Appendix~\ref{app:ChPT} for more technical details.
The corresponding amplitudes $\mathbf{g}(t)$ are shown in Fig.~\ref{fig:Amp_isovector_NNbar}. 
The real part of $h^{0,1}_+(t)$ exhibits two distinct logarithmic divergences below 1~GeV that correspond to the pole-term contributions from the $\Lambda$ (large peak at 0.82~GeV) and $\Sigma$ (small peak at 0.73~GeV) hyperons, strictly analogous to the branch point from the nucleon pole in $g^0_+(t)$ that emerges slightly below the $\pi\eta$ threshold.\footnote{Note that here we display only the imaginary part of $h^{0,1}_+(t)$ induced by unitarity [cf.~Eq.~\eqref{eq:unitarity_v3} ], while the imaginary part arising from the LHC associated with the hyperon pole is not shown.}

\begin{figure}[t]
    \centering
    \includegraphics[width=.65\linewidth]{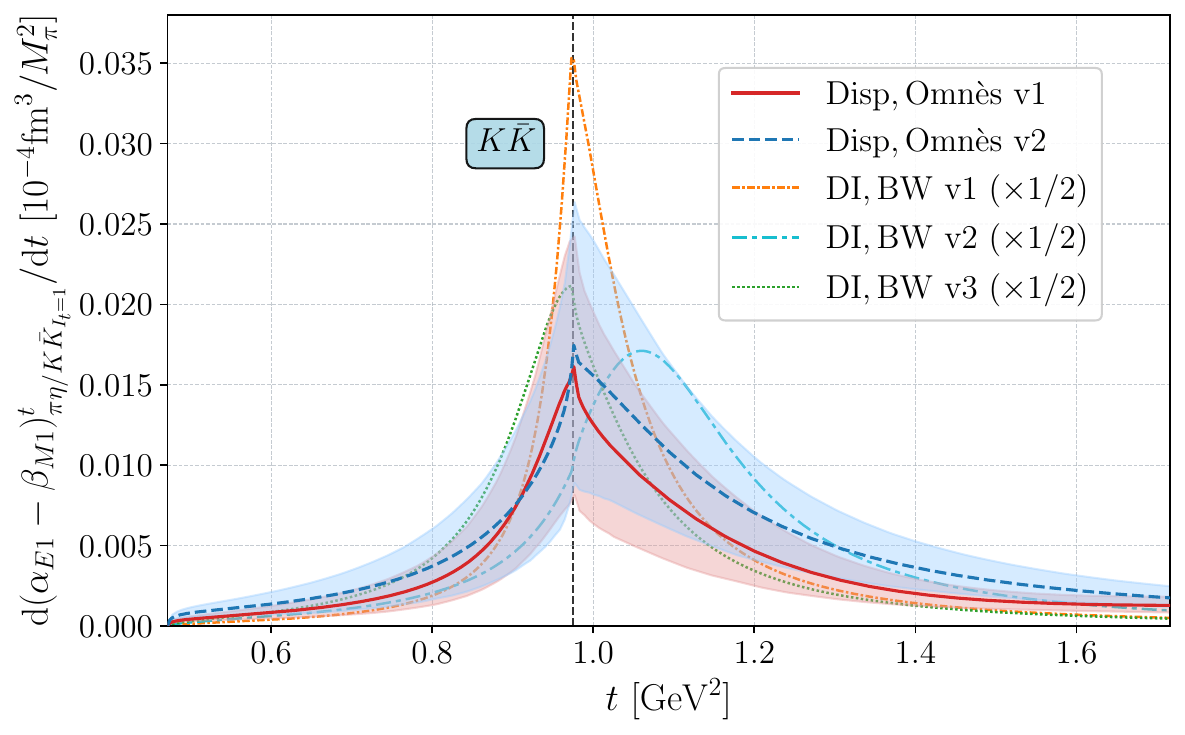}
    \caption{The integrand for the $t$-channel isovector contribution. Here we also show, for comparison, the dispersively improved BW (``DI, BW") parametrizations corresponding to three $a_0$ parameter sets: v1~\cite{Lu:2020qeo}, v2~\cite{Deineka:2024mzt}, and v3~\cite{Kuang-Dai}.}
    \label{fig:t_a0}
\end{figure}

Once the $I=1$ $\gamma\gamma\to \pi\eta/K\bar{K}$ amplitude and the $\pi\eta/K\bar{K}\to N\bar{N}$ amplitudes are determined, the $S$-wave contribution to the $I=1$ BEFT sum rule follows directly. 
The corresponding spectral function is shown in Fig.~\ref{fig:t_a0}. 
Surprisingly, despite sizable differences between the two versions of the $\gamma\gamma\to \pi\eta/K\bar K$ and $\mathbf{g}(t)$ amplitudes derived from the corresponding two versions of the Omn\`es matrices, the resulting spectral functions share a similar cusp-like profile, which indicates that the $a_0(980)$ pole is located in a region of the unphysical Riemann sheet that is connected to the physical boundary only by circling the $K\bar K$ threshold in the complex $t$-plane. 
In this sense, we are confident that the saturation of the (S-wave) $I=1$ BEFT sum rule is moderately sensitive to the details of the respective dispersive representations, provided that unitarity and analyticity are properly implemented. 
Consequently, one can obtain a data-driven dispersive result for the $t$-channel contribution,
\begin{align}\label{eq:BEFT_I=1_value}
    \left(\alpha_{E 1}-\beta_{M 1}\right)_{\pi\eta/K\bar{K}_{I_t=1}}^t=0.19(3)_{\rm MO}\,.
\end{align}
The $I=1$ $D$-wave and higher partial waves can be safely neglected in view of the minor contribution of Eq.~\eqref{eq:BEFT_I=1_value}.

A comparison of the dispersive MO representation with the (narrow) pole approximation is instructive, since the $a_0(980)$ resonance almost dominates the isovector channel.\footnote{Reference~\cite{Lu:2020qeo} discusses the $a_0(1450)$ and finds that the narrow $a_0(1450)$ claimed in Belle Collaboration~\cite{Belle:2009xpa} could be unphysical; instead, it favors a somewhat broad $a_0'$ resonance whose contribution is minor and difficult to isolate from the continuum background. }
The pole approximation suffers from two obvious deficiencies. 
First, its contribution is extremely sensitive to the pole position and to the coupling [cf.~\eqref{eq:pole_form}], both of which are rather poorly determined for $a_0(980)$. 
Second, the sign of the product $g_{a_0 NN}\,F_{a_0\to\gamma\gamma}$ (proportional to the $a_0(980)$ contribution) is not known a priori in the case of the pole approximation. 
With these specifics in mind, we also employ a more realistic, dispersively improved BW parameterization (details in Appendix~\ref{app:BW}).
As shown in Fig.~\ref{fig:t_a0}, the shapes of the spectral function corresponding to the three parameter choices differ visibly near the $a_0$ pole, in stark contrast to the dispersive MO representations.  
Based on the quark model, Ref.~\cite{Schumacher:2007ez} predicted that the $a_0(980)$ contribution to the BEFT sum rule would have the sign opposite to that of $f_0(980)$.
The rigorous dispersive analysis actually confirms this point [cf.~Figs.~\ref{fig:t_Swave} and~\ref{fig:t_a0}].

\section{Results and Discussion}\label{sec:Results}

\subsection{Polarizabilities and uncertainty estimations}

The central result of our work is a dispersive determination of the nucleon backward polarizabilities $(\alpha_{E1}-\beta_{M1})^{p,n}$ using the BEFT sum rule [cf.~Eqs.~\eqref{eq:s_piN_values},~\eqref{eq:s_pipiN_values},~\eqref{eq:Swave_value},~\eqref{eq:Dwave_value}, and~\eqref{eq:BEFT_I=1_value}], which yields
\begin{align}
    \label{eq:p_a-b}(\alpha_{E1}-\beta_{M1})^{p}= {}& (\alpha_{E1}-\beta_{M1})^t_S+(\alpha_{E1}-\beta_{M1})^t_D+(\alpha_{E1}-\beta_{M1})^t_{\pi\eta/K\bar{K}_{I_t=1}} \nonumber\\
    & +(\alpha_{E1}-\beta_{M1})^{s,p}_{\pi N}+(\alpha_{E1}-\beta_{M1})^{s,p}_\mathrm{inel} \nonumber\\
     = {}& 12.82(96)-0.44(5)+0.19(11)-4.96(33)+[0.56+0.16]=8.33(1.02)\,,\\
    \label{eq:n_a-b}(\alpha_{E1}-\beta_{M1})^{n}= {}& (\alpha_{E1}-\beta_{M1})^t_S+(\alpha_{E1}-\beta_{M1})^t_D-(\alpha_{E1}-\beta_{M1})^t_{\pi\eta/K\bar{K}_{I=1_t}} \nonumber\\
    & +(\alpha_{E1}-\beta_{M1})^{s,n}_{\pi N}+(\alpha_{E1}-\beta_{M1})^{s,n}_\mathrm{inel} \nonumber\\
    = {}& 12.82(96)-0.44(5)-0.19(11)-3.69(52)+[0.56+0.09]=9.15(1.10)\,,
\end{align}
where the isovector contribution carries opposite signs for protons and neutrons.

Let us discuss in more detail the uncertainty estimates as shown in Eqs.~\eqref{eq:p_a-b} and~\eqref{eq:n_a-b}, especially for the $t$-channel integrals.
In the isoscalar channel, we set the $S$-wave cutoff at $\sqrt{t}=1.3$~GeV, since the Roy--Steiner equation solutions and the modified MO representation used here are reliable only below this energy. 
To estimate the corresponding systematic uncertainty we assign the associated uncertainty (``cut'') by varying it from $1.2~{\rm GeV}$ to $1.5~{\rm GeV}$, which yields $(\alpha_{E1}-\beta_{M1})^{t}_{S}=12.82(87)_\text{MO}(41)_\text{cut}=12.82(96)$. 
To fully include the $f_2(1270)$ contribution in the $D$-wave, we set the corresponding cutoff at $\sqrt{t}=1.5$~GeV. 
The corresponding uncertainty is estimated by varying it from $1.4~{\rm GeV}$ to $1.7~{\rm GeV}$ and the total $D$-wave uncertainty is $(\alpha_{E1}-\beta_{M1})^{t}_{D}=-0.44(2)_\text{MO}(5)_\text{cut}=-0.44(5)$.
The higher precision of the $\pi\pi/K\bar K_{I_t=0}$ amplitude from the $\pi N$ Roy--Steiner equation analysis renders the associated uncertainty negligibly minor.

The uncertainty of the isovector channel is estimated from three sources. 
First, we use $\sqrt{t}=1.4$~GeV as the cutoff and estimate the associated uncertainty (`cut') by varying the cutoff from 1.3~GeV to 1.6~GeV in the BEFT integral~\eqref{eq:BEFT_t_I=1}. 
Second, there is a small discrepancy between the two dispersive MO representations (`MO') shown in Fig.~\ref{fig:t_a0}. 
Third, for the uncertainty from the ChPT matching (`ChPT'), we quantify it by propagating the errors of the ChPT low-energy constants (LECs) and decay constant $F_0$ with a bootstrap procedure (see the light error bands in Fig.~\ref{fig:t_a0}).
The quoted errors are added in quadrature to give the value in Eq.~\eqref{eq:BEFT_I=1_value}, that is, $\left(\alpha_{E 1}-\beta_{M 1}\right)_{\pi\eta/K\bar{K}_{I_t=1}}^t=0.19(02)_\text{cut}(03)_\text{MO}(10)_\text{ChPT}=0.19(11)$.

For the $s$-channel integral, the dominant source of uncertainty is the single pion CGLN amplitudes, which is roughly estimated by three well-established SAID, MAID and BnGa partial-wave data sets. 
Since the model estimate of the non-$\pi N$ inelastic contribution is numerically about one order of magnitude smaller than the $\pi N$ contribution, we neglect its associated uncertainty in this work. 
We anticipate that future partial-wave analyses of $\gamma N\to\pi\pi N$ will provide more reliable input to the BEFT sum rule.
As mentioned above, the uncertainties associated with the inelastic contributions are currently inaccessible. 
Since their central values are significantly smaller than the total uncertainties, it is reasonable to neglect their contributions to the error estimates.

\subsection{Comparison with previous estimates based on BEFT sum rule}

Although Bernab{\'e}u and Tarrach~\cite{Bernabeu:1977hp} first formulated the BEFT sum rule in its accepted form, they lacked the necessary partial-wave amplitudes for a reliable numerical evaluation. 
The first calculation by Guia{\c{s}}u and Radescu~\cite{Guiasu:1976pq,Guiasu:1976wu} used only the Born approximation, which neglects the $S$-wave $\pi\pi$ correlations associated with the elusive $\sigma$ state. 

Model-dependent calculations incorporating $S$-wave $\pi\pi$ correlations first appeared in Refs.~\cite{Guiasu:1978vg,Budnev:1978ke}.
With the rapid development of low-energy two-photon fusion and pion photoproduction experiments, Holstein and Nathan~\cite{Holstein:1994tw} provided the first modern determination using $\gamma\gamma\to \pi\pi$ amplitudes obtained from the Omn\`es representation supplemented by ChPT~\cite{Donoghue:1993kw}.
They obtained the value $(-5.42)_{s\text{-channel}}+(10.3)_S+(-1.7)_D$ for the proton.
However, Drechsel et al.~\cite{Drechsel:2002ar} provided a quite different value, $(-5.56)_{s\text{-channel}}+(16.46)_{S+D}$, employing $\gamma\gamma\to \pi\pi$ partial-wave amplitudes extracted from dispersive fits to the experimental cross-section data.
The discrepancy between the two results is driven almost exclusively by the differences in the $t$-channel $S$-wave $\pi\pi$ contribution. 
Incidentally, the value quoted in the review~\cite{Schumacher:2005an} is basically combining the Holstein--Nathan and Drechsel et al. values, yielding 14(2) for $(\alpha_{E1}-\beta_{M1})^t_S$. 
Consequently, the central value and the assigned uncertainty should be regarded as a compromise and not an independent and definite prediction.

Our results, Eqs.~\eqref{eq:p_a-b} and~\eqref{eq:n_a-b}, resolve the long-standing tension among previous $t$-channel determinations and provide an unambiguous value $12.82(96)$ for the dominant $S$-wave contribution, including both the $\pi\pi$ and $K\bar K$ rescattering effects. 
In particular, we employ state-of-the-art $\pi\pi/K\bar{K}\to N\bar N$ partial-wave amplitudes from Roy--Steiner equations, valid up to energies below the $N\bar N$ threshold. 
The $\gamma\gamma$ amplitudes are taken from the latest coupled-channel MO dispersive analyses.
For the first time, the contribution of the isovector $a_0(980)$ is constructed using the MO approach to estimate its subleading effect. 
In addition, the $s$-channel contribution remains moderately unchanged and we only update the numerical input using the most recent CGLN amplitudes.

\subsection{Comparison with other results}

\begin{figure}[t]
    \centering
    \includegraphics[width=.8\linewidth]{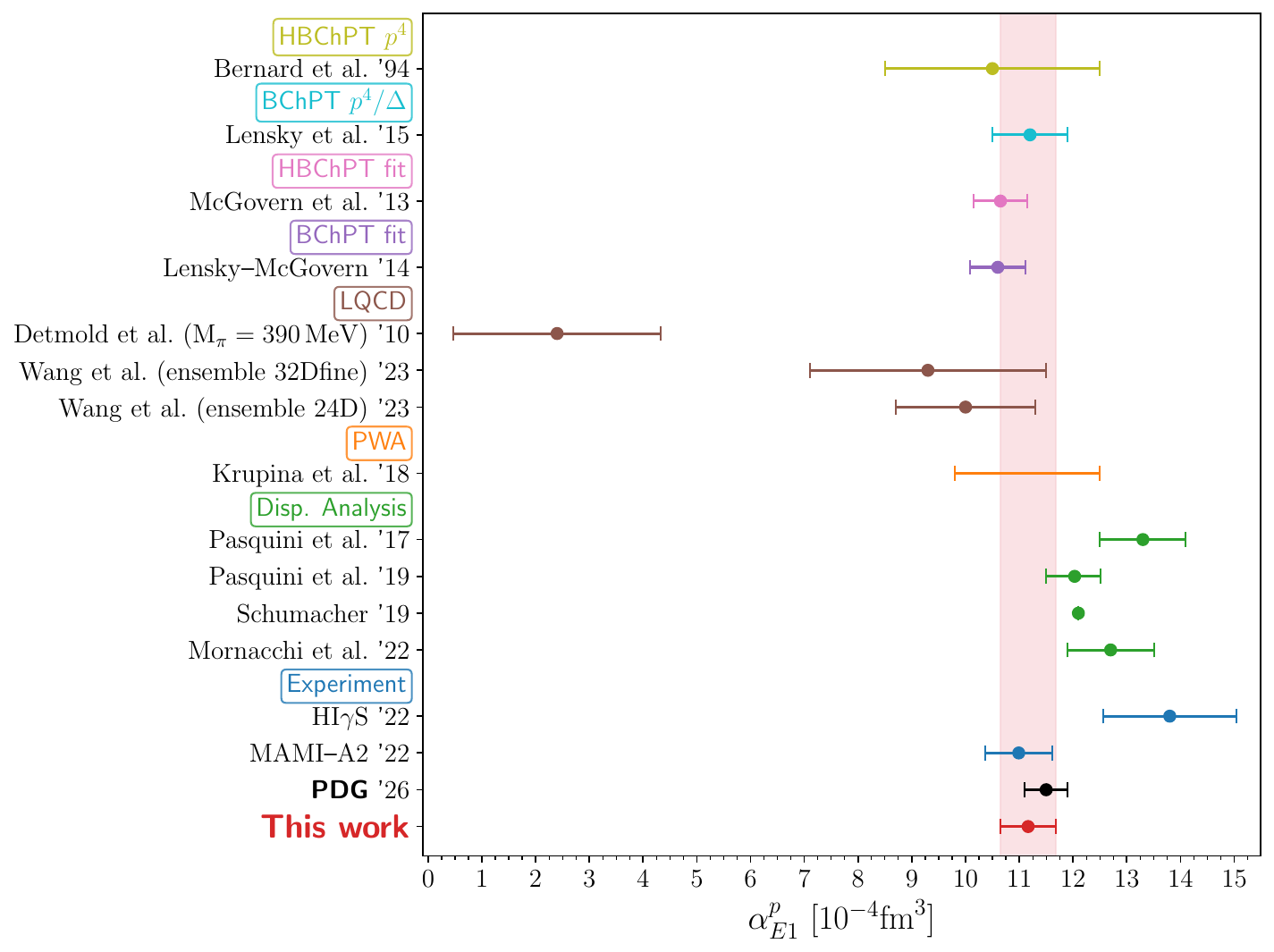}
    \caption{Comparison of our result for the proton electric dipole polarizability, $\alpha_{E1}^p$, with experimental determinations~\cite{A2:2021wmm,Li:2022vnz} and various phenomenological calculations/fits, including dispersive analysis~\cite{Pasquini:2017ehj,Pasquini:2019nnx,Schumacher:2019ikn,Mornacchi:2022cln}, partial wave analysis (PWA)~\cite{Krupina:2017pgr}, LQCD~\cite{Detmold:2010ts,Wang:2023omf}, covariant baryon ChPT (BChPT)~\cite{Lensky:2014efa,Lensky:2015awa}, and heavy-baryon ChPT (HBChPT)~\cite{Bernard:1993ry,McGovern:2012ew}.}
    \label{fig:alpha_p}
\end{figure}
\begin{figure}[t]
    \centering
    \includegraphics[width=.8\linewidth]{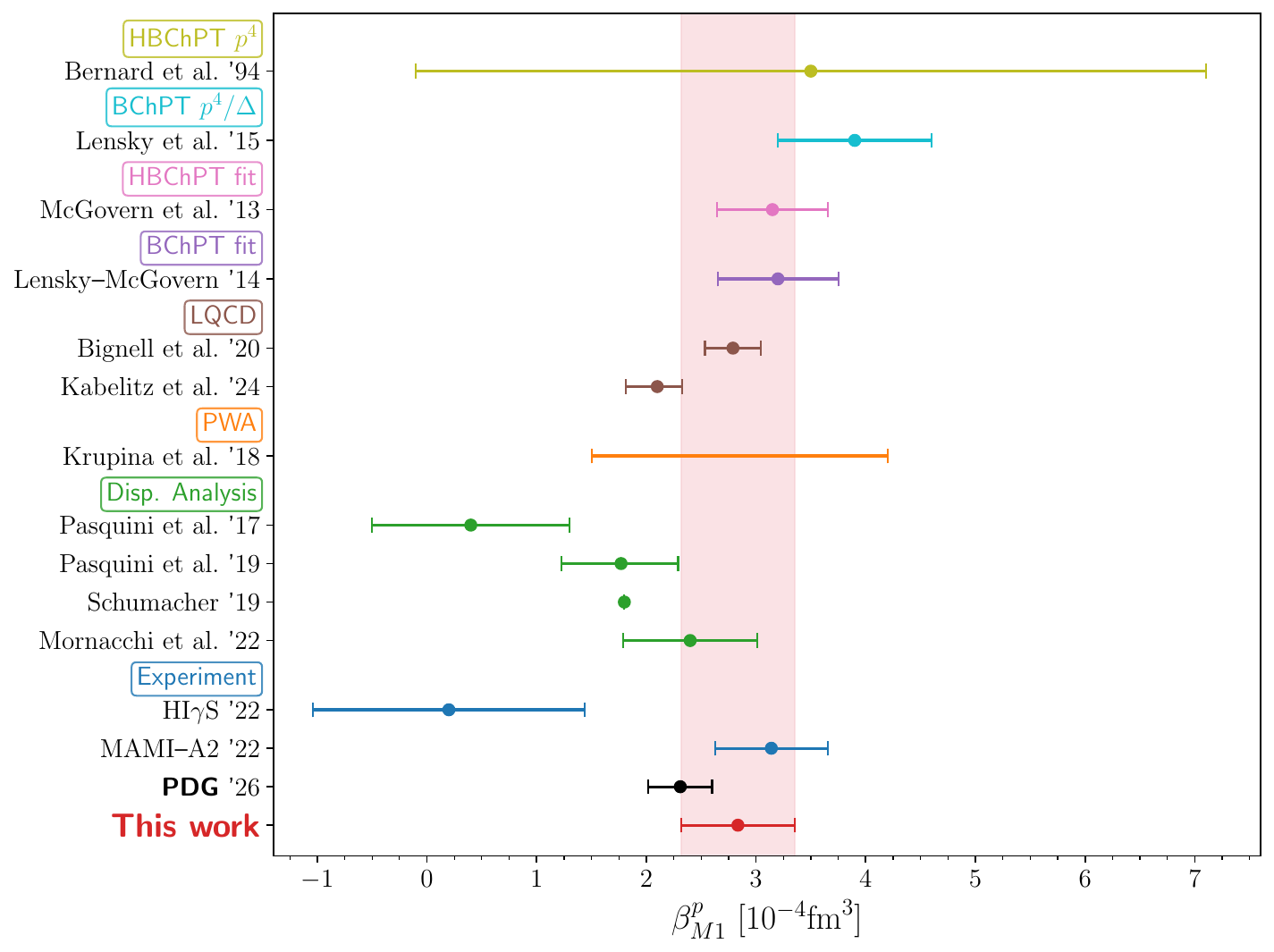}
    \caption{Comparison of our result for the proton magnetic dipole polarizability, $\beta_{M1}^p$, with experimental determinations~\cite{A2:2021wmm,Li:2022vnz} and various phenomenological calculations/fits, including dispersive analysis~\cite{Pasquini:2017ehj,Pasquini:2019nnx,Schumacher:2019ikn,Mornacchi:2022cln}, PWA~\cite{Krupina:2017pgr}, LQCD~\cite{Bignell:2020xkf,Kabelitz:2024aye}, BChPT~\cite{Lensky:2014efa,Lensky:2015awa}, and HBChPT~\cite{Bernard:1993ry,McGovern:2012ew}.}
    \label{fig:beta_p}
\end{figure}
\begin{figure}[t]
    \centering
    \includegraphics[width=.8\linewidth]{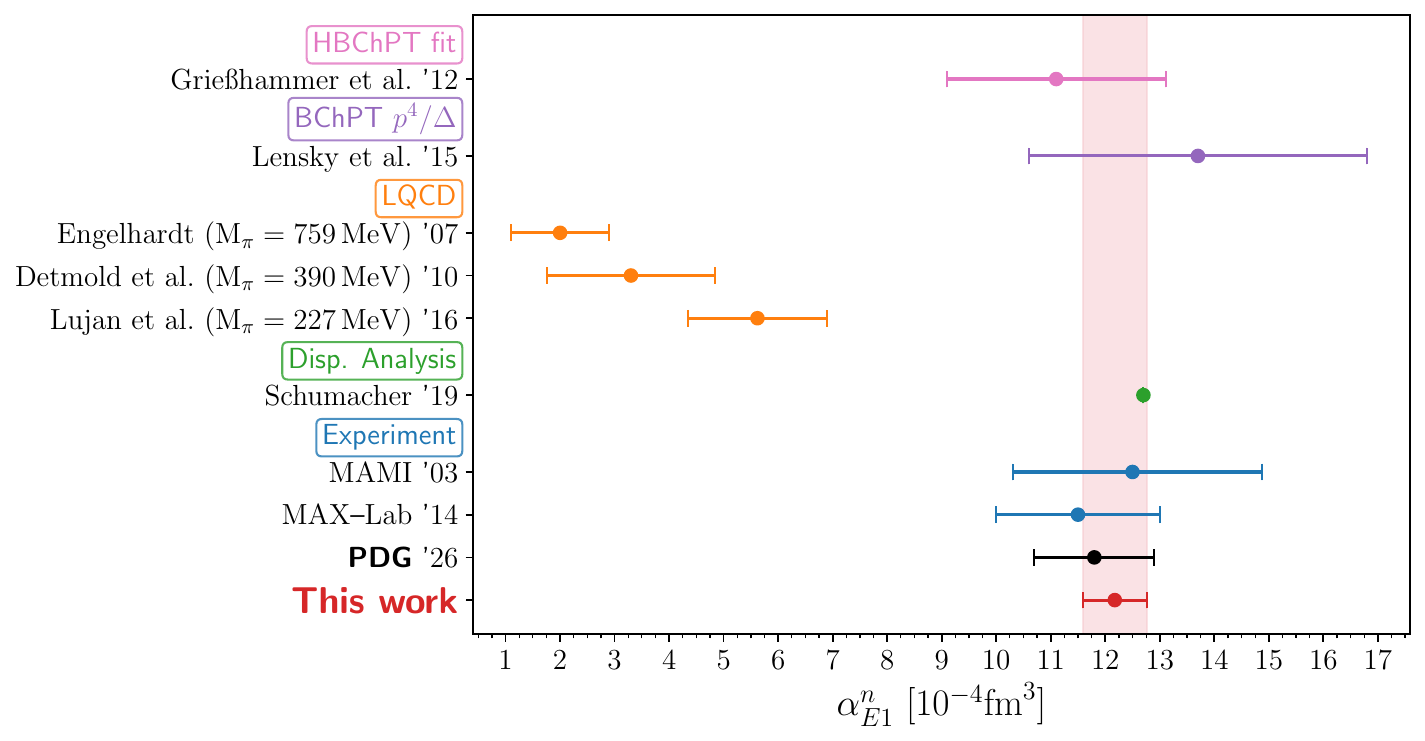}
    \caption{Comparison of our result for the neutron electric dipole polarizability, $\alpha_{E1}^n$, with experimental determinations~\cite{Kossert:2002jc,Kossert:2002ws,COMPTONMAX-lab:2014cve} and various phenomenological calculations/fits, including dispersive analysis~\cite{Schumacher:2019ikn}, LQCD~\cite{Engelhardt:2007ub,Detmold:2010ts,Lujan:2016ffj}, BChPT~\cite{Lensky:2015awa}, and HBChPT~\cite{Griesshammer:2012we}.}
    \label{fig:alpha_n}
\end{figure}
\begin{figure}[t]
    \centering
    \includegraphics[width=.8\linewidth]{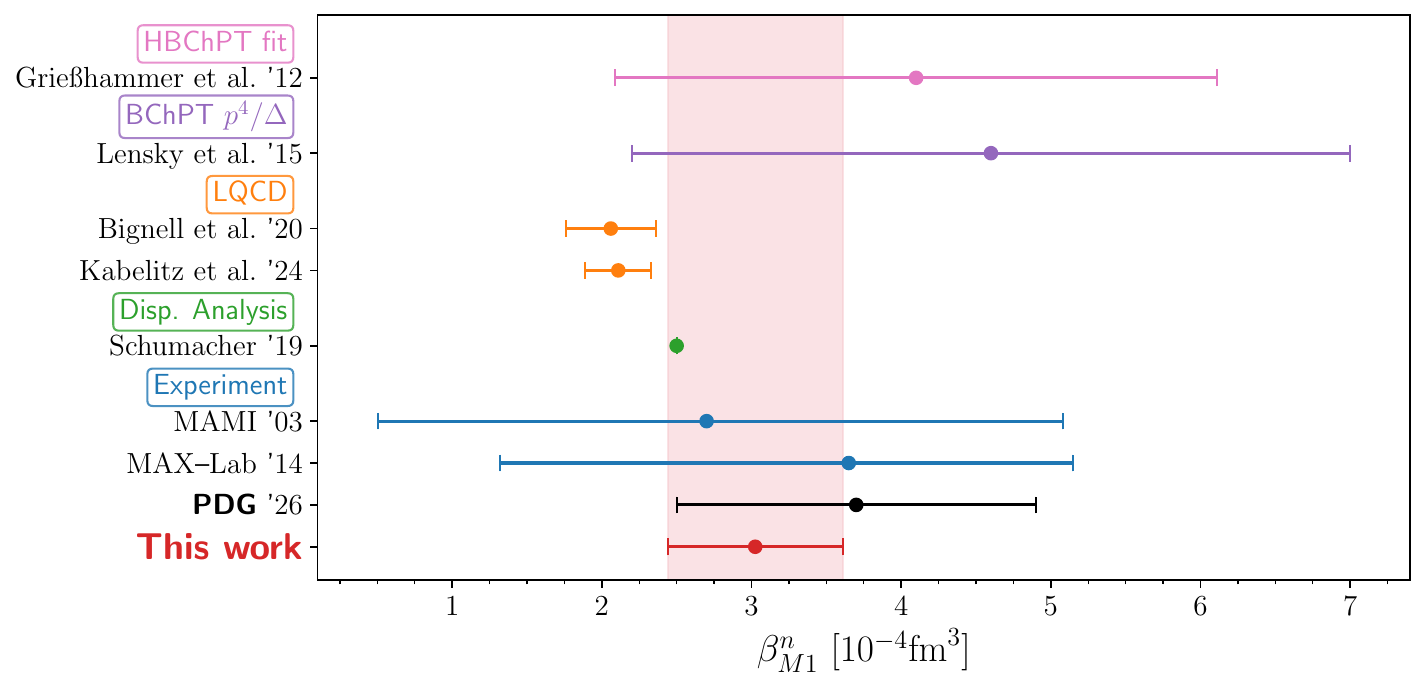}
    \caption{Comparison of our result for the neutron magnetic dipole polarizability, $\beta_{M1}^n$, with experimental determinations~\cite{Kossert:2002jc,Kossert:2002ws,COMPTONMAX-lab:2014cve} and various phenomenological calculations/fits, including dispersive analysis~\cite{Schumacher:2019ikn}, LQCD~\cite{Bignell:2020xkf,Kabelitz:2024aye}, BChPT~\cite{Lensky:2015awa}, and HBChPT~\cite{Griesshammer:2012we}.}
    \label{fig:beta_n}
\end{figure}

Empirical evaluations of the Baldin sum rule [cf.~\eqref{eq:BL_SR}] based on the total photoabsorption cross section yield the most precise sums for the proton~\cite{Gryniuk:2015eza} and neutron~\cite{Levchuk:1999zy} dipole polarizabilities,
\begin{align}\label{eq:BL_SR_values}
\begin{aligned}
& (\alpha_{E 1}+\beta_{M 1})^p=14.0(2)\,, \\
& (\alpha_{E 1}+\beta_{M 1})^n=15.2(4)\,.
\end{aligned}
\end{align}
By combining Eqs.~\eqref{eq:p_a-b},~\eqref{eq:n_a-b} and~\eqref{eq:BL_SR_values}, we can finally obtain the predictions of the BEFT sum rule for the individual dipole polarizabilities that do not directly rely on any low-energy CS empirical data,
\begin{align}\label{eq:polarizability_values}
    \begin{aligned}
    \alpha_{E1}^p & =11.17(51)_\text{BEFT}(10)_\text{Baldin}=11.17(52)\,,\\
    \beta_{M1}^p & =2.84(51)_\text{BEFT}(10)_\text{Baldin}=2.84(52)\,,\\
    \alpha_{E1}^n & =12.18(55)_\text{BEFT}(20)_\text{Baldin}=12.18(59)\,,\\
    \beta_{M1}^n & =3.03(55)_\text{BEFT}(20)_\text{Baldin}=3.03(59)\,.\\
    \end{aligned}
\end{align}
The quoted errors from the BEFT and Baldin sum rules are added in quadrature.
The nucleon polarizability values reported in Eq.~\eqref{eq:polarizability_values} are shown as red points in Figs.~\ref{fig:alpha_p},~\ref{fig:beta_p},~\ref{fig:alpha_n} and~\ref{fig:beta_n} (bottom row in each figure).
The vertical red bands represent the errors evaluated, as explained above.
In the same figures, the new results of $\alpha_{E1}$ and $\beta_{M1}$ from the BEFT sum rule are compared with some of the averaged values of the PDG~\cite{ParticleDataGroup:2026mpi}, the experimental extractions~\cite{A2:2021wmm,Li:2022vnz,Kossert:2002jc,Kossert:2002ws,COMPTONMAX-lab:2014cve} and existing theoretical calculations using DRs~\cite{Pasquini:2017ehj,Pasquini:2019nnx,Schumacher:2019ikn,Mornacchi:2022cln}, PWA~\cite{Krupina:2017pgr}, LQCD~\cite{Detmold:2010ts,Wang:2023omf,Bignell:2020xkf,Kabelitz:2024aye,Engelhardt:2007ub,Lujan:2016ffj}, BChPT~\cite{Lensky:2014efa,Lensky:2015awa} and HBChPT~\cite{Bernard:1993ry,McGovern:2012ew}, respectively.

For the proton, modern low-energy CS data now enable the extraction of the dipole polarizabilities within two state-of-the-art theoretical frameworks: DRs and ChPT. 
The quoted uncertainties now appear controlled below the $10\%$ level. 
Unfortunately, the central values obtained from the DR and ChPT fits are separated by about $2\sigma$, indicating a tension between the two approaches for $\alpha_{E1}^p$ and $\beta_{M1}^p$, cf.~Figs.~\ref{fig:alpha_p} and~\ref{fig:beta_p}. 
Inconsistencies among the experimental data subsets used in the analyses, and partly also different theoretical estimates of the higher-order spin-polarizability contributions to the RCS cross section, are the principal cause of this discrepancy, not an intrinsic ``model dependence" of the theoretical approaches~\cite{Griesshammer:2012we,Krupina:2017pgr,Pasquini:2019nnx}. 
To date, the scientific community has not reached a consensus on the definition of the proton RCS data set below the pion-production threshold~\cite{Pasquini:2019nnx}.
Our new dispersive determination of $\alpha_{E1}^p$~\eqref{eq:polarizability_values} slightly favors the ChPT extractions within errors and is also fully consistent with the recent result $\alpha_{E1}^p=10.99 \pm 0.16 \pm 0.47 \pm 0.17 \pm 0.34$ from MAMI~\cite{A2:2021wmm}. 
Likewise, for the magnetic polarizability $\beta_{M1}^p$, our result exhibits a mild preference for the ChPT results while remaining in good agreement with both the latest MAMI value, $3.14 \pm 0.21 \pm 0.24 \pm 0.20 \pm 0.35$~\cite{A2:2021wmm}, and the recent simultaneous (fixed-$t$) dispersion extraction, $2.4\pm 0.6 \pm 0.1$~\cite{Mornacchi:2022cln}.
Currently, the precision of the results $\alpha_{E1}^p$ and $\beta_{M1}^p$ of the BEFT sum rule is comparable to that achieved by the state-of-the-art DR and ChPT fits.
More importantly, we stress that our dispersive determination does not rely on low-energy CS data and thus provides an independent, model-insensitive benchmark for future proton CS work.

Predicting the neutron dipole polarizabilities is the most important application of the BEFT sum rule, yet these values are somewhat contentious owing to the absence of direct neutron CS experiments; see Ref.~\cite{Schumacher:2005an} for a comprehensive historical review.
In Figs.~\ref{fig:alpha_n} and~\ref{fig:beta_n} we display the analyses of the two most advanced deuteron CS experiments.
Compared to the proton, the neutron polarizabilities are far less precisely determined, a direct consequence of the lack of free neutron targets.
Achieving a precision comparable to that of the proton will require additional high-accuracy $\gamma$-deuteron~\cite{Kossert:2002jc,Kossert:2002ws,COMPTONMAX-lab:2014cve} or $\gamma$-${}^3\text{He}/{}^4\text{He}$ measurements~\cite{Li:2019irp,Zhou:2026bgr}, together with sophisticated theoretical analyses~\cite{Margaryan:2018opu,Griesshammer:2024twu}.
Our results provide an independent validation of the extraction procedure and report the first dispersive determination of neutron polarizabilities reaching a precision comparable to that of the proton with well-controlled errors.
\begin{figure}[t]
    \centering
    \includegraphics[width=.6\linewidth]{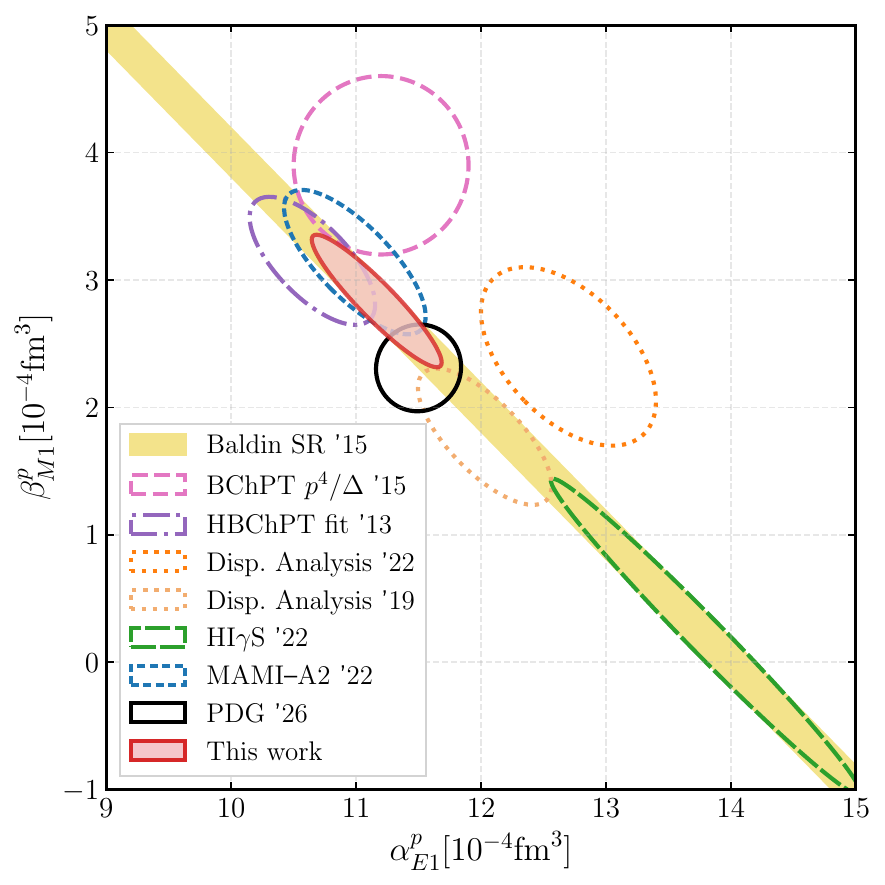}
    \caption{Plot of $\alpha^p_{E1}$ versus $\beta^p_{M1}$ for the proton, obtained from different experiments and theories. References and declarations are given in the main text, cf.~also Figs.~\ref{fig:alpha_p} and~\ref{fig:beta_p}.}
    \label{fig:2Dplot_p}
\end{figure}
\begin{figure}[t]
    \centering
    \includegraphics[width=.6\linewidth]{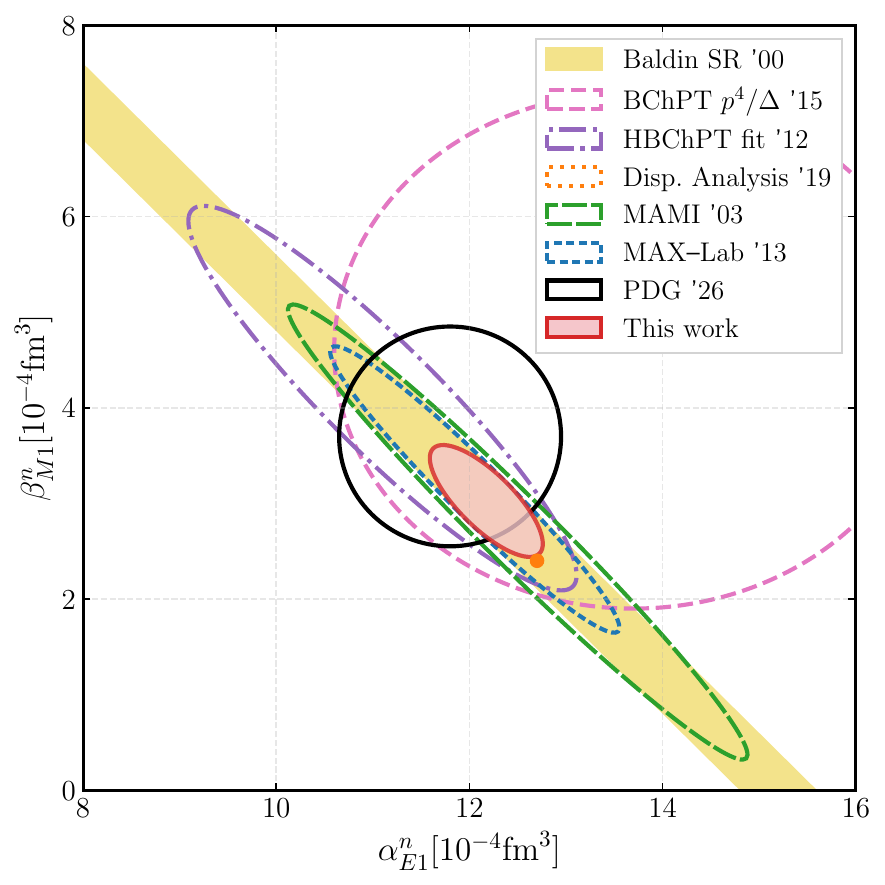}
    \caption{Plot of $\alpha^n_{E1}$ versus $\beta^n_{M1}$ for the neutron, obtained from different experiments and theories. References and declarations are given in the main text, cf.~also Figs.~\ref{fig:alpha_n} and~\ref{fig:beta_n}.}
    \label{fig:2Dplot_n}
\end{figure}

Figures~\ref{fig:2Dplot_p} and~\ref{fig:2Dplot_n} show the scalar polarizabilities obtained in this work as red full ellipses.
The orange bands represent the constraint by the Baldin sum rule, cf.~\eqref{eq:BL_SR_values}.
Also shown are various previously published fit results and predictions of these two parameters.

Here we briefly comment on recent LQCD calculations of the nucleon magnetic polarizabilities, with particular attention to the neutron. 
Calculations employing the background-field method in LQCD~\cite{Bignell:2020xkf,Kabelitz:2024aye} have reached uncertainties that are competitive with current experimental determinations, particularly for the neutron, where the reported LQCD errors are smaller than the present experimental uncertainties. 
It should be stressed that Refs.~\cite{Bignell:2020xkf,Kabelitz:2024aye} perform calculations mainly at rather large pion masses, so the extracted values at the physical mass point depend on chiral extrapolations. 
However, the perturbative chiral expansion for all polarizabilities becomes intrinsically unreliable once $M_\pi$ approaches roughly 300~MeV~\cite{Griesshammer:2015ahu}.
Since the $\pi\pi$ $S$-wave is important, the situation is similar to the observation in Ref.~\cite{Cao:2025dkv} that the pion-mass dependence can be abrupt when the pions are so heavy that the $\sigma$ becomes a bound state. Explicit calculation of the chiral extrapolation using a nonperturbative method is desirable.
Furthermore, these determinations have also been hindered by the light-quark exceptional configuration problem near the physical pion mass~\cite{Kabelitz:2024cig}. 
Previous works~\cite{Bignell:2020xkf,Kabelitz:2024aye,Kabelitz:2024cig} neglect or remove such exceptional configurations to stabilize the background-field extraction, but there is no first-principles justification for this procedure. 
Further study, e.g., using the four-point function approach near the physical pion mass point~\cite{Wilcox:2021rtt,Wang:2023omf}, is required to address the challenges of light-quark dynamics in nucleon four-point correlators and to achieve reliable \emph{ab initio} LQCD results for nucleon magnetic polarizabilities in the physical situation.

\subsection{A sum rule for nucleon isovector polarizabilities}

\begin{figure}[t]
    \centering
    \includegraphics[width=.6\linewidth]{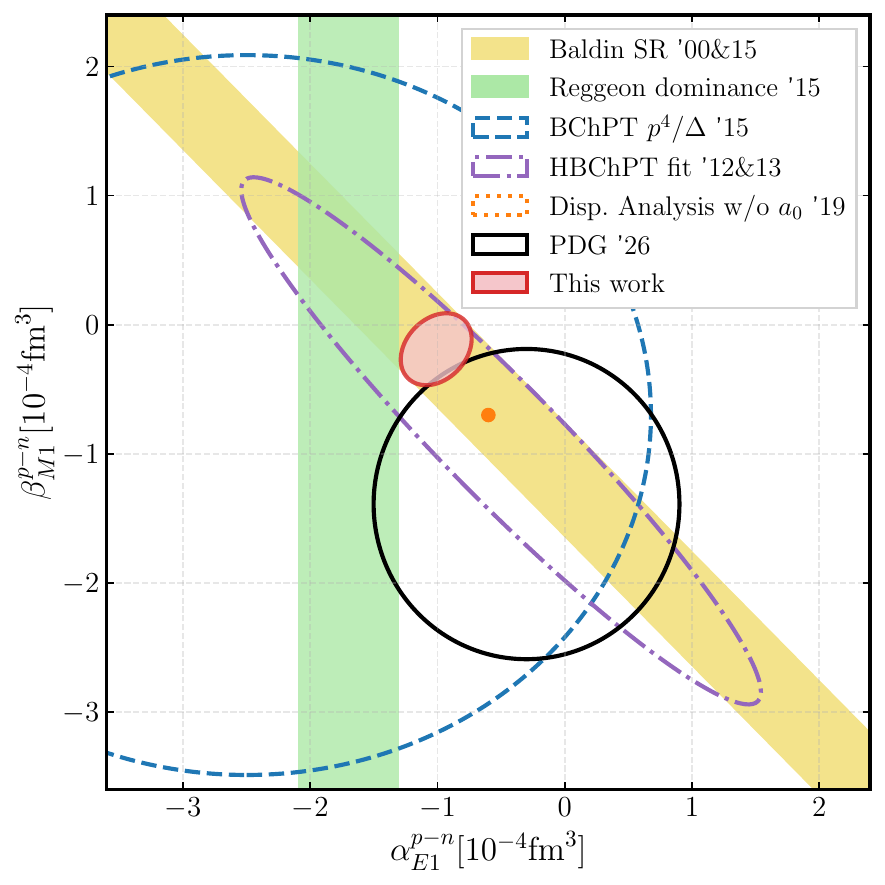}
    \caption{Plot of $\alpha^{p-n}_{E1}$ versus $\beta^{p-n}_{M1}$ for the nucleon, obtained from different theories. References and declarations are given in the main text, cf.~also Figs.~\ref{fig:alpha_p},~\ref{fig:beta_p},~\ref{fig:alpha_n} and~\ref{fig:beta_n}. Note that Ref.~\cite{Schumacher:2019ikn} does not involve the `$a_0$' contribution.}
    \label{fig:2Dplot_p-n}
\end{figure}
The electromagnetic part of the proton--neutron mass difference, $\delta m^{\mathrm{em}} \equiv m_p^{\mathrm{em}} - m_n^{\mathrm{em}}$, is conventionally connected to the isovector magnetic polarizability $\beta_{M1}^{p-n} \equiv \beta_{M1}^{p} - \beta_{M1}^{n}$ via the Cottingham sum rule~\cite{Cottingham:1963zz}. 
Although there is some scale-dependence in assigning strong and electromagnetic self-energy differences~\cite{Gasser:2003hk}, there is broad agreement that $\delta m^{\mathrm{em}}$ and $\beta_{M1}^{p-n}$ are connected because $\beta_{M1}^{p-n}$ corresponds to a key component of the integrand (subtraction function) in the Cottingham sum rule~\cite{Walker-Loud:2012ift,Thomas:2014dxa,Erben:2014hza,Tomalak:2018dho}.
However, the quantitative strength of the relation depends on assumptions about the behavior of the integrand, a point that remains debated at present~\cite{Gasser:1974wd,Walker-Loud:2012ift,Thomas:2014dxa,Erben:2014hza,Tomalak:2018dho,Gasser:2015dwa,Gasser:2020hzn,Gasser:2020mzy}.

By combining Eqs.~\eqref{eq:p_a-b} and~\eqref{eq:n_a-b} with the Baldin sum rule~\eqref{eq:BL_SR_values}, one can readily write two sum rules for the isovector nucleon polarizabilities,
\begin{align}\label{eq:isovector_SR}
    \begin{aligned}
    \alpha^{p-n}_{E1} & =\frac{1}{2}\left[\left(\alpha_{E1}+\beta_{M1}\right)^{p-n}+\left(\alpha_{E1}-\beta_{M1}\right)^{p-n}\right] \\
    & =\frac{1}{2}\left[\left(\alpha_{E1}+\beta_{M1}\right)^{p-n}+\left(\alpha_{E1}-\beta_{M1}\right)^{p-n}_{\pi N}+\left(\alpha_{E1}-\beta_{M1}\right)^{p-n}_\mathrm{inel}+\left(\alpha_{E1}-\beta_{M1}\right)^{p-n}_{\pi\eta/K\bar{K}_{I_t=1}}\right]\,, \\
    \beta^{p-n}_{M1} & =\frac{1}{2}\left[\left(\alpha_{E1}+\beta_{M1}\right)^{p-n}-\left(\alpha_{E1}-\beta_{M1}\right)^{p-n}\right] \\
    & =\frac{1}{2}\left[\left(\alpha_{E1}+\beta_{M1}\right)^{p-n}-\left(\alpha_{E1}-\beta_{M1}\right)^{p-n}_{\pi N}-\left(\alpha_{E1}-\beta_{M1}\right)^{p-n}_\mathrm{inel}-\left(\alpha_{E1}-\beta_{M1}\right)^{p-n}_{\pi\eta/K\bar{K}_{I_t=1}}\right]\,,
    \end{aligned}
\end{align}
which yields 
\begin{gather}
    \alpha^{p-n}_{E1}=-1.01(28)\,,\quad \beta^{p-n}_{M1}=-0.19(28)\,,
\end{gather} 
Notice that this sum rule is sensitive to the details of the isovector channel, because the subtle isoscalar contribution from the $t$-channel BEFT sum rule is completely excluded here. 
Interestingly, the nucleon isovector magnetic polarizability provides a test of the Reggeon dominance ansatz. 
Under the assumption of Reggeon dominance and employing the Baldin sum rule, Ref.~\cite{Gasser:2015dwa} predicted the isovector magnetic polarizability.\footnote{The corresponding sum rule for $\alpha_{E1}^{p-n}$ used in Ref.~\cite{Gasser:2015dwa} is indeed an improved version of the Bernab\'eu--Tarrach sum rule~\cite{Bernabeu:1974mb}. For a discussion of the convergence issues of the original Bernab\'eu--Tarrach sum rule, see recent Ref.~\cite{Biloshytskyi:2023fyv}.}
Our data-driven results~\eqref{eq:p_a-b} and~\eqref{eq:n_a-b} differ from the results in Ref.~\cite{Gasser:2015dwa}, $\alpha^{p-n}_{E1}=-1.7(4)$, by about $1.4\sigma$; see Fig.~\ref{fig:2Dplot_p-n}.
This consequence warrants careful interpretation. 
This result introduces a mild tension with the standard Reggeon dominance hypothesis in real and virtual CS~\cite{Elitzur:1970yy,Gasser:1974wd,Gasser:2020hzn}.
In turn, it appears to revitalize the long-standing conjecture~\cite{Brodsky:1971zh,Brodsky:1972vv} (see in particular~\cite{Brodsky:2008qu,Muller:2015vha}) that $T_1(\nu, Q^2)$ contains a non-negligible $J=0$ fixed pole in the angular momentum plane, arising from local coupling of two photons to the fundamental quark currents of a hadron and appearing as a real, $\nu$-independent asymptotic constant.

\section{Summary and Outlook}\label{sec:Summary}

In this work, we have presented a comprehensive study of the Bernab\'eu--Ericson--Ferro Fontan--Tarrach sum rule based on a hyperbolic dispersion relation for the nucleon static dipole polarizability difference $\alpha_{E1}-\beta_{M1}$. 
Our analysis includes $s$-channel contributions from single-pion photoproduction and non-$\pi N$ inelastic states, as well as $t$-channel contributions from the isoscalar ($I_t=0$) $\pi\pi/K\bar{K}$ $S$- and $D$-waves and the isovector ($I_t=1$) $\pi\eta/K\bar{K}$ channel, which is evaluated dispersively here for the first time.
Considerable recent progress has been made in this regard. 
In the case of the isoscalar $t$-channel, successful Muskhelishvili–Omn\`es dispersive analyses of the $\gamma \gamma \rightarrow \pi \pi/K\bar K$ and Roy--Steiner equation analyses of the $\pi \pi/K\bar{K} \rightarrow N\bar N$ have allowed a reliable data-driven estimate of this contribution.
Using chiral perturbation theory, we have for the first time extended the approach to the isovector channel, thereby obtaining a robust estimate of the $a_0(980)$ contribution.
In the case of the $s$-channel, a modern multipole analysis of the $\pi N$ intermediate state enables a robust calculation of this piece.
 
Our main goal is to provide self-consistent  benchmark results for future experimental and theoretical analyses. The results are derived from a detailed data-driven evaluation of the $t$-channel intermediate states rather than global fits to the world CS database, avoiding the ambiguity introduced by the inconsistent database~\cite{Krupina:2017pgr}. 
For protons, our numerical precision is comparable to state-of-the-art chiral perturbation theory and dispersion relation fits; for neutrons, our work yields the most precise data-driven theoretical constraint available to date.

Regarding $\pi\eta/K\bar{K}_{I_t=1} \to N\bar N$, a complete account of the crossed-channel contributions would require an extension of the $\pi N$ Roy--Steiner equation analysis to $K N$ and $\bar{K} N$ systems, which we leave for future work.
In addition, the forthcoming two-photon fusion data at BESIII may soon provide a considerable improvement of the database, in both total and differential cross sections~\cite{Kussner:2022dft,Kussner:2024ryb}.
More interestingly, the calculations presented here will improve the dispersive evaluation of the quark-sector of the nucleon gravitational $D$-term form factor in Ref.~\cite{Pasquini:2014vua}, while a dispersive analysis of the total gravitational form factors has been carried out in Refs.~\cite{Cao:2024zlf,Cao:2025dkv}. Relevant work is in progress~\cite{Cao2026DispDVCS}.

\begin{acknowledgements}

We are grateful to Bachir Moussallam for kindly sending the results of Refs.~\cite{Garcia-Martin:2010kyn,Lu:2020qeo} and to Xiu-Lei Ren for sharing the corresponding correlation matrix of Ref.~\cite{Ren:2012aj}.
In particular, XHC thanks Shi-Qing Kuang and Bing Wu for useful discussions.
This work is supported in part by National Natural Science Foundation of China under Grants No.~12125507, 12447101, 12322502, and 12335002, by the Chinese Academy of Sciences under Grant No.~YSBR-101, and by the Hunan Province Major Basic Research Projects with Grant No.~2026JC0006.

\end{acknowledgements}

\appendix

\section{Derivation of the Backward Optical Theorem}\label{app:backward_optical}

In this appendix, we present a step-by-step derivation of the backward unitarity relation~\eqref{eq:fpi_Im}, explicitly tracking the nucleon spin degrees of freedom, the canonical-to-helicity frame transformations, which are missing in Ref.~\cite{Schumacher:2005an}.
Let $\omega$ and $k$ denote the incoming photon energy in the lab and c.m. frames, respectively. 
With the total c.m. energy $W=\sqrt{s}$, the Mandelstam variable $s$ and the c.m. photon momentum are given by $s = W^2 = m_N^2 + 2 m_N \omega$ and $k = (s - m_N^2)/(2W) = m_N \omega / W$, respectively, while the scattered photon energy in the lab frame at backward angle ($\theta_{\mathrm{lab}}=\theta=\pi$) is $\omega' = \omega m_N^2 / s$. 
At the collinear angles ($\theta = 0, \pi$), the CS invariant matrix element $T_{fi}$ in Eq.~\eqref{eq:Tfi} between canonical (Pauli) two-component nucleon spinors may be written in the form~\cite{Babusci:1998ww}
\begin{align}
    \frac{T_{fi}}{8\pi m_N} &= f_{0,\pi}(\omega)\,\boldsymbol{\epsilon}^{\prime*}\cdot\boldsymbol{\epsilon} + i g_{0,\pi}(\omega)\,\boldsymbol{\sigma}\cdot\left(\boldsymbol{\epsilon}^{\prime*}\times\boldsymbol{\epsilon}\right)\,, \nonumber\\
    \frac{T_{fi}}{8\pi W} &= \hat{f}_{0,\pi}(W)\,\boldsymbol{\epsilon}^{\prime*}\cdot\boldsymbol{\epsilon} + i \hat{g}_{0,\pi}(W)\,\boldsymbol{\sigma}\cdot\left(\boldsymbol{\epsilon}^{\prime*}\times\boldsymbol{\epsilon}\right)\,.
\end{align}
Because the Lorentz boost relating the lab and c.m. frames at $\theta=0,\pi$ is collinear with both the initial and final nucleon three-momenta, the associated Wigner rotation on the canonical spin states reduces to the identity transformation, and the transverse photon polarization vectors $\boldsymbol{\epsilon}, \boldsymbol{\epsilon}'$ remain invariant. 
Matching the two expressions above immediately yields
\begin{align}\label{eq:app_lab_cm_f}
    f_{0,\pi}(\omega) = \frac{W}{m_N}\hat{f}_{0,\pi}(W)\,, \quad g_{0,\pi}(\omega) = \frac{W}{m_N}\hat{g}_{0,\pi}(W)\,,
\end{align}
which is fully consistent with the relation between the unpolarized differential cross sections in the two frames, $(\omega'/\omega)^2 \overline{|T_{fi}|^2}/(64\pi^2 m_N^2)$ and $\overline{|T_{fi}|^2}/(64\pi^2 s)$, given the backward solid-angle Jacobian $\mathrm{d}\Omega_{\mathrm{cm}}/\mathrm{d}\Omega_{\mathrm{lab}} = s/m_N^2 (\omega'/\omega)^2 = m_N^2/s$.

To perform the partial-wave expansion in the c.m. frame, we use the two-body Jacob--Wick helicity basis~\cite{Jacob:1959at}. 
Taking the incoming photon along the $+z$-axis with helicity $\lambda=\pm 1$ and the incoming nucleon along the $-z$-axis with helicity $h=\pm 1/2$, we label the two-body state by $a=(\lambda,h)$ with net helicity projection $\mu_a = \lambda - h$ along $+z$. The single-particle states along the $z$-axis are fixed as~\cite{Walker:1968xu}
\begin{align}
    \boldsymbol{\epsilon}_\lambda(+z) = -\frac{\lambda\hat{\mathbf{x}} + i\hat{\mathbf{y}}}{\sqrt{2}}\,, \quad
    \psi_h^\mathrm{in}(-z) = e^{-i\pi \sigma_y / 2} \chi_h(+z) = \tau_h \chi_{-h}(+z)\,,
\end{align}
where $\chi_h(+z)$ is the canonical nucleon spinor quantized along $+z$ and $\tau_h = (-1)^{1/2-h}$. 
Rotating the two-body state to a general scattering direction $\Omega=(\theta,\phi)$ via $|\Omega;a\rangle = R_z(\phi)R_y(\theta)|+z,\lambda;-z,h\rangle$, the angular-normalized eigenstates of half-integer total angular momentum $J \ge |\mu_a|$ are defined as~\cite{Jacob:1959at}
\begin{align}
    |JM;a\rangle = \sqrt{\frac{2J+1}{4\pi}} \int \mathrm{d}\Omega\, D^{J*}_{M\mu_a}(\phi,\theta,0)|\Omega;a\rangle\,.
\end{align}
With $d^J_{m'm}(\theta) = \langle J m'|e^{-i\theta J_y}|J m\rangle$, the c.m. helicity amplitude $\hat{F}_{ba}(\theta)$ is expanded as
\begin{align}
    \hat{F}_{ba}(\theta) = \sum_{J} (2J+1) d^J_{\mu_a \mu_b}(\theta) \hat{F}^J_{ba}(W)\,.
\end{align}
At the backward angle $\theta=\pi$ (setting $\phi=0$), the Wigner $d$-function reduces to $d^J_{\mu_a \mu_b}(\pi) = (-1)^{J+\mu_a} \delta_{\mu_b, -\mu_a}$. Only the configuration where both photon and nucleon helicities flip, $b = \bar{a} \equiv (-\lambda, -h)$ with $\mu_{\bar{a}} = -\mu_a$, contributes. 
Under the backward rotation $R_y(\pi)$, the outgoing photon polarization vector becomes $\boldsymbol{\epsilon}'_{-\lambda}(-z) = R_y(\pi)\boldsymbol{\epsilon}_{-\lambda}(+z) = \boldsymbol{\epsilon}_\lambda(+z)$, yielding $\boldsymbol{\epsilon}^{\prime*}_{-\lambda}(-z)\cdot\boldsymbol{\epsilon}_\lambda(+z) = 1$. 
Meanwhile, the outgoing nucleon spinor is rotated into 
\begin{align}
    \psi^{\mathrm{out}}_{-h}(+z) = R_y(\pi)\psi_{-h}(-z) = R_y(\pi)R_y(\pi)\chi_{-h}(+z) = -\chi_{-h}(+z),
\end{align}
where the minus sign arises from the $2\pi$ rotation of a spin-$1/2$ fermion. 
Consequently, the projection coefficient between the initial and final canonical nucleon spinors is 
\begin{align}
    B_a \equiv [\psi^{\mathrm{out}}_{-h}(+z)]^\dagger \mathbf{1} \psi^{\mathrm{in}}_h(-z) = -\tau_h = -(-1)^{1/2-h}.
\end{align}
Multiplying $\hat{F}_{\bar{a}a}(\pi)$ by $B_a$ (with $B_a^2=1$) and averaging over the four initial helicity states $(\lambda=\pm 1, h=\pm 1/2)$, the spin-dependent $\hat{g}_\pi$ terms cancel out, leaving
\begin{align}\label{eq:app_fpi_proj}
    \hat{f}_\pi(W) = \frac{1}{4}\sum_{a} B_a \hat{F}_{\bar{a}a}(\pi) = \frac{1}{4}\sum_{J,a} (2J+1) B_a d^J_{\mu_a,-\mu_a}(\pi) \hat{F}^J_{\bar{a}a}(W)\,.
\end{align}

To resolve the parity content of the partial waves, we introduce the electric ($X=E$) and magnetic ($X=M$) photon multipole states of integer angular momentum $L \ge 1$, $|L m; E/M\rangle = (|L m; +\rangle \pm |L m; -\rangle)/\sqrt{2}$, which carry definite parities $P_E = (-1)^L$ and $P_M = (-1)^{L+1}$, respectively.
Coupling these photon multipoles with the positive-parity nucleon spin $1/2$ yields total angular momentum $J = L \pm 1/2$. The unitary transformation from the multipole-spin basis $|JM; L, X\rangle$ to the two-body helicity basis $|JM; a\rangle$ reads~\cite{Babusci:1998ww}
\begin{align}
    U^J_{LX, a} = \tau_h \sqrt{\frac{2L+1}{2J+1}} C^{J\mu_a}_{L\lambda, \frac{1}{2}, -h}\, v_X(\lambda), \qquad v_E(\lambda) = \frac{1}{\sqrt{2}}, \quad v_M(\lambda) = \frac{\lambda}{\sqrt{2}}\,,
\end{align}
satisfying $\sum_a U^J_{LX, a} U^{J*}_{L'X', a} = \delta_{LL'}\delta_{XX'}$ (with four allowed states for $J \ge 3/2$ and two $L=1$ states for $J=1/2$, as transverse real photons have no $L=0$ state). 
Using $\tau_{-h} = -\tau_h$, $v_X(-\lambda) = P_X (-1)^L v_X(\lambda)$, and the Clebsch--Gordan reflection relation $C^{J,-\mu_a}_{L,-\lambda,\frac{1}{2},h} = (-1)^{L+1/2-J} C^{J\mu_a}_{L\lambda,\frac{1}{2},-h}$, one finds $U^J_{LX, \bar{a}} = P_X (-1)^{J+1/2} U^J_{LX, a}$, which establishes the parity transformation of the two-body helicity partial-wave states:
\begin{align}\label{eq:app_parity_helicity}
    \mathcal{P}|JM; a\rangle = \xi_J |JM; \bar{a}\rangle\,, \quad \xi_J = (-1)^{J+1/2}\,.
\end{align}

Finally, we invoke $s$-channel unitarity $S^{J\dagger}S^J = \mathbf{1}$ with the normalization $S^J = \mathbf{1} + 2 i k \hat{F}^J$, and decompose the absorptive matrix $w^{JP}_{ba}(W) \equiv (\operatorname{Im}\hat{F}^J(W))^P_{ba}$ according to the total parity $P=\pm 1$ of the intermediate states $|X, JP\rangle$. 
By definition, the diagonal elements satisfy $w^{JP}_{aa}(W) \ge 0$ and determine the unpolarized photoabsorption cross section into the parity-$P$ channel via the (forward) optical theorem
\begin{align}\label{eq:app_sigma_P}
    \sigma(W, P) = \frac{4\pi}{k}\frac{1}{4}\sum_{J,a} (2J+1) w^{JP}_{aa}(W)\,.
\end{align}
Meanwhile, parity conservation of the transition operator ($\mathcal{P}^\dagger T \mathcal{P} = T$) together with Eq.~\eqref{eq:app_parity_helicity} implies $\langle X, JP | T | JM; \bar{a} \rangle = P \xi_J \langle X, JP | T | JM; a \rangle$, which locks the helicity-flip absorptive part to the diagonal one:
\begin{align}\label{eq:app_w_flip}
    w^{JP}_{\bar{a}a}(W) = P \xi_J w^{JP}_{aa}(W)\,.
\end{align}
Substituting Eq.~\eqref{eq:app_w_flip} into the absorptive part of Eq.~\eqref{eq:app_fpi_proj}, the overall backward phase product simplifies to
\begin{align}
    B_a d^J_{\mu_a,-\mu_a}(\pi) \xi_J = \left[-(-1)^{1/2-h}\right] (-1)^{J+\lambda-h} (-1)^{J+1/2} = -(-1)^{2J+1+\lambda-2h} = -1\,,
\end{align}
because $2J$, $\lambda$, and $2h$ are all odd integers for any physical state, making the exponent $2J+1+\lambda-2h$ an even integer. Inserting this universal $-1$ factor into Eq.~\eqref{eq:app_fpi_proj} and comparing with Eq.~\eqref{eq:app_sigma_P}, we obtain in the c.m. frame
\begin{align}
    \operatorname{Im}\hat{f}_\pi(W) = \frac{1}{4}\sum_{J,a,P} (2J+1) (-P) w^{JP}_{aa}(W) = \frac{k}{4\pi}\left[\sigma(W, P=-1) - \sigma(W, P=+1)\right].
\end{align}
Converting back to the lab frame via Eq.~\eqref{eq:app_lab_cm_f} and using $(W/m_N)k = \omega$, the $W/m_N$ recoil factor cancels against $k/\omega$, yielding
\begin{align}
    \operatorname{Im} f_\pi(\omega) = \frac{W}{m_N}\operatorname{Im}\hat{f}_\pi(W) = \frac{\omega}{4\pi}\left[\sigma(\omega, P=-1) - \sigma(\omega, P=+1)\right].
\end{align}
For a positive-parity ground-state nucleon ($P_N=+1$), intermediate states with $P=-1$ change parity ($\Delta P = \text{yes}$) while those with $P=+1$ preserve parity ($\Delta P = \text{no}$), which completes the derivation of Eq.~\eqref{eq:fpi_Im}.

\section{Conventions for $\pi \eta$ and $KN$ scattering}\label{app:isospin}

In this appendix, we briefly introduce the additional formalism necessary to arrive at a closed system for $\eta N$ and $K N$ scattering.

We define the nucleon and antinucleon states
\begin{align}
    |p\rangle=\left|\frac{1}{2}, \frac{1}{2}\right\rangle_N\,, \quad|n\rangle=\left|\frac{1}{2},-\frac{1}{2}\right\rangle_N\,, \quad|\bar{n}\rangle=-\left|\frac{1}{2}, \frac{1}{2}\right\rangle_{\bar{N}}\,, \quad|\bar{p}\rangle=\left|\frac{1}{2},-\frac{1}{2}\right\rangle_{\bar{N}}\,,
\end{align}
and the pion and kaon states
\begin{align}
    \left|\pi^{+}\right\rangle & =-|1,1\rangle_{\pi}\,, \quad\left|\pi^{-}\right\rangle=|1,-1\rangle_{\pi}\,, \quad\left|\pi^0\right\rangle=|1,0\rangle_{\pi}\,,\\
    \left|K^{+}\right\rangle & =\left|\frac{1}{2}, \frac{1}{2}\right\rangle_K\,, \quad\left|K^0\right\rangle=\left|\frac{1}{2},-\frac{1}{2}\right\rangle_K\,, \quad\left|\bar{K}^0\right\rangle=\left|\frac{1}{2}, \frac{1}{2}\right\rangle_{\bar K}\,, \quad\left|K^{-}\right\rangle=-\left|\frac{1}{2},-\frac{1}{2}\right\rangle_{\bar K}\,,
\end{align}
with crossing properties under charge conjugation $\mathcal{C}$
\begin{align*}
    \mathcal{C}|p\rangle=|\bar{p}\rangle\,, \quad \mathcal{C}|n\rangle=|\bar{n}\rangle\, \quad \mathcal{C}|\pi^\pm\rangle=|\pi^\mp\rangle,\quad \mathcal{C}|\pi^0\rangle=|\pi^0\rangle,\quad\mathcal{C}\left|K^{+}\right\rangle=\left|K^{-}\right\rangle\,, \quad \mathcal{C}\left|K^0\right\rangle=\left|\bar{K}^0\right\rangle\,,
\end{align*}
while $\left|\eta\right\rangle=\left|0,0\right\rangle$.
This convention is widely adopted in the DR community~\cite{Garcia-Martin:2010kyn,Lu:2020qeo,Dai:2014zta,Danilkin:2018qfn,Deineka:2024mzt}.

We choose the isospin decomposition of the $\pi^a N\to \eta N$ and $KN\to KN$ amplitude as
\begin{align}
    T^a=T_\eta \tau^a\,,\quad T=T^{+}-\boldsymbol{\tau}_N \cdot \boldsymbol{\tau}_K T^{-}=T^{+}-2\left(I_s\left(I_s+1\right)-\frac{3}{2}\right) T^{-}\,,
\end{align}
respectively. 
In this section, $I_s, I_t$, and $I_u$ denote $s$-, $t$-, and $u$-channel isospin, respectively, and $\boldsymbol{\tau}_N$ and $\boldsymbol{\tau}_K$ are the Pauli matrices associated with the nucleon and kaon isospin operators. 
For $\pi N\to \eta N$ reactions, it leads to
\begin{align}
    \mA(\pi^+ n\to \eta p)&=\mA(\pi^- p\to \eta n)=\sqrt{2}\mA_\eta=-\sqrt{\frac{2}{3}}\mA^{1/2}\,,\\
    \mA(\pi^0 p\to \eta p)&=-\mA(\pi^0 n\to \eta n)=\mA_\eta=-\frac{1}{\sqrt{3}}\mA^{1/2}\,.
\end{align}
We can readily obtain the $t$-channel $\pi \eta\to \bar{N} N$ amplitudes from the $s$-channel ones via crossing
\begin{align}
    \mA(\pi^+ \eta\to\bar{n}p)&=\mA(\pi^- \eta\to \bar{p}n)=\sqrt{2}\mA_\eta=\mA^1\,,\\
    \mA(\pi^0 \eta\to\bar{p}p)&=-\mA(\pi^0 \eta\to \bar{n}n)=\mA_\eta=\frac{1}{\sqrt{2}}\mA^1\,,
\end{align}
which yields the isospin crossing coefficients $c_J^{\eta}$ 
\begin{align}\label{eq:cJ_etaN}
    \mA_\eta=c_J^{\eta}\mA^1\,,\quad c^\eta_J=\frac{1}{\sqrt{2}}\,,
\end{align}
for all angular momentum $J$.
For $s$-channel $K N\to K N$ reactions, we can also obtain the relations between the $KN$ isospin amplitudes
\begin{align}
    \mA(K^+ p\to K^+ p)&=\mA(K^0 n\to K^0 n)=\mA^+-\mA^-=\mA^{I_s=1}\,,\nonumber\\
    \mA(K^+ n\to K^+ n)&=\mA(K^0 p\to K^0 p)=\mA^++\mA^-=\frac{1}{2}\mA^{I_s=1}+\frac{1}{2}\mA^{I_s=0}\,,\nonumber\\
    \mA(K^+ n\to K^0 p)&=\mA(K^0 p\to K^+ n)=-2\mA^-=\frac{1}{2}\mA^{I_s=1}-\frac{1}{2}\mA^{I_s=0}\,.
\end{align}
The relations for the isospin even/odd amplitudes with $I=+/-$ and the amplitudes in the $s$-channel isospin basis $I_s \in\{0,1\}$ can be summarized in matrix notation
\begin{align}
    \binom{\mathcal{A}^{+}}{\mathcal{A}^{-}}=C_{\nu s}\binom{\mathcal{A}^{I_s=0}}{\mathcal{A}^{I_s=1}}\,, \quad\binom{\mathcal{A}^{I_s=0}}{\mathcal{A}^{I_s=1}}=C_{s \nu}\binom{\mathcal{A}^{+}}{\mathcal{A}^{-}}\,, \quad C_{\nu s}=\frac{1}{4} C_{s \nu}=\frac{1}{4}\left(\begin{array}{cc}
1 & 3 \\
1 & -1
\end{array}\right)\,.
\end{align}
For $u$-channel $\bar K N\to \bar K N$ reactions, the relations between the $\bar K N$ isospin amplitudes are the following
\begin{align}
    \mA(\bar K^0 p\to \bar K^0 p)&=\mA(K^- n\to K^- n)=\mA^++\mA^-=\mA^{I_u=1}\,,\nonumber\\
    \mA(\bar K^0 n\to \bar K^0 n)&=\mA(K^- p\to K^- p)=\mA^+-\mA^-=\frac{1}{2}\mA^{I_u=1}+\frac{1}{2}\mA^{I_u=0}\,,\nonumber\\
    \mA(\bar K^0 n\to K^- p)&=\mA(K^- p\to \bar K^0 n)=-2\mA^-=-\frac{1}{2}\mA^{I_u=1}+\frac{1}{2}\mA^{I_u=0}\,.
\end{align}
The corresponding isospin relations read
\begin{align}
    \binom{\mathcal{A}^{+}}{\mathcal{A}^{-}} & =C_{\nu u}\binom{\mathcal{A}^{I_u=0}}{\mathcal{A}^{I_u=1}}\,, \quad\binom{\mathcal{A}^{I_u=0}}{\mathcal{A}^{I_u=1}}=C_{u \nu}\binom{\mathcal{A}^{+}}{\mathcal{A}^{-}}\,,\\ 
    C_{\nu u} & =\frac{1}{4}\left(\begin{array}{cc}
1 & 3 \\
-1 & 1
\end{array}\right)\,, \quad C_{u \nu}=\left(\begin{array}{cc}
1 & -3 \\
1 & 1
\end{array}\right)\,.
\end{align}
The $t$-channel $K\bar K\to \bar N N$ amplitudes can be obtained from the crossing consideration
\begin{align}\label{eq:iso_KK->NN}
    \mA(K^+\bar K^0\to\bar n p)&=\mA(K^0 K^-\to\bar p n)=-2\mA^-=-\mA^{I_t=1}\,,\nonumber\\
    \mA(K^+ K^-\to \bar n n)&=\mA(K^0 \bar K^0\to \bar p p)=\mA^++\mA^-=\frac{1}{2}\mA^{I_t=1}+\frac{1}{2}\mA^{I_t=0}\,,\nonumber\\
    \mA(K^+ K^-\to \bar p p)&=\mA(K^0 \bar K^0\to \bar n n)=\mA^+-\mA^-=-\frac{1}{2}\mA^{I_t=1}+\frac{1}{2}\mA^{I_t=0}\,.
\end{align}
Thereby we can easily deduce the crossing isospin relations
\begin{align}\label{eq:cJ_KN}
    \binom{\mathcal{A}^{+}}{\mathcal{A}^{-}}=C_{\nu t}\binom{\mathcal{A}^{I_t=0}}{\mathcal{A}^{I_t=1}}\,, \quad C_{\nu t}=c_J^{KN}\left(\begin{array}{cc}
1 & 0 \\
0 & 1
\end{array}\right)\,,\quad c_J^{KN}=\frac{1}{2}\,,
\end{align}
which is exactly consistent with Ref.~\cite{Ditsche:2012fv}.

In complete analogy to $\pi N$ scattering, the amplitude for the process $K(p_1)+N(p_2) \rightarrow K\left(p_3\right)+N\left(p_4\right)$ can be decomposed as
\begin{align}
    T^I=\bar{u}\left(p_4\right)\left\{A^I+\frac{\slashed{p}_1+\slashed{p}_3}{2} B^I\right\} u(p_2)\,, \quad I \in\{+,-\}\,.
\end{align}
The $\bar{K} K \rightarrow \bar{N} N$ partial waves can be obtained from the invariant amplitudes using the projection formula (with the $t$-channel scattering angle $z_t=\cos \theta_t^{K N}$)\cite{Frazer:1960zza,Hoferichter:2012wf}
\begin{align}\label{eq:PW_KKNN}
    \begin{aligned}
    & h_{+}^{J, I=(0,1)}(t)=-\frac{1}{8 \pi} \int_{-1}^1 \mathrm{~d} z_t P_J\left(z_t\right)\left\{\frac{p_t^2}{\left(p_t k_t\right)^J} A^{ \pm}\left(t, z_t\right)-\frac{m_N}{\left(p_t k_t\right)^{J-1}} z_t B^{ \pm}\left(t, z_t\right)\right\}\,, \\
    & h_{-}^{J, I=(0,1)}(t)=\frac{1}{8 \pi} \frac{\sqrt{J(J+1)}}{2 J+1} \frac{1}{\left(p_t k_t\right)^{J-1}} \int_{-1}^1 \mathrm{~d} z_t\left(P_{J-1}\left(z_t\right)-P_{J+1}\left(z_t\right)\right) B^{ \pm}\left(t, z_t\right)\,.
    \end{aligned}
\end{align}
The projection formula for $\pi N\to \eta N$ scattering is closely analogous to that of $K N$ scattering, while the main difference is that the $\pi N\to \eta N$ reaction has only one independent isospin amplitude (corresponding to $I=1$).
In addition, the required partial-wave projection formula can be obtained by means of substituting $k_t \to l_t$ in Eq.~\eqref{eq:PW_KKNN}, where $k_t=\sqrt{\frac{t}{4}-M_K^2}$ and $l_t=\frac{\lambda^{1/2}(t,M_\pi^2,M_\eta^2)}{2\sqrt{t}}$.

\section{Chiral matching scheme of $\mathbf{g(t)}$}\label{app:ChPT}

In this section, we calculate the $\pi\eta\to\bar N N$ and $K\bar K\to\bar N N$ amplitudes within the framework of BChPT, and discuss in detail the chiral matching of the inhomogeneous MO dispersive representation, cf.~Eq.~\eqref{eq:g_DR}.

Since the SU(3) pseudoscalar meson octet-baryon Lagrangian is well established, see, e.g., Refs.~\cite{Krause:1990xc,Oller:2006yh,Frink:2006hx}, we do not reproduce it here. 
The relevant terms up to second order and corresponding conventions at this order to which we work are given in Refs.~\cite{Ren:2012aj,Wu:2023uva,Wu:2024xwy}. 
If one also includes the decuplet states as active degrees of freedom in ChPT, then the relevant decuplet- and octet-baryons with the pseudoscalar meson interaction part of the LO chiral Lagrangian reads~\cite{Geng:2009hh}
\begin{align}\label{eq:naive}
    \mathcal{L}_{\phi B T}^{(1)}=\frac{h_A}{2\sqrt{2}} \epsilon^{a b c} \bar{T}_\mu^{a d e}\left(g^{\mu \nu}+z \gamma^\mu \gamma^\nu\right) B^{eb} u_\nu^{dc}+\text{h.c.}\,,
\end{align}
with the octet-baryons collected in
\begin{align}
    B=\left(\begin{array}{ccc}
\frac{1}{\sqrt{2}} \Sigma^0+\frac{1}{\sqrt{6}} \Lambda & \Sigma^{+} & p \\
\Sigma^{-} & -\frac{1}{\sqrt{2}} \Sigma^0+\frac{1}{\sqrt{6}} \Lambda & n \\
\Xi^{-} & \Xi^0 & -\frac{2}{\sqrt{6}} \Lambda
\end{array}\right)\,,
\end{align}
the pseudo Nambu-Goldstone bosons encoded in
\begin{align}
    \Phi&=\left(\begin{array}{ccc}
\pi^0+\frac{1}{\sqrt{3}} \eta & \sqrt{2} \pi^{+} & \sqrt{2} K^{+} \\
\sqrt{2} \pi^{-} & -\pi^0+\frac{1}{\sqrt{3}} \eta & \sqrt{2} K^0 \\
\sqrt{2} K^{-} & \sqrt{2} \bar{K}^0 & -\frac{2}{\sqrt{3}} \eta
\end{array}\right)\,,\\
    u^2 &\equiv U=\exp \left(i \Phi / F_0\right)\,, \quad u_\mu=i u^{\dagger}\partial_\mu u-i u\partial_\mu u^{\dagger}=u_\mu^{\dagger}\,,
\end{align}
the spin-$3/2$ decuplet-baryons expressed by a totally flavor symmetric Rarita--Schwinger tensor field $T_{abc}\equiv T^\mu_{abc}$ (each element of $T^\mu_{abc}$ is a four-component Dirac spinor) with
\begin{align}
    T_{111} & =\Delta^{++}\,, \quad T_{112}=\frac{1}{\sqrt{3}} \Delta^{+}\,, \nonumber\\
    T_{122} & =\frac{1}{\sqrt{3}} \Delta^0\,, \quad T_{222}=\Delta^{-}\,, \nonumber\\
    T_{113} & =\frac{1}{\sqrt{3}} \Sigma^{*+}\,, \quad T_{123}=\frac{1}{\sqrt{6}} \Sigma^{* 0}\,, \quad T_{223}=\frac{1}{\sqrt{3}} \Sigma^{*-}\,, \nonumber\\
    T_{133} & =\frac{1}{\sqrt{3}} \Xi^{* 0}\,, \quad T_{233}=\frac{1}{\sqrt{3}} \Xi^{*-}\,, \quad T_{333}=\Omega\,,
\end{align}
where $F_0$ is the meson decay constant in the chiral limit, $h_A$ denotes the $\phi BT$ coupling. 
The constant $z$ is an off-shell parameter that is relevant only for loop computations; thus we fix it to zero for definiteness.
A subtle ambiguity in the treatment of relativistic spin-$3/2$ Rarita--Schwinger fields concerns the spurious spin-$1/2$ degrees of freedom. 
Here, the interaction of Eq.~\eqref{eq:naive} gives rise not only to the desired spin-$3/2$ resonance exchange but also to an additional spin-$1/2$ term~\cite{Pascalutsa:2000kd}.\footnote{Further discussion can be found in Appendix~B of Ref.~\cite{Cao:2024nxm} and references therein.} 
The spurious spin-$1/2$ components are off-shell effects and can be absorbed into contact terms, such that they do not affect the physical observables.

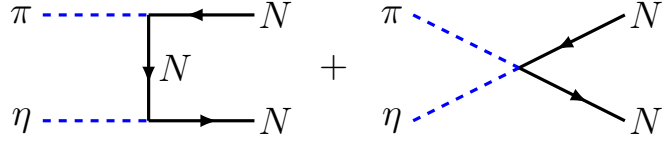
\begin{figure}
    \centering
\begin{tikzpicture}[scale=0.7]
    
    \draw[pion, dashed, very thick] (-1,2) to (1,2);
    \draw[nucleon, very thick] (3,2) to (1,2);
    \draw[pion, dashed, very thick] (-1,0) to (1,0);
    \draw[nucleon, very thick] (1,0) to (3,0);
    \draw[nucleon, very thick] (1,2) to (1,0);
    \node[font=\Large] at (-1.4,2) {$\pi$};
    \node[font=\Large] at (-1.4,0) {$\eta$};
    \node[font=\Large] at (3.4,2) {$N$};
    \node[font=\Large] at (3.4,0) {$N$};
    \node[font=\Large] at (1.5,1) {$N$};
    \node[font=\Large] at (4.5,1) {$+$};

    \begin{scope}[yshift=0cm, xshift=7cm]
    \draw[pion, dashed, very thick] (-1,2) to (1,1);
    \draw[nucleon, very thick] (3,2) to (1,1);
    \draw[pion, dashed, very thick] (-1,0) to (1,1);
    \draw[nucleon, very thick] (1,1) to (3,0);
    \node[font=\Large] at (-1.4,2) {$\pi$};
    \node[font=\Large] at (-1.4,0) {$\eta$};
    \node[font=\Large] at (3.4,2) {$N$};
    \node[font=\Large] at (3.4,0) {$N$};
    \end{scope}
\end{tikzpicture}
\caption{Pictorial representation of the $\pi\eta\to N\bar N$ amplitude obtained up to chiral NLO. Note that the crossed Born diagram is not shown.}
\label{fig:Feyn_pieta}
\end{figure}
\begin{figure}
    \centering
\begin{tikzpicture}[scale=0.7]
    
    \draw[pion, dashed, very thick] (-1,2) to (1,2);
    \draw[nucleon, very thick] (3,2) to (1,2);
    \draw[pion, dashed, very thick] (-1,0) to (1,0);
    \draw[nucleon, very thick] (1,0) to (3,0);
    \draw[nucleon, very thick] (1,2) to (1,0);
    \node[font=\Large] at (-1.4,2) {$\bar K$};
    \node[font=\Large] at (-1.4,0) {$K$};
    \node[font=\Large] at (3.4,2) {$N$};
    \node[font=\Large] at (3.4,0) {$N$};
    \node[font=\Large] at (2.5,1) {$\Lambda/\Sigma/\Sigma^*$};
    \node[font=\Large] at (4.5,1) {$+$};

    \begin{scope}[yshift=0cm, xshift=7cm]
    \draw[pion, dashed, very thick] (-1,2) to (1,1);
    \draw[nucleon, very thick] (3,2) to (1,1);
    \draw[pion, dashed, very thick] (-1,0) to (1,1);
    \draw[nucleon, very thick] (1,1) to (3,0);
    \node[font=\Large] at (-1.4,2) {$\bar K$};
    \node[font=\Large] at (-1.4,0) {$K$};
    \node[font=\Large] at (3.4,2) {$N$};
    \node[font=\Large] at (3.4,0) {$N$};
    \end{scope}
\end{tikzpicture}
\caption{Pictorial representation of the $K\bar K\to N\bar N$ amplitude obtained up to chiral NLO.}
\label{fig:Feyn_KK}
\end{figure}
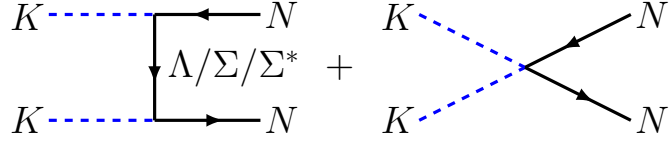
The $\pi\eta/K\bar K \to N\bar N$ amplitudes are described as a Born term in the LO plus a NLO contact term within SU(3) ChPT shown in Figs.~\ref{fig:Feyn_pieta} and~\ref{fig:Feyn_KK}.
For the $t$-channel $\pi\eta\to\bar N N$ reactions, the invariant functions $A, B$ read
\begin{align}\label{eq:Amp_pieta}
    A&=-\frac{m_N(D+F)(D-3F)}{\sqrt{3} F_0^2}-\frac{2M_\pi^2(b_D+b_F)+(t-M_\pi^2-M_\eta^2)(3b_1-b_2+b_3)}{\sqrt{3}F_0^2}\,,\nonumber\\
    B&=\frac{m_N^2(D+F)(D-3F)}{\sqrt{3} F_0^2}\left(\frac{1}{s-m_N^2}-\frac{1}{u-m_N^2}\right)+\frac{(s-u)(3b_5+b_6-b_7)}{\sqrt{3}F_0^2}\,,
\end{align}
while for the $t$-channel $K\bar K\to\bar N N$ reactions, 
\begin{align}\label{eq:Amp_KK}
    A^-= {}&\frac{1}{8F_0^2}\left[\frac{(D-F)^2(m_N+m_\Sigma)(m_\Sigma^2-m_N^2)}{s-m_\Sigma^2}-\frac{1}{3}\frac{(D+3F)^2(m_N+m_\Lambda)(m_\Lambda^2-m_N^2)}{s-m_\Lambda^2}\right] \nonumber\\
    & +\frac{h_A^2}{288 F_0^2 m_{\Sigma^*}^2}\frac{1}{s-m_{\Sigma^*}^2}\Biggl[-2 m_{\Sigma^*}^5 - 3 m_{\Sigma^*}^4 m_N + m_{\Sigma^*}^3 \left(4 m_K^2 + 2 m_N^2 - 3 t\right) \nonumber\\
    & + m_{\Sigma^*}^2 m_N \left(4 (m_K^2 + m_N^2) - 3 t\right) + 2 m_{\Sigma^*} m_K^2 (m_N^2 - m_K^2) - m_N (m_K^2 - m_N^2)^2 \Biggr] \nonumber\\
    & +\frac{1}{8F_0^2}\left[(D-F)^2(m_N+m_\Sigma)-\frac{1}{3}(D+3F)^2(m_N+m_\Lambda)\right] \nonumber\\
    & +\frac{1}{2 F_0^2}\left[2M_K^2(b_D+b_F)+(t-2M_K^2)(b_1+b_2+b_3)\right] \nonumber\\
    & -\frac{h_A^2}{288 F_0^2 m_{\Sigma^*}^2}\left[2 m_{\Sigma^*}^3 + 3 m_{\Sigma^*}^2 m_N + 2 m_{\Sigma^*} m_K^2+ 2 m_K^2 m_N - 2 m_N^3 + m_N s\right]\,,\nonumber\\
    B^-= {}&-\frac{1}{8F_0^2}\left[\frac{(D-F)^2(m_\Sigma^2+m_N^2+2m_N m_\Sigma)}{s-m_\Sigma^2}-\frac{1}{3}\frac{(D+3F)^2(m_\Lambda^2+m_N^2+2m_N m_\Lambda)}{s-m_\Lambda^2}\right] \nonumber\\
    & -\frac{h_A^2}{288 F_0^2 m_{\Sigma^*}^2}\frac{1}{s-m_{\Sigma^*}^2}\left\{\left[m_{\Sigma^*}(m_{\Sigma^*} - m_N) - (m_K^2 - m_N^2)\right]^2 + 3 m_{\Sigma^*}^2 (t - 3 m_N^2)\right\} \nonumber\\
    & -\frac{1}{8F_0^2}\left[(D-F)^2-\frac{1}{3}(D+3F)^2\right]-\frac{t-m_N^2-M_K^2}{F_0^2}\left(b_5+b_6+b_7\right) \nonumber\\
    & -\frac{h_A^2}{288F_0^2 m_{\Sigma^*}^2}\left[s+m_{\Sigma^*}^2+2 m_{\Sigma^*}m_N-2m_N^2+2M_K^2\right]\,,
\end{align}
where $b_{D/F}$ and $b_i$ are the chiral NLO LECs.
The LECs values used here are taken from Fit~II in Ref.~\cite{Ren:2012aj}, which are $D=0.80, F=0.46, b_D=0.222(20)~\mathrm{GeV}^{-1}, b_F=-0.428(12)~\mathrm{GeV}^{-1}, b_1=0.515(132)~\mathrm{GeV}^{-1}, b_2=0.148(48)~\mathrm{GeV}^{-1}, b_3=-0.663(155)~\mathrm{GeV}^{-1}, b_5=-0.643(246)~\mathrm{GeV}^{-2}, b_6=-0.268(334)~\mathrm{GeV}^{-2}, b_7=0.176(72)~\mathrm{GeV}^{-2}$. 
$F_0$ in the decay constant of the pseudo-Nambu-Goldstone bosons in the chiral limit. At the NLO for SU(3) ChPT considered here, it is the same for all mesons.
However, one may also take the physical values, with difference only starting from the next-to-next-to-leading order (i.e., the leading one-loop order). There are three different values, namely $F_\pi=92.4~\mathrm{MeV}, F_K=113.0$~MeV and $F_\eta=120.1$~MeV~\cite{Mai:2009ce}. 
Often, one chooses the average of these, that is, $F_0=\left(F_\pi+F_K+F_\eta\right) / 3=108.5$~MeV. 
Here, we take $F_0=109(15)$~MeV with the error estimating SU(3) breaking effects.
The $\phi B T$ coupling $h_A\simeq 2.3$ is obtained by matching $\Sigma^* \rightarrow \Sigma \pi$ decay width.\footnote{In the SU(3) context, the value of $h_A$ can vary depending on the decay channel, with the one obtained from the $\Delta \rightarrow N \pi$ decay being the largest ($\simeq 2.8$). 
Since we include only the $\Sigma^*$ resonance, we take the determination $h_A= 2.3$ from the $\Sigma^*$ decays as the more reasonable ones for our purposes.}
For $K\bar K\to\bar N N$ scattering, we present here only the relevant $I_t=1$ ($I=-$) part that we used.

In these conventions, the chiral NLO amplitudes are given via Eq.~\eqref{eq:PW_KKNN}
\begin{align}
    \left(g_+^0\right)^\text{LO,\,P}(t)= {}&\frac{(D+F)(D-3F)m_N^3}{4\sqrt{3} \pi F_0^2}\left[1-y^{\pi\eta}Q_0(y^{\pi\eta})\right]\,,\\
    \left(g_+^0\right)^\text{LO,\,NP}(t)= {}&(t-4m_N^2) \frac{(D+F)(D-3F)m_N}{16\sqrt{3} \pi F_0^2}\,,\\
    \left(g_+^0\right)^\text{NLO}(t)= {}&(t-4m_N^2)\left\{\frac{1}{16\sqrt{3}\pi F_0^2}\left[2 M_\pi^2\left(b_D+b_F\right)+\left(t-M_\pi^2-M_\eta^2\right)\left(3 b_1-b_2+b_3\right)\right]\right. \nonumber\\
    & \left.+\frac{m_N l_t^2}{12\sqrt{3}\pi F_0^2}(3b_5+b_6-b_7)\right\}\,,\label{eq:g+0_NLO}\\
    \left(h_+^{0,1}\right)^\text{LO,\,P}(t)= {}&-\frac{(D-F)^2(m_N+m_\Sigma)^2}{64\pi F_0^2}\left[m_N-\frac{m_\Sigma t-2 M_K^2 m_N+2m_N(m_N-m_\Sigma)^2}{4p_t k_t}Q_0(y_\Sigma^{K\bar K})\right]\,,\nonumber\\
    & +\frac{(D+3F)^2(m_N+m_\Lambda)^2}{192\pi F_0^2}\left[m_\Sigma \to m_\Lambda\right]\,,\\
    \left(h_+^{0,1}\right)^\text{LO,\,NP}(t)= {}&(t-4m_N^2)\left[-\frac{(D-F)^2(m_N+m_\Sigma)}{128\pi F_0^2}+\frac{(D+3F)^2(m_N+m_\Lambda)}{384\pi F_0^2}\right]\,,\\
    \left(h_+^{0,1}\right)^\text{NLO}(t)= {}&(t-4m_N^2)\left\{-\frac{1}{32\pi F_0^2}\left[2M_K^2(b_D+b_F)+(t-2M_K^2)(b_1+b_2+b_3)+\frac{4}{3}m_N k_t^2\left(b_5+b_6+b_7\right)\right]\right\}\,,\\
    \left(h_+^{0,1}\right)^\text{$\Sigma^*$,\,P}(t)= {}&\frac{h_A^2}{2304\pi F_0^2 m_{\Sigma^*}^2}\frac{p_t}{k_t}\left[-2 m_{\Sigma^*}^5 - 3 m_{\Sigma^*}^4 m_N + m_{\Sigma^*}^3 \left(4 m_K^2 + 2 m_N^2 - 3 t\right)\right. \nonumber\\
    & \left. + m_{\Sigma^*}^2 m_N \left(4 (m_K^2 + m_N^2) - 3 t\right) + 2 m_{\Sigma^*} m_K^2 (m_N^2 - m_K^2) - m_N (m_K^2 - m_N^2)^2\right]Q_0\left(y_{\Sigma^*}^{K\bar K}\right) \nonumber\\
    & -\frac{h_A^2 m_N}{2304\pi F_0^2 m_{\Sigma^*}^2}\left\{\left[m_{\Sigma^*}(m_{\Sigma^*} - m_N) - (m_K^2 - m_N^2)\right]^2 + 3 m_{\Sigma^*}^2 (t - 3 m_N^2)\right\} \nonumber\\
    & \times\left[1-y_{\Sigma^*}^{K\bar K} Q_0\left(y_{\Sigma^*}^{K\bar K}\right)\right]\,,\label{eq:Sigma*_P}\\
    \left(h_+^{0,1}\right)^\text{$\Sigma^*$,\,NP}(t)= {}&(t-4m_N^2)\Biggl\{\frac{h_A^2}{13824\pi F_0^2 m_{\Sigma^*}^2}\left[6 m_{\Sigma^*}^3 + 9 m_{\Sigma^*}^2 m_N + 6 m_{\Sigma^*} m_K^2\right. \nonumber\\
    & \left.+ 11 m_K^2 m_N - 3 m_N^3 - 2 m_N t\right]\Biggl\}\,,\label{eq:Sigma*_NP}
\end{align}
with
\begin{align*}
    y^{\pi\eta}=\frac{t-M_\pi^2-M_\eta^2}{4 p_t l_t}\,,\quad y_Y^{K\bar K}=\frac{t-2 M_K^2+2\left(m_Y^2-m_N^2\right)}{4 p_t k_t}\,,\quad Y=\Sigma\,,\Lambda\,,\Sigma^*\,.
\end{align*}
We separate the pure pole term (``P") from the non-pole term (``NP"), which will prove convenient later.
The Legendre functions of the second kind $Q_\ell(z)$ can be expressed for general complex argument $z$ in the Neumann representation as
\begin{align}
    Q_\ell(z)=\frac{1}{2} \int_{-1}^1 \mathrm{d} x \frac{P_\ell(x)}{z-x}=(-1)^{\ell+1} Q_\ell(-z)\,,
\end{align}
which for $\ell=0$ reduces to
\begin{align}
    Q_0(z)=\frac{1}{2} \int_{-1}^1 \frac{\mathrm{d} x}{z-x}\,, \quad Q_0(z \pm i \epsilon)=\frac{1}{2} \log \left|\frac{1+z}{1-z}\right| \mp i \frac{\pi}{2} \theta\left(1-z^2\right)\,.
\end{align}
We also need the analytic continuation for purely imaginary argument $z=iy, y>1$ in the pseudophysical region of the $t$-channel reaction
\begin{align}
    Q_0(i y)=\frac{1}{2} \log \frac{i y+1}{i y-1}=\frac{1}{2} \log \frac{1+i y}{1-i y}-i \frac{\pi}{2}\,.
\end{align}
The partial-wave amplitudes $g_+^0(t)$ ($\left(g_+^0\right)^\text{LO,P}(t)$) have a logarithmic singularity (originating from the projection of the nucleon pole terms in the scattering amplitudes) located at
\begin{align}
    M_\eta^2+M_\pi^2+\frac{M_\pi M_\eta\left(-M_\pi M_\eta+\sqrt{4 m_N^2-M_\eta^2} \sqrt{4 m_N^2-M_\pi^2}\right)}{2 m_N^2}\simeq(M_\pi+M_\eta)^2-0.53M_\pi^2\,,
\end{align}
very close to the lowest physical threshold at $(M_\pi+M_\eta)^2$.
This singularity gives rise to a modest enhancement of the spectral function near threshold, a feature absent in the pole approximation.

Several remarks are in order.
In ChPT the LO contribution, i.e., the Born term can be decomposed into a pure pole term and a polynomial term. 
The latter stems from the derivative coupling structure of ChPT, whereas the pure pole piece corresponds only to a pseudoscalar coupling. 
Notice that the pole term is invariant under field redefinitions, while the constant term generally shifts and, in the effective field theory language, is fixed by a set of definite LEC values correspondingly.
The pole term contributes to the LHC part, that is, to the inhomogeneity $\boldsymbol{\Delta}(t)$ in Eq.~\eqref{eq:g_DR} after partial-wave projection.
However, there is an inherent freedom in whether the constant term should be included in $\boldsymbol{\Delta}(t)$ (also called the ``polynomial ambiguity'' in the MO representation~\cite{Garcia-Martin:2010kyn,Hoid:2026atr}). 
Because the inhomogeneous term is only required to be free of right-hand cuts, it can contain additional regular terms, leading to different subtraction schemes. 
We chose the inhomogeneity $\boldsymbol{\Delta}(t)$ to incorporate only the pure pole component of the Born term, and match the residual subtraction polynomial $\boldsymbol{P}_1(t)$ at $t=0$ to the corresponding polynomial terms of the chiral NLO amplitudes, which are free of singularities.

How to treat the $\Sigma^*$ contribution demands careful consideration. 
The $s$-channel $\Sigma^*$-exchange amplitude can be decomposed into a genuine pole term $\left(h_+^{0,1}\right)^\text{$\Sigma^*$,\,P}$~\eqref{eq:Sigma*_P} and a non-pole piece $\left(h_+^{0,1}\right)^\text{$\Sigma^*$,\,NP}$~\eqref{eq:Sigma*_NP}. 
It is stressed that the non-pole term represents local contact interactions that are already taken into account by the NLO contact terms.
Since Ref.~\cite{Ren:2012aj} does not include explicit decuplet baryons, its LECs implicitly contain part of the decuplet contributions.
With this in mind, the non-pole part must be removed by hand to prevent double counting.
The inclusion of the $\Sigma^*$-pole term must be accompanied by imposing the correct chiral low-energy behavior: near $t=0$, the corresponding amplitude should be identical to that derived in the absence of an explicit $\Sigma^*$.
However, a LHC remains on the real $t$ axis that ranges from $-\infty$ up to
\begin{align}
    t_{\Sigma^*}\equiv -\frac{\lambda\left(m_{\Sigma^*}^2, m_N^2, M_K^2\right)}{m_{\Sigma^*}^2}\simeq 0.122~\text{GeV}^2\,,
\end{align}
which prevents a reasonable matching at $t = 0$. 
One practical choice is to perform the matching at a small but finite value, e.g. $t=0.13~\text{GeV}^2\gtrsim t_{\Sigma^*}$. 
Alternatively, one may consider an appropriate theoretical limit. 
In the chiral limit $M_K \to 0$, the cut between $-\infty$ and $t_{\Sigma^*}>0$ becomes $-\infty$ and $t_{\Sigma^*}<0$, so matching can be carried out at $t=0$. 
Therefore, in this limit, the $\Sigma^*$-pole contribution evaluated at $t=0$ up to $\mathcal{O}(t)$ yields 
\begin{align}
    &\left(h_+^{0,1}\right)^\text{$\Sigma^*$,\,P\,,low}(t) \nonumber\\
    &\equiv (t-4m_N^2)\lim_{t\to 0}\,\lim_{M_K\to 0}\frac{\left(h_+^{0,1}\right)^\text{$\Sigma^*$,\,P}(t)}{t-4m_N^2}\nonumber\\
    &=(t-4m_N^2) \frac{h_A^2}{4608\pi F_0^2 m_{\Sigma^*}^2}\left[ (m_N-2 m_{\Sigma^*})(m_{\Sigma^*}+m_N)^2+\frac{6 m_{\Sigma^*}^2-2 m_{\Sigma^*} m_N+m_N^2}{3(m_N-m_{\Sigma^*})}t+\mathcal{O}(t^2)\right]\,,
\end{align}
where the factor $(t-4m_N^2)$ is included to enforce the proper threshold behavior. 
Numerically, the ratio between $\left(h_+^{0,1}\right)^\text{$\Sigma^*$,\,P}$ and $\left(h_+^{0,1}\right)^\text{$\Sigma^*$,\,P\,,low}$ at $t=0.13~\text{GeV}^2$ is 0.87, close to 1, demonstrating that matching at this reasonable point is consistent with the calculation in low-energy limit.

Accordingly, the matching between the MO dispersive representation~\eqref{eq:g_DR} and the NLO chiral amplitudes is performed by expanding the dispersive expression except $\boldsymbol{\Delta}(t)$ around the subtraction point $t=0$ up to $\mathcal{O}(t)$. 
Crucially, this matching procedure guarantees that the resulting dispersive amplitude $\mathbf{g}(t)$ is uniquely determined and invariant under any polynomial redistribution between the inhomogeneity $\boldsymbol{\Delta}(t)$ and the subtraction term. 
Specifically, we identify the inhomogeneity $\boldsymbol{\Delta}(t)$ as
\begin{align}
    \boldsymbol{\Delta}(t)=\begin{pmatrix}\left(g_+^0\right)^\text{LO,\,P}(t) \\ \sqrt{2}\left[\left(h_+^{0,1}\right)^\text{LO,\,P}(t)+\left(h_+^{0,1}\right)^\text{$\Sigma^*$,\,P}(t)\right]\end{pmatrix}\,,
\end{align}
while the subtraction function $\boldsymbol{P}_1(t)$ takes the Laurent-polynomial form
\begin{align}\label{eq:P1_Laurent}
    \boldsymbol{P}_1(t) = \frac{\boldsymbol{p}_{-1}}{t} + \boldsymbol{p}_0 + \boldsymbol{p}_1 t\,,
\end{align}
where the kinematically singular part $\propto 1/t$ stems from the $\pi\eta$ mass difference in $l_t^2$ and will be discussed in detail below. 
Matching $\boldsymbol{\Omega}_0^1(t)\boldsymbol{P}_1(t)$ at $t=0$ to the remaining non-pole chiral amplitude,
\begin{align}
    \boldsymbol{P}_1^\text{ChPT}(t) &\equiv \frac{1}{t-4m_N^2}\begin{pmatrix}\left(g_+^0\right)^\text{LO,\,NP}(t)+\left(g_+^0\right)^\text{NLO}(t) \\ \sqrt{2}\left[\left(h_+^{0,1}\right)^\text{LO,\,NP}(t)+\left(h_+^{0,1}\right)^\text{NLO}(t)-\left(h_+^{0,1}\right)^\text{$\Sigma^*$,\,P,\,low}(t)\right]\end{pmatrix} \nonumber\\
    &= \frac{\boldsymbol{p}_{-1}^\text{ChPT}}{t} + \boldsymbol{p}_0^\text{ChPT} + \boldsymbol{p}_1^\text{ChPT} t\,,
\end{align}
yields the subtraction coefficients
\begin{align}\label{eq:p_matching}
    \boldsymbol{p}_{-1} &= \boldsymbol{p}_{-1}^\text{ChPT}\,, \nonumber\\
    \boldsymbol{p}_0 &= \boldsymbol{p}_0^\text{ChPT} - \dot{\boldsymbol{\Omega}}_0^1(0)\,\boldsymbol{p}_{-1}^\text{ChPT}\,, \nonumber\\
    \boldsymbol{p}_1 &= \boldsymbol{p}_1^\text{ChPT} - \dot{\boldsymbol{\Omega}}_0^1(0)\,\boldsymbol{p}_0^\text{ChPT} + \left(\left[\dot{\boldsymbol{\Omega}}_0^1(0)\right]^2 - \frac{1}{2}\ddot{\boldsymbol{\Omega}}_0^1(0)\right)\boldsymbol{p}_{-1}^\text{ChPT}\,.
\end{align}
Here, $\dot{\boldsymbol{\Omega}}_0^1(0)$ and $\ddot{\boldsymbol{\Omega}}_0^1(0)$ denote the first and second derivatives of the coupled-channel isovector $S$-wave Omn\`es matrix at $t=0$, which are evaluated as
\begin{align}
    \text{\cite{Lu:2020qeo}}:\quad \dot{\boldsymbol{\Omega}}_0^1(0) &= \begin{pmatrix} -0.22 & -0.41 \\ -0.44 & 0.16 \end{pmatrix}\,\text{GeV}^{-2}\,,\quad \ddot{\boldsymbol{\Omega}}_0^1(0)= \begin{pmatrix} -0.06 & -0.34 \\ -0.67 & 0.39 \end{pmatrix}\,\text{GeV}^{-4}\,,\\
    \text{\cite{Deineka:2024mzt}}:\quad \dot{\boldsymbol{\Omega}}_0^1(0) &= \begin{pmatrix} -0.03 & -0.31 \\ -0.32 & 0.29 \end{pmatrix}\,\text{GeV}^{-2}\,,\quad \ddot{\boldsymbol{\Omega}}_0^1(0)= \begin{pmatrix} -0.20 & -0.44 \\ -0.61 & 0.53 \end{pmatrix}\,\text{GeV}^{-4}\,,
\end{align}
respectively.
Writing the chiral coefficient vectors in component form as $\boldsymbol{p}_n^\text{ChPT} = \left(p_{n,1}^\text{ChPT},\, \sqrt{2}\,p_{n,2}^\text{ChPT}\right)^T$ ($n=-1,0,1$), the explicit NLO chiral expressions for the $\pi\eta$ ($i=1$) and $K\bar{K}$ ($i=2$) channels read
\begin{align}
    p_{-1,1}^\text{ChPT} ={}& \frac{m_N\left(M_\pi^2-M_\eta^2\right)^2}{48\sqrt{3}\pi F_0^2}\left(3b_5+b_6-b_7\right)\,, \nonumber\\
    p_{0,1}^\text{ChPT} ={}& \frac{1}{16\sqrt{3}\pi F_0^2}\bigg[(D+F)(D-3F)m_N + 2M_\pi^2(b_D+b_F) \nonumber\\
    & - (M_\pi^2+M_\eta^2)(3b_1-b_2+b_3) - \frac{2}{3}m_N(M_\pi^2+M_\eta^2)(3b_5+b_6-b_7)\bigg]\,, \nonumber\\
    p_{1,1}^\text{ChPT} ={}& \frac{1}{16\sqrt{3}\pi F_0^2}\left[3b_1-b_2+b_3 + \frac{1}{3}m_N(3b_5+b_6-b_7)\right]\,,
\end{align}
and
\begin{align}
    p_{-1,2}^\text{ChPT} ={}& 0\,, \nonumber\\
    p_{0,2}^\text{ChPT} ={}& - \frac{(D-F)^2(m_N+m_\Sigma)}{128\pi F_0^2} + \frac{(D+3F)^2(m_N+m_\Lambda)}{384\pi F_0^2} \nonumber\\
    & -\frac{M_K^2}{16\pi F_0^2}\left[b_D+b_F-(b_1+b_2+b_3)-\frac{2}{3}m_N(b_5+b_6+b_7)\right] \nonumber\\
    & - \frac{h_A^2(m_N-2m_{\Sigma^*})(m_{\Sigma^*}+m_N)^2}{4608\pi F_0^2 m_{\Sigma^*}^2}\,, \nonumber\\
    p_{1,2}^\text{ChPT} ={}& -\frac{1}{32\pi F_0^2}\left[b_1+b_2+b_3+\frac{1}{3}m_N(b_5+b_6+b_7)\right] \nonumber\\
    & - \frac{h_A^2}{4608\pi F_0^2 m_{\Sigma^*}^2}\frac{6m_{\Sigma^*}^2-2m_{\Sigma^*}m_N+m_N^2}{3(m_N-m_{\Sigma^*})}\,.
\end{align}

A subtlety in this prescription concerns the kinematically singular part $\propto 1/t$ of the subtraction function $\boldsymbol{P}_1(t)$, which stems from the term $C(t-4m_N^2)l_t^2$ ($C\equiv m_N(3b_5+b_6-b_7)/(12\sqrt{3}\pi F_0^2)$) in $\left(g_+^0\right)^{\text{NLO}}(t)$ (see Eq.~\eqref{eq:g+0_NLO}).
Due to the unequal meson masses $M_\pi \neq M_\eta$ in $l_t^2 = \lambda(t, M_\pi^2, M_\eta^2)/(4t)$, the subtraction function $\boldsymbol{P}_1(t)$ in Eq.~\eqref{eq:g_DR} is not actually a pure subtraction polynomial. 
A similar issue was carefully discussed in the single-channel context in Ref.~\cite{Chen:2015jgl}; here, we generalize the argument to the coupled-channel framework.
To justify this representation~\eqref{eq:g_DR} rigorously, we isolate the corresponding contribution to $\boldsymbol{P}_1(t)$,
\begin{align}
    \delta \boldsymbol{P}_1(t) = \begin{pmatrix} C l_t^2 \\ 0 \end{pmatrix} = \mathbf{m}_1 t + \mathbf{m}_0 + \frac{\mathbf{m}_{-1}}{t}\,,
\end{align}
with $\mathbf{m}_1 = (C/4, 0)^T$, $\mathbf{m}_0 = (-C(M_\pi^2+M_\eta^2)/2, 0)^T$, and $\mathbf{m}_{-1} = (C(M_\pi^2-M_\eta^2)^2/4, 0)^T$, and treat the entire term $\delta\boldsymbol{\Delta}(t) \equiv (t-4m_N^2)\delta\boldsymbol{P}_1(t)$ as an additional inhomogeneity on top of $\boldsymbol{\Delta}(t)$. 
With a Laurent-polynomial inhomogeneity, the dispersive integral can be performed analytically using the dispersive representations of the inverse of the Omn\`es matrix. 
Defining
\begin{align}
    \mathbf{I}_n(t) = \frac{1}{\pi}\int_{(M_\pi+M_\eta)^2}^{\infty} \mathrm{d}t'\,\frac{\operatorname{Im}\left[\mathbf{\Omega}_0^1(t')\right]^{-1}}{t'^n (t' - t)}\,,
\end{align}
we have
\begin{align}\label{eq:Omega_inv_expansion}
    \left[\mathbf{\Omega}_0^1(t)\right]^{-1} &= \mathbf{1} + t\,\mathbf{I}_1(t) \nonumber\\
    &= \mathbf{1} - t\,\dot{\mathbf{\Omega}}_0^1(0) + t^2 \mathbf{I}_2(t) \nonumber\\
    &= \mathbf{1} - t\,\dot{\mathbf{\Omega}}_0^1(0) - \frac{t^2}{2}\left[\ddot{\mathbf{\Omega}}_0^1(0) - 2\left(\dot{\mathbf{\Omega}}_0^1(0)\right)^2\right] + t^3 \mathbf{I}_3(t)\,,
\end{align}
which can be solved for the $\mathbf{I}_n(t)$. 
The full contribution of the additional inhomogeneity $\delta\boldsymbol{\Delta}(t)$ to $\mathbf{g}(t)$ is then given by
\begin{align}
    \delta \mathbf{g}(t) ={}& \delta\boldsymbol{\Delta}(t) - (t - 4m_N^2)\mathbf{\Omega}_0^1(t)\frac{t^2}{\pi}\int_{(M_\pi+M_\eta)^2}^{\infty} \mathrm{d}t'\,\frac{\operatorname{Im}\left[\mathbf{\Omega}_0^1(t')\right]^{-1}\delta\boldsymbol{\Delta}(t')}{t'^2 (t' - 4m_N^2)(t' - t)} \nonumber\\
    ={}& (t - 4m_N^2)\mathbf{\Omega}_0^1(t)\Bigg\{ \mathbf{m}_1 t+\Big[\mathbf{1} - t\,\dot{\mathbf{\Omega}}_0^1(0)\Big] \mathbf{m}_0 \nonumber\\
    &+ \left(\mathbf{1} - t\,\dot{\mathbf{\Omega}}_0^1(0) - \frac{t^2}{2}\left[\ddot{\mathbf{\Omega}}_0^1(0) - 2\left(\dot{\mathbf{\Omega}}_0^1(0)\right)^2\right]\right) \frac{\mathbf{m}_{-1}}{t}\Bigg\} \nonumber\\
    ={}& (t - 4m_N^2)\mathbf{\Omega}_0^1(t)\Big\{ \delta\boldsymbol{P}_1(t) + \text{[linear subtraction polynomial]} \Big\}\,,
\end{align}
where the terms involving $\left[\mathbf{\Omega}_0^1(t)\right]^{-1}$ in the solutions of Eq.~\eqref{eq:Omega_inv_expansion} for $\mathbf{I}_n(t)$ have exactly canceled the leading $\delta\boldsymbol{\Delta}(t)$. 
The first part in the curly brackets actually acts as a subtraction function in the dispersion integral, including the singular term in $l_t^2 \propto 1/t$.

\section{Dispersively improved Breit--Wigner parameterization}\label{app:BW}

Even for channels where no precise spectral function as in Eq.~\eqref{eq:BEFT_t_I=1} is available, one can construct an ansatz with the correct analytic properties by replacing the energy-dependent BW propagator with a dispersion representation~\cite{Moussallam:2013una,Cao:2025ncx} [cf.~\eqref{eq:HDR_diff_t}]:
\begin{align}\label{eq:BEFT_a0_Dis.BW}
    \left(\alpha_{E 1}-\beta_{M 1}\right)_{a_0}^t=\frac{1}{2 \pi^2} \int_{\left(M_\pi+M_\eta\right)^2}^{\infty}\frac{\mathrm{d} t^{\prime}}{t^{\prime}}~\frac{g_{a_0 N N} F_{a_0 \gamma \gamma} \sqrt{t'} \Gamma_{a_0}(t')}{\left(t'-M_{a_0}^2\right)^2+t' \left(\Gamma_{a_0}(t')\right)^2}\,,
\end{align}
with the energy-dependent width of $a_0(980)$
\begin{align}
    \begin{aligned}
    \Gamma_{a_0}(t) & =\Gamma_{a_0\to\pi \eta} \frac{\rho_{\pi \eta}(t)}{\rho_{\pi \eta}\left(M_{a_0}^2\right)}+\Gamma_{a_0\to K \bar{K}} \frac{\rho_{K \bar{K}}(t)}{\rho_{K \bar{K}}\left(M_{a_0}^2\right)} \Theta\left(t-4 M_K^2\right)\,,\\
    \Gamma_{a_0\to i} & = \frac{|g_{a_0\to i}|^2 \rho_i\!\left(M_{a_0}^2\right)}{|g_{a_0\to\pi\eta}|^2 \rho_{\pi\eta}\!\left(M_{a_0}^2\right) + |g_{a_0\to K\bar{K}}|^2 \rho_{K\bar{K}}\!\left(M_{a_0}^2\right)}\,\Gamma_{a_0}\,, \quad i = \pi\eta, K\bar{K}\,,
    \end{aligned}
\end{align}
where $\rho_{AB}(t)\equiv \lambda^{1/2}(t,M_A^2,M_B^2)/ t$. 
Here, $g_{a_0\to\pi\eta/K\bar K}$ indicates the strength of the coupling of $a_0(980)$ to each channel and can be extracted as residues of the hadronic partial-wave amplitude poles, $g_{a_0 NN}$ is the coupling between $a_0(980)$ mesons and nucleons.
For the coupling $|g_{a0 NN}|$, one can employ values determined within the phenomenological meson-exchange nucleon-nucleon potential, fitted to low-energy nucleon-nucleon scattering data, that is, 5.95 (Bonn model)~\cite{Machleidt:1987hj}, 6.04 (Bochum model)~\cite{Pluemper:1994bs}, 4.95 (Nijmegen model)~\cite{Stoks:1999bz}, and 4.22 (Hamburg model)~\cite{Jade:1998xs}.\footnote{Notice that ``$a_0$" is usually denoted as ``$\delta$" in nuclear physics.} 
We take the averaged value 
\begin{align}\label{eq:a0NN}
    |g_{a0 NN}|=5.29.
\end{align}
Regarding the pole position of $a_0(980)$, we note that the pole roughly lies in the range estimated by PDG~\cite{ParticleDataGroup:2026mpi}, $\sqrt{s_{a_0}^{\mathrm{PDG}}}=(970-1020)-i(30-70)~$MeV.
The wide range of values implies the current lack of consensus on the precise nature of $a_0(980)$.
As a pragmatic compromise, we adopt three sets of pole positions and couplings (on the second sheet) determined from dispersive analyses as inputs for our calculations,
\begin{align}\label{eq:3_a0}
    \begin{aligned}
    \text{\cite{Lu:2020qeo}}: \sqrt{s_{a_0}} & =1000.7-i36.6~\mathrm{MeV}, &&\frac{|g_{a_0 \to\pi\eta}|}{|g_{a_0 \to K \bar K}|}=0.54,\quad \Gamma(a_0\to \gamma\gamma)=0.52~\mathrm{keV},\\
    \text{\cite{Deineka:2024mzt}}: \sqrt{s_{a_0}} & =1047-i72~\mathrm{MeV}, &&\frac{|g_{a_0 \to\pi\eta}|}{|g_{a_0 \to K \bar K}|}=0.74,\quad \Gamma(a_0\to \gamma\gamma)=1.00~\mathrm{keV}\,,\\
    \text{\cite{Kuang-Dai}}: \sqrt{s_{a_0}} & =991.4-i50.9~\mathrm{MeV}, &&\frac{|g_{a_0 \to\pi\eta}|}{|g_{a_0 \to K \bar K}|}=1.04,\quad \Gamma(a_0\to \gamma\gamma)=0.41~\mathrm{keV}\,.
    \end{aligned}
\end{align}

As shown in Fig.~\ref{fig:t_a0}, the integrand curves in Eq.~\eqref{eq:BEFT_a0_Dis.BW} also show a visible cusp near the $K\bar K$ threshold due to the non-negligible coupling between the $a_0(980)$ meson and the $K\bar K$ continua.
The results of the $I=1$ BEFT sum rule corresponding to the three $a_0(980)$ pole parameter sets are, 
\begin{align}
    \left|\left(\alpha_{E 1}-\beta_{M 1}\right)_{\pi \eta / K \bar{K}_{I_t=1}}^t\right|\simeq \left|\left(\alpha_{E 1}-\beta_{M 1}\right)_{a_0}^t\right|= 0.62\,, 0.66\,, 0.57\,,
\end{align}
respectively. 
In contrast to the full dispersive result~\eqref{eq:BEFT_I=1_value}, even the dispersively improved pole approximation overestimates the contribution by a factor of two to three.

\newpage
\bibliography{refs}

@article{ParticleDataGroup:2026mpi,
    author = "Takahashi, F. and others",
    collaboration = "Particle Data Group",
    title = "{Review of Particle Physics}",
    doi = "10.1142/s0217751x26300115",
    journal = "Int. J. Mod. Phys. A",
    volume = "41",
    number = "22",
    pages = "2630011",
    year = "2026"
}

@article{Kibble:1960zz,
    author = "Kibble, T. W. B.",
    title = "{Kinematics of General Scattering Processes and the Mandelstam Representation}",
    doi = "10.1103/PhysRev.117.1159",
    journal = "Phys. Rev.",
    volume = "117",
    pages = "1159--1162",
    year = "1960"
}

@article{Drechsel:1999rf,
    author = "Drechsel, D. and Gorchtein, M. and Pasquini, B. and Vanderhaeghen, M.",
    title = "{Fixed-$t$ subtracted dispersion relations for Compton Scattering off the nucleon}",
    eprint = "hep-ph/9904290",
    archivePrefix = "arXiv",
    reportNumber = "MKPH-T-99-09",
    doi = "10.1103/PhysRevC.61.015204",
    journal = "Phys. Rev. C",
    volume = "61",
    pages = "015204",
    year = "1999"
}

@book{Itzykson:1980rh,
    author = "Itzykson, C. and Zuber, J. B.",
    title = "{Quantum Field Theory}",
    isbn = "978-0-486-44568-7",
    publisher = "McGraw-Hill",
    address = "New York",
    series = "International Series In Pure and Applied Physics",
    year = "1980"
}

@article{PhysRev.110.240,
  title = {Dispersion Relations for Compton Scattering},
  author = {Prange, R. E.},
  journal = {Phys. Rev.},
  volume = {110},
  issue = {1},
  pages = {240--252},
  numpages = {0},
  year = {1958},
  month = {Apr},
  publisher = {American Physical Society},
  doi = {10.1103/PhysRev.110.240},
  url = {https://link.aps.org/doi/10.1103/PhysRev.110.240}
}

@article{Lvov:1980wp,
    author = "L'vov, A. I.",
    title = "{Compton Scattering on Proton at Energies up to 400 MeV and Finite Energy Sum Rules}",
    reportNumber = "LEBEDEV-80-203",
    journal = "Sov. J. Nucl. Phys.",
    volume = "34",
    pages = "597",
    year = "1981"
}

@article{Lvov:1996rmi,
    author = "L'vov, A. I. and Petrun'kin, V. A. and Schumacher, Martin",
    title = "{Dispersion theory of proton Compton scattering in the first and second resonance regions}",
    doi = "10.1103/PhysRevC.55.359",
    journal = "Phys. Rev. C",
    volume = "55",
    pages = "359--377",
    year = "1997"
}

@article{Gell-Mann:1954ttj,
    author = "Gell-Mann, Murray and Goldberger, M. L. and Thirring, Walter E.",
    title = "{Use of causality conditions in quantum theory}",
    doi = "10.1103/PhysRev.95.1612",
    journal = "Phys. Rev.",
    volume = "95",
    pages = "1612--1627",
    year = "1954"
}

@article{Babusci:1998ww,
    author = "Babusci, D. and Giordano, G. and L'vov, A. I. and Matone, G. and Nathan, A. M.",
    title = "{Low-energy Compton scattering of polarized photons on polarized nucleons}",
    eprint = "hep-ph/9803347",
    archivePrefix = "arXiv",
    doi = "10.1103/PhysRevC.58.1013",
    journal = "Phys. Rev. C",
    volume = "58",
    pages = "1013--1041",
    year = "1998"
}

@article{Schumacher:2005an,
    author = "Schumacher, Martin",
    title = "{Polarizability of the nucleon and Compton scattering}",
    eprint = "hep-ph/0501167",
    archivePrefix = "arXiv",
    reportNumber = "HEP-PH-0501167",
    doi = "10.1016/j.ppnp.2005.01.033",
    journal = "Prog. Part. Nucl. Phys.",
    volume = "55",
    pages = "567--646",
    year = "2005"
}

@article{Drechsel:2002ar,
    author = "Drechsel, D. and Pasquini, B. and Vanderhaeghen, M.",
    title = "{Dispersion relations in real and virtual Compton scattering}",
    eprint = "hep-ph/0212124",
    archivePrefix = "arXiv",
    doi = "10.1016/S0370-1573(02)00636-1",
    journal = "Phys. Rept.",
    volume = "378",
    pages = "99--205",
    year = "2003"
}

@article{baldin1960polarizability,
  title={Polarizability of nucleons},
  author={Baldin, A. M.},
  journal={Nucl. Phys.},
  volume={18},
  pages={310--317},
  year={1960},
  publisher={Elsevier}
}

@article{lapidus1963scattering,
  title={Scattering of gamma quanta and polarizability of nuclei and nucleons},
  author={Lapidus, L. I.},
  journal={Sov. Phys. JETP},
  volume={16},
  pages={964--966},
  year={1963}
}

@article{Hite:1973pm,
    author = "Hite, G. E. and Steiner, F.",
    title = "{New dispersion relations and their application to partial-wave amplitudes}",
    doi = "10.1007/BF02722827",
    journal = "Nuovo Cim. A",
    volume = "18",
    pages = "237--270",
    year = "1973"
}

@book{Hohler:1983,
  title={Pion-Nukleon-Streuung: Methoden und Ergebnisse ph{\"a}nomenologischer Analysen. Teil 2},
  author={H{\"o}hler, G.},
  isbn={9780387112824},
  series={in Landolt-B\" ornstein 9b2, H. Schopper eds., Springer Verlag, Berlin Germany},
  url={https://books.google.com.hk/books?id=reqlDAEACAAJ},
  year={1983},
  publisher={Springer}
}

@article{Cao:2025hqm,
    author = "Cao, Xiong-Hui and Guo, Feng-Kun and Guo, Zhi-Hui and Li, Qu-Zhi",
    title = "{Revisiting Roy-Steiner-equation analysis of pion-kaon scattering from lattice QCD data}",
    eprint = "2506.10619",
    archivePrefix = "arXiv",
    primaryClass = "hep-ph",
    doi = "10.1103/hn8j-95vn",
    journal = "Phys. Rev. D",
    volume = "112",
    number = "3",
    pages = "034042",
    year = "2025"
}

@article{Cao:2022zhn,
    author = "Cao, Xiong-Hui and Li, Qu-Zhi and Zheng, Han-Qing",
    title = "{A possible subthreshold pole in $S_{11}$ channel from $\pi N$ Roy-Steiner equation analyses}",
    eprint = "2207.09743",
    archivePrefix = "arXiv",
    primaryClass = "hep-ph",
    doi = "10.1007/JHEP12(2022)073",
    journal = "JHEP",
    volume = "12",
    pages = "073",
    year = "2022"
}

@article{Hoferichter:2015hva,
    author = "Hoferichter, Martin and Ruiz de Elvira, Jacobo and Kubis, Bastian and Mei{\ss}ner, Ulf-G.",
    title = "{Roy-Steiner equation analysis of pion-nucleon scattering}",
    eprint = "1510.06039",
    archivePrefix = "arXiv",
    primaryClass = "hep-ph",
    reportNumber = "INT-PUB-15-050",
    doi = "10.1016/j.physrep.2016.02.002",
    journal = "Phys. Rept.",
    volume = "625",
    pages = "1--88",
    year = "2016"
}

@article{Ditsche:2012fv,
	author = "Ditsche, C. and Hoferichter, M. and Kubis, B. and Mei\ss{}ner, Ulf-G.",
	title = "{Roy-Steiner equations for pion-nucleon scattering}",
	eprint = "1203.4758",
	archivePrefix = "arXiv",
	primaryClass = "hep-ph",
	doi = "10.1007/JHEP06(2012)043",
	journal = "JHEP",
	volume = "06",
	pages = "043",
	year = "2012"
}

@article{Buettiker:2003pp,
    author = "B{\"u}ettiker, Paul and Descotes-Genon, S. and Moussallam, B.",
    title = "{A new analysis of $\pi K$ scattering from Roy and Steiner type equations}",
    eprint = "hep-ph/0310283",
    archivePrefix = "arXiv",
    reportNumber = "HISKP-TH-03-18, IPNO-DR-03-08, LPT-ORSAY-03-76",
    doi = "10.1140/epjc/s2004-01591-1",
    journal = "Eur. Phys. J. C",
    volume = "33",
    pages = "409--432",
    year = "2004"
}

@article{Pelaez:2020gnd,
    author = "Pel{\'a}ez, Jos{\'e} R. and Rodas, Arkaitz",
    title = {Dispersive $\pi {K} \rightarrow \pi {K}$ and $\pi \pi \rightarrow {K} \bar {K}$ amplitudes from scattering data, threshold parameters, and the lightest strange resonance $\kappa$ or ${K}_0^*(700)$},
     eprint = "2010.11222",
    archivePrefix = "arXiv",
    primaryClass = "hep-ph",
    reportNumber = "JLAB-THY-20-3276",
    doi = "10.1016/j.physrep.2022.03.004",
    journal = "Phys. Rept.",
    volume = "969",
    pages = "1--126",
    year = "2022"
}

@article{Hearn:1962zz,
    author = "Hearn, A. C. and Leader, E.",
    title = "{Fixed-Angle Dispersion Relations for Nucleon Compton Scattering. I}",
    doi = "10.1103/PhysRev.126.789",
    journal = "Phys. Rev.",
    volume = "126",
    pages = "789--805",
    year = "1962"
}

@article{Koberle:1968zz,
    author = "K{\"o}berle, R.",
    title = "{Low-Energy Nucleon Compton Scattering}",
    doi = "10.1103/PhysRev.166.1558",
    journal = "Phys. Rev.",
    volume = "166",
    pages = "1558--1567",
    year = "1968"
}

@article{Schumacher:2007ez,
    author = "Schumacher, Martin",
    title = "{Properties of the $\pi^0, \eta, \eta', \sigma, f_0(980)$ and $a_0(980)$ mesons and their relevance for the polarizabilities of the nucleon}",
    eprint = "0712.1417",
    archivePrefix = "arXiv",
    primaryClass = "hep-ph",
    doi = "10.1140/epja/i2007-10506-3",
    journal = "Eur. Phys. J. A",
    volume = "34",
    pages = "293--301",
    year = "2007"
}

@article{Lehmann:1958ita,
	author = "Lehmann, H.",
	title = "{Analytic properties of scattering amplitudes as functions of momentum transfer}",
	doi = "10.1007/bf02859794",
	journal = "Nuovo Cim.",
	volume = "10",
	number = "4",
	pages = "579--589",
	year = "1958"
}

@article{Martin:1965jj,
    author = "Martin, Andre",
    title = "{Extension of the axiomatic analyticity domain of scattering amplitudes by unitarity I}",
    doi = "10.1007/BF02720568",
    journal = "Nuovo Cim. A",
    volume = "42",
    pages = "930--953",
    year = "1965"
}

@article{Martin:1966zsy,
    author = "Martin, Andre",
    title = "{Extension of the axiomatic analyticity domain of scattering amplitudes by unitarity II}",
    doi = "10.1007/BF02719361",
    journal = "Nuovo Cim. A",
    volume = "44",
    pages = "1219--1244",
    year = "1966"
}

@article{Bernabeu:1974zu,
    author = "Bernab\'eu, J. and Ericson, Torleif Erik Oskar and Ferro Fontan, C.",
    title = "{The Nucleon Electromagnetic Polarizabilities}",
    reportNumber = "CERN-TH-1804",
    doi = "10.1016/0370-2693(74)90186-5",
    journal = "Phys. Lett. B",
    volume = "49",
    pages = "381--384",
    year = "1974"
}

@article{Chew:1957tf,
    author = "Chew, G. F. and Goldberger, M. L. and Low, F. E. and Nambu, Yoichiro",
    title = "{Relativistic dispersion relation approach to photomeson production}",
    doi = "10.1103/PhysRev.106.1345",
    journal = "Phys. Rev.",
    volume = "106",
    pages = "1345--1355",
    year = "1957"
}

@article{Drechsel:1992pn,
    author = "Drechsel, D. and Tiator, L.",
    title = "{Threshold pion photoproduction on nucleons}",
    doi = "10.1088/0954-3899/18/3/004",
    journal = "J. Phys. G",
    volume = "18",
    pages = "449--497",
    year = "1992"
}

@article{Huang:2010aro,
    author = "Huang, F. and Sibirtsev, A. and Haidenbauer, J. and Krewald, S. and Mei{\ss}ner, Ulf-G.",
    title = "{Backward pion-nucleon scattering}",
    eprint = "0910.4275",
    archivePrefix = "arXiv",
    primaryClass = "nucl-th",
    reportNumber = "FZJ-IKP-TH-2009-30, JLAB-THY-09-1095",
    doi = "10.1140/epja/i2010-10930-2",
    journal = "Eur. Phys. J. A",
    volume = "44",
    pages = "81--92",
    year = "2010"
}

@article{Dai:2014zta,
    author = "Dai, Ling-Yun and Pennington, Michael R.",
    title = "{Comprehensive amplitude analysis of $\gamma\gamma \rightarrow \pi^+\pi^-, \pi^0\pi^0$ and $\overline{K} K$ below 1.5 GeV}",
    eprint = "1404.7524",
    archivePrefix = "arXiv",
    primaryClass = "hep-ph",
    reportNumber = "JLAB-THY-14-1842",
    doi = "10.1103/PhysRevD.90.036004",
    journal = "Phys. Rev. D",
    volume = "90",
    number = "3",
    pages = "036004",
    year = "2014"
}

@article{Dai:2014lza,
    author = "Dai, Ling-Yun and Pennington, M. R.",
    title = "{Two photon couplings of the lightest isoscalars from BELLE data}",
    eprint = "1403.7514",
    archivePrefix = "arXiv",
    primaryClass = "hep-ph",
    reportNumber = "JLAB-THY-14-1869",
    doi = "10.1016/j.physletb.2014.07.005",
    journal = "Phys. Lett. B",
    volume = "736",
    pages = "11--15",
    year = "2014"
}

@article{Dai:2016ytz,
    author = "Dai, Ling-Yun and Pennington, Michael R.",
    title = "{Pion polarizabilities from $\gamma\gamma\to\pi\pi$ analysis}",
    eprint = "1611.04441",
    archivePrefix = "arXiv",
    primaryClass = "hep-ph",
    reportNumber = "JLAB-THY-16-2386",
    doi = "10.1103/PhysRevD.94.116021",
    journal = "Phys. Rev. D",
    volume = "94",
    number = "11",
    pages = "116021",
    year = "2016"
}

@article{Frazer:1960zza,
    author = "Frazer, William R. and Fulco, Jose R.",
    title = "{Partial-Wave Dispersion Relations for the Process $\pi+\pi \to N+\bar{N}$}",
    doi = "10.1103/PhysRev.117.1603",
    journal = "Phys. Rev.",
    volume = "117",
    pages = "1603--1609",
    year = "1960"
}

@article{Frazer:1960zzb,
    author = "Frazer, William R. and Fulco, Jose R.",
    title = "{Effect of a Pion-Pion Scattering Resonance on Nucleon Structure. II}",
    doi = "10.1103/PhysRev.117.1609",
    journal = "Phys. Rev.",
    volume = "117",
    pages = "1609--1614",
    year = "1960"
}

@article{Cutkosky:1960sp,
    author = "Cutkosky, R. E.",
    title = "{Singularities and discontinuities of Feynman amplitudes}",
    doi = "10.1063/1.1703676",
    journal = "J. Math. Phys.",
    volume = "1",
    pages = "429--433",
    year = "1960"
}

@article{gourdin1960pion,
  title={Pion photon scattering},
  author={Gourdin, M and Martin, A},
  doi = "10.1007/BF02732027",
  journal={Il Nuovo Cimento (1955-1965)},
  volume={17},
  number={2},
  pages={224--243},
  year={1960},
  publisher={Springer}
}

@article{Bernabeu:1977hp,
    author = "Bernab{\'e}u, J. and Tarrach, R.",
    title = "{The Electromagnetic Polarizabilities of the Proton and the Scalar-Isoscalar $\gamma \gamma \to \pi \pi$ Amplitude}",
    reportNumber = "CERN-TH-2331",
    doi = "10.1016/0370-2693(77)90851-6",
    journal = "Phys. Lett. B",
    volume = "69",
    pages = "484--488",
    year = "1977"
}

@article{Cao:2024zlf,
    author = "Cao, Xiong-Hui and Guo, Feng-Kun and Li, Qu-Zhi and Yao, De-Liang",
    title = "{Dispersive determination of nucleon gravitational form factors}",
    eprint = "2411.13398",
    archivePrefix = "arXiv",
    primaryClass = "hep-ph",
    doi = "10.1038/s41467-025-62278-9",
    journal = "Nature Commun.",
    volume = "16",
    pages = "6979",
    year = "2025"
}

@article{Cao:2025dkv,
    author = "Cao, Xiong-Hui and Guo, Feng-Kun and Li, Qu-Zhi and Wu, Bo-Wen and Yao, De-Liang",
    title = "{Gravitational form factors of pions, kaons and nucleons from dispersion relations}",
    eprint = "2507.05375",
    archivePrefix = "arXiv",
    primaryClass = "hep-ph",
    doi = "10.1140/epjs/s11734-025-02025-9",
    journal = "Eur. Phys. J. ST",
    volume = "235",
    number = "9",
    pages = "2121--2154",
    year = "2026"
}

@article{Hoferichter:2012wf,
    author = "Hoferichter, M. and Ditsche, C. and Kubis, B. and Mei{\ss}ner, Ulf-G.",
    title = "{Dispersive analysis of the scalar form factor of the nucleon}",
    eprint = "1204.6251",
    archivePrefix = "arXiv",
    primaryClass = "hep-ph",
    doi = "10.1007/JHEP06(2012)063",
    journal = "JHEP",
    volume = "06",
    pages = "063",
    year = "2012"
}

@article{Moussallam:2011zg,
    author = "Moussallam, B.",
    title = "{Couplings of light $I=0$ scalar mesons to simple operators in the complex plane}",
    eprint = "1110.6074",
    archivePrefix = "arXiv",
    primaryClass = "hep-ph",
    doi = "10.1140/epjc/s10052-011-1814-z",
    journal = "Eur. Phys. J. C",
    volume = "71",
    pages = "1814",
    year = "2011"
}

@article{Danilkin:2020pak,
    author = "Danilkin, Igor and Deineka, Oleksandra and Vanderhaeghen, Marc",
    title = "{Data-driven dispersive analysis of the $\pi \pi$ and $\pi K$ scattering}",
    eprint = "2012.11636",
    archivePrefix = "arXiv",
    primaryClass = "hep-ph",
    doi = "10.1103/PhysRevD.103.114023",
    journal = "Phys. Rev. D",
    volume = "103",
    number = "11",
    pages = "114023",
    year = "2021"
}

@article{Hoferichter:2023mgy,
    author = "Hoferichter, Martin and de Elvira, Jacobo Ruiz and Kubis, Bastian and Mei{\ss}ner, Ulf-G.",
    title = "{Nucleon resonance parameters from Roy{\textendash}Steiner equations}",
    eprint = "2312.15015",
    archivePrefix = "arXiv",
    primaryClass = "hep-ph",
    reportNumber = "IPARCOS-UCM-23-139",
    doi = "10.1016/j.physletb.2024.138698",
    journal = "Phys. Lett. B",
    volume = "853",
    pages = "138698",
    year = "2024"
}

@article{Garcia-Martin:2011nna,
    author = "Garc{\'i}a-Mart{\'i}n, R. and Kami{\'n}ski, R. and Pel{\'a}ez, J. R. and Ruiz de Elvira, J.",
    title = "{Precise determination of the $f_0(600)$ and $f_0(980)$ pole parameters from a dispersive data analysis}",
    eprint = "1107.1635",
    archivePrefix = "arXiv",
    primaryClass = "hep-ph",
    doi = "10.1103/PhysRevLett.107.072001",
    journal = "Phys. Rev. Lett.",
    volume = "107",
    pages = "072001",
    year = "2011"
}

@article{Pennington:2006dg,
    author = "Pennington, M. R.",
    title = "{Sigma coupling to photons: Hidden scalar in $\gamma \gamma \to \pi^0 \pi^0$}",
    doi = "10.1103/PhysRevLett.97.011601",
    journal = "Phys. Rev. Lett.",
    volume = "97",
    pages = "011601",
    year = "2006"
}

@article{Schumacher:2013hu,
    author = "Schumacher, Martin and Scadron, Michael D.",
    title = "{Dispersion theory of nucleon Compton scattering and polarizabilities}",
    eprint = "1301.1567",
    archivePrefix = "arXiv",
    primaryClass = "hep-ph",
    doi = "10.1002/prop.201300007",
    journal = "Fortsch. Phys.",
    volume = "61",
    pages = "703--741",
    year = "2013"
}

@article{Schumacher:2019ikn,
    author = "Schumacher, Martin",
    title = "{Polarizability of the nucleon}",
    eprint = "1907.05434",
    archivePrefix = "arXiv",
    primaryClass = "hep-ph",
    doi = "10.31526/lhep.4.2019.136",
    journal = "LHEP",
    volume = "2",
    number = "4",
    pages = "4",
    year = "2019"
}

@article{Holstein:2013kia,
    author = "Holstein, Barry R. and Scherer, Stefan",
    title = "{Hadron Polarizabilities}",
    eprint = "1401.0140",
    archivePrefix = "arXiv",
    primaryClass = "hep-ph",
    reportNumber = "MITP-13-086",
    doi = "10.1146/annurev-nucl-102313-025555",
    journal = "Ann. Rev. Nucl. Part. Sci.",
    volume = "64",
    pages = "51--81",
    year = "2014"
}

@article{Schumacher:2011gi,
    author = "Schumacher, Martin",
    title = "{Structure of Scalar Mesons and the Higgs Sector of Strong Interaction}",
    eprint = "1106.1015",
    archivePrefix = "arXiv",
    primaryClass = "hep-ph",
    doi = "10.1088/0954-3899/38/8/083001",
    journal = "J. Phys. G",
    volume = "38",
    pages = "083001",
    year = "2011"
}

@article{Schumacher:2006cy,
    author = "Schumacher, Martin",
    title = "{Electromagnetic polarizabilities of the nucleon and properties of the $\sigma$-meson pole contribution}",
    eprint = "hep-ph/0609040",
    archivePrefix = "arXiv",
    reportNumber = "HEP-PH-0609040",
    doi = "10.1140/epja/i2006-10103-0",
    journal = "Eur. Phys. J. A",
    volume = "30",
    pages = "413--422",
    year = "2006",
    note = "[Erratum: Eur.Phys.J.A 32, 121 (2007)]"
}

@article{Lu:2020qeo,
    author = "Lu, Junxu and Moussallam, B.",
    title = "{The $\pi \eta $ interaction and $a_0$ resonances in photon--photon scattering}",
    eprint = "2002.04441",
    archivePrefix = "arXiv",
    primaryClass = "hep-ph",
    doi = "10.1140/epjc/s10052-020-7969-8",
    journal = "Eur. Phys. J. C",
    volume = "80",
    number = "5",
    pages = "436",
    year = "2020"
}

@article{Deineka:2024mzt,
    author = "Deineka, Oleksandra and Danilkin, Igor and Vanderhaeghen, Marc",
    title = "{Dispersive estimate of the $a_0(980)$ contribution to $(g-2)_\mu$}",
    eprint = "2410.12894",
    archivePrefix = "arXiv",
    primaryClass = "hep-ph",
    doi = "10.1103/PhysRevD.111.034009",
    journal = "Phys. Rev. D",
    volume = "111",
    number = "3",
    pages = "034009",
    year = "2025"
}

@article{Briscoe:2023gmb,
    author = "Briscoe, William J. and Schmidt, Axel and Strakovsky, Igor and Workman, Ron L. and Svarc, Alfred",
    collaboration = "SAID Group",
    title = "{Extended SAID partial-wave analysis of pion photoproduction}",
    eprint = "2309.06631",
    archivePrefix = "arXiv",
    primaryClass = "hep-ph",
    doi = "10.1103/PhysRevC.108.065205",
    journal = "Phys. Rev. C",
    volume = "108",
    number = "6",
    pages = "065205",
    year = "2023"
}

@article{Burkert:2025coj,
    author = "Burkert, Volker and Eichmann, Gernot and Klempt, Eberhard",
    title = "{The impact of $\gamma N$ and $\gamma^* N$ interactions on our understanding of nucleon excitations}",
    eprint = "2506.16482",
    archivePrefix = "arXiv",
    primaryClass = "hep-ph",
    reportNumber = "JLAB-PHY-25-4380",
    doi = "10.1016/j.ppnp.2025.104214",
    journal = "Prog. Part. Nucl. Phys.",
    volume = "146",
    pages = "104214",
    year = "2026"
}

@article{Doring:2025sgb,
    author = {D{\"o}ring, Michael and Haidenbauer, Johann and Mai, Maxim and Sato, Toru},
    title = "{Dynamical coupled-channel models for hadron dynamics}",
    eprint = "2505.02745",
    archivePrefix = "arXiv",
    primaryClass = "nucl-th",
    reportNumber = "JLAB-THY-25-4298",
    doi = "10.1016/j.ppnp.2025.104213",
    journal = "Prog. Part. Nucl. Phys.",
    volume = "146",
    pages = "104213",
    year = "2026"
}

@article{Drechsel:2007if,
    author = "Drechsel, D. and Kamalov, S. S. and Tiator, L.",
    title = "{Unitary Isobar Model - MAID2007}",
    eprint = "0710.0306",
    archivePrefix = "arXiv",
    primaryClass = "nucl-th",
    doi = "10.1140/epja/i2007-10490-6",
    journal = "Eur. Phys. J. A",
    volume = "34",
    pages = "69--97",
    year = "2007"
}

@article{Guo:2022hud,
    author = "Guo, Dan and Liu, Zhan-Wei",
    title = "{Pion photoproduction off nucleon with Hamiltonian effective field theory}",
    eprint = "2201.11555",
    archivePrefix = "arXiv",
    primaryClass = "hep-ph",
    doi = "10.1103/PhysRevD.105.114039",
    journal = "Phys. Rev. D",
    volume = "105",
    number = "11",
    pages = "114039",
    year = "2022"
}

@article{Zhuge:2024iuw,
    author = "Zhuge, Yu and Liu, Zhan-Wei and Leinweber, Derek B. and Thomas, Anthony W.",
    title = "{Pion photoproduction of nucleon excited states with Hamiltonian effective field theory}",
    eprint = "2407.05334",
    archivePrefix = "arXiv",
    primaryClass = "hep-ph",
    reportNumber = "ADP-24-11/T1250",
    doi = "10.1103/PhysRevD.110.094015",
    journal = "Phys. Rev. D",
    volume = "110",
    number = "9",
    pages = "094015",
    year = "2024"
}

@article{Cao:2021kvs,
    author = "Cao, Xiong-Hui and Ma, Yao and Zheng, Han-Qing",
    title = "{Dispersive analysis of low energy $\gamma^* N\rightarrow\pi N$ process}",
    eprint = "2101.12576",
    archivePrefix = "arXiv",
    primaryClass = "nucl-th",
    doi = "10.1103/PhysRevD.103.114007",
    journal = "Phys. Rev. D",
    volume = "103",
    number = "11",
    pages = "114007",
    year = "2021"
}

@article{Kamano:2016bgm,
    author = "Kamano, H. and Nakamura, S. X. and Lee, T. S. H. and Sato, T.",
    title = "{Isospin decomposition of $\gamma N \to N^*$ transitions within a dynamical coupled-channels model}",
    eprint = "1605.00363",
    archivePrefix = "arXiv",
    primaryClass = "nucl-th",
    reportNumber = "KEK-TH-1902, J-PARC-TH-0051",
    doi = "10.1103/PhysRevC.94.015201",
    journal = "Phys. Rev. C",
    volume = "94",
    number = "1",
    pages = "015201",
    year = "2016"
}

@article{Osmanovic:2021rck,
    author = "Osmanovi{\'c}, H. and Had{\v{z}}imehmedovi{\'c}, M. and Omerovi{\'c}, R. and Stahov, J. and Kashevarov, V. and Ostrick, M. and Tiator, L. and {\v{S}}varc, A.",
    title = "{Single-energy partial-wave analysis for pion photoproduction with fixed-$t$ analyticity}",
    eprint = "2107.06513",
    archivePrefix = "arXiv",
    primaryClass = "nucl-th",
    doi = "10.1103/PhysRevC.104.034605",
    journal = "Phys. Rev. C",
    volume = "104",
    number = "3",
    pages = "034605",
    year = "2021"
}

@article{Ronchen:2022hqk,
    author = {R{\"o}nchen, Deborah and D{\"o}ring, Michael and Mei{\ss}ner, Ulf-G. and Shen, Chao-Wei},
    title = "{Light baryon resonances from a coupled-channel study including $\mathbf{K\Sigma }$ photoproduction}",
    eprint = "2208.00089",
    archivePrefix = "arXiv",
    primaryClass = "nucl-th",
    reportNumber = "JLAB-THY-22-3667",
    doi = "10.1140/epja/s10050-022-00852-1",
    journal = "Eur. Phys. J. A",
    volume = "58",
    number = "11",
    pages = "229",
    year = "2022"
}

@article{Ma:2020hpe,
    author = "Ma, Yao and Niu, Wen-Qi and Yao, De-Liang and Zheng, Han-Qing",
    title = "{Dispersive analysis of low energy $\gamma N\to\pi N$ process and studies on the $N^*(890)$ resonance}",
    eprint = "2005.10695",
    archivePrefix = "arXiv",
    primaryClass = "hep-ph",
    doi = "10.1088/1674-1137/abc169",
    journal = "Chin. Phys. C",
    volume = "45",
    number = "1",
    pages = "014104",
    year = "2021"
}

@article{Gao:2025loz,
    author = "Gao, Yu-Sheng and Zhang, Zhao-Long and Feng, Xu and Jin, Lu-Chang and Liu, Chuan and Mei{\ss}ner, Ulf-G.",
    title = "{Lattice QCD Study of Pion Electroproduction and Weak Production from a Nucleon}",
    eprint = "2502.12074",
    archivePrefix = "arXiv",
    primaryClass = "hep-lat",
    doi = "10.1103/PhysRevLett.134.171904",
    journal = "Phys. Rev. Lett.",
    volume = "134",
    number = "17",
    pages = "171904",
    year = "2025"
}

@article{Hunt:2018wqz,
    author = "Hunt, B. C. and Manley, D. M.",
    title = "{Updated determination of $N^*$ resonance parameters using a unitary, multichannel formalism}",
    eprint = "1810.13086",
    archivePrefix = "arXiv",
    primaryClass = "nucl-ex",
    doi = "10.1103/PhysRevC.99.055205",
    journal = "Phys. Rev. C",
    volume = "99",
    number = "5",
    pages = "055205",
    year = "2019"
}

@misc{GWU,
  title = {{INS Data Analysis Center}},
  howpublished = {\url{https://gwdac.phys.gwu.edu/}}
}

@misc{BG,
  title = {{Bonn-Gatchina Partial Wave Analysis}},
  howpublished = {\url{https://pwa.hiskp.uni-bonn.de//}}
}

@misc{Mainz,
  title = {{MAID: Photo- and Electroproduction of Pions, Eta, Etaprime and Kaons on the Nucleon}},
  howpublished = {\url{https://maid.kph.uni-mainz.de}}
}

@article{Walker:1968xu,
    author = "Walker, R. L.",
    title = "{Phenomenological analysis of single pion photoproduction}",
    doi = "10.1103/PhysRev.182.1729",
    journal = "Phys. Rev.",
    volume = "182",
    pages = "1729--1748",
    year = "1969"
}

@article{Workman:2012jf,
    author = "Workman, Ron L. and Paris, Mark W. and Briscoe, William J. and Strakovsky, Igor I.",
    title = "{Unified Chew-Mandelstam SAID analysis of pion photoproduction data}",
    eprint = "1202.0845",
    archivePrefix = "arXiv",
    primaryClass = "hep-ph",
    doi = "10.1103/PhysRevC.86.015202",
    journal = "Phys. Rev. C",
    volume = "86",
    pages = "015202",
    year = "2012"
}

@article{Workman:2011vb,
    author = "Workman, Ron L. and Briscoe, William J. and Paris, Mark W. and Strakovsky, Igor I.",
    title = "{Updated SAID analysis of pion photoproduction data}",
    eprint = "1109.0722",
    archivePrefix = "arXiv",
    primaryClass = "hep-ph",
    doi = "10.1103/PhysRevC.85.025201",
    journal = "Phys. Rev. C",
    volume = "85",
    pages = "025201",
    year = "2012"
}

@article{Holstein:1994tw,
    author = "Holstein, Barry R. and Nathan, A. M.",
    title = "{Dispersion relations and the nucleon polarizability}",
    eprint = "hep-ph/9402248",
    archivePrefix = "arXiv",
    reportNumber = "UMHEP-404",
    doi = "10.1103/PhysRevD.49.6101",
    journal = "Phys. Rev. D",
    volume = "49",
    pages = "6101--6108",
    year = "1994"
}

@article{Donoghue:1993kw,
    author = "Donoghue, John F. and Holstein, Barry R.",
    title = "{Photon-photon scattering, pion polarizability and chiral symmetry}",
    eprint = "hep-ph/9302203",
    archivePrefix = "arXiv",
    reportNumber = "UMHEP-383",
    doi = "10.1103/PhysRevD.48.137",
    journal = "Phys. Rev. D",
    volume = "48",
    pages = "137--146",
    year = "1993"
}

@article{Danilkin:2019mhd,
    author = "Danilkin, Igor and Redmer, Christoph Florian and Vanderhaeghen, Marc",
    title = "{The hadronic light-by-light contribution to the muon{\textquoteright}s anomalous magnetic moment}",
    eprint = "1901.10346",
    archivePrefix = "arXiv",
    primaryClass = "hep-ph",
    doi = "10.1016/j.ppnp.2019.05.002",
    journal = "Prog. Part. Nucl. Phys.",
    volume = "107",
    pages = "20--68",
    year = "2019"
}

@article{Aoyama:2020ynm,
    author = "Aoyama, T. and others",
    title = "{The anomalous magnetic moment of the muon in the Standard Model}",
    eprint = "2006.04822",
    archivePrefix = "arXiv",
    primaryClass = "hep-ph",
    reportNumber = "FERMILAB-PUB-20-207-T, INT-PUB-20-021, KEK Preprint 2020-5,
  MITP/20-028, KEK Preprint 2020-5, MITP/20-028, CERN-TH-2020-075, IFT-UAM/CSIC-20-74, LMU-ASC 18/20, LTH 1234,
  LU TP 20-20, LTH 1234, LU TP 20-20, MAN/HEP/2020/003, PSI-PR-20-06, UWThPh 2020-14, ZU-TH 18/20",
    doi = "10.1016/j.physrep.2020.07.006",
    journal = "Phys. Rept.",
    volume = "887",
    pages = "1--166",
    year = "2020"
}

@article{Mao:2009cc,
    author = "Mao, Yu and Wang, Xuan-Gong and Zhang, Ou and Zheng, H. Q. and Zhou, Z. Y.",
    title = "{A Dispersive Analysis on the $f_0(600)$ and $f_0(980)$ Resonances in $\gamma \gamma \to \pi^+ \pi^-, \pi^0 \pi^0$ Processes}",
    eprint = "0904.1445",
    archivePrefix = "arXiv",
    primaryClass = "hep-ph",
    doi = "10.1103/PhysRevD.79.116008",
    journal = "Phys. Rev. D",
    volume = "79",
    pages = "116008",
    year = "2009"
}

@article{Oller:2008kf,
    author = "Oller, J. A. and Roca, L.",
    title = "{Two photons into $\pi^0 \pi^0$}",
    eprint = "0804.0309",
    archivePrefix = "arXiv",
    primaryClass = "hep-ph",
    doi = "10.1140/epja/i2008-10600-0",
    journal = "Eur. Phys. J. A",
    volume = "37",
    pages = "15--32",
    year = "2008"
}

@article{Garcia-Martin:2010kyn,
    author = "Garcia-Martin, R. and Moussallam, B.",
    title = "{MO analysis of the high statistics Belle results on $\gamma\gamma\to \pi^+\pi^-,\pi^0\pi^0$ with chiral constraints}",
    eprint = "1006.5373",
    archivePrefix = "arXiv",
    primaryClass = "hep-ph",
    doi = "10.1140/epjc/s10052-010-1471-7",
    journal = "Eur. Phys. J. C",
    volume = "70",
    pages = "155--175",
    year = "2010"
}

@article{Hoferichter:2011wk,
    author = "Hoferichter, Martin and Phillips, Daniel R. and Schat, Carlos",
    title = "{Roy-Steiner equations for $\gamma \gamma \to \pi \pi$}",
    eprint = "1106.4147",
    archivePrefix = "arXiv",
    primaryClass = "hep-ph",
    doi = "10.1140/epjc/s10052-011-1743-x",
    journal = "Eur. Phys. J. C",
    volume = "71",
    pages = "1743",
    year = "2011"
}

@article{Danilkin:2018qfn,
    author = "Danilkin, Igor and Vanderhaeghen, Marc",
    title = "{Dispersive analysis of the $\gamma\gamma^{*} \to \pi \pi$ process}",
    eprint = "1810.03669",
    archivePrefix = "arXiv",
    primaryClass = "hep-ph",
    doi = "10.1016/j.physletb.2018.12.047",
    journal = "Phys. Lett. B",
    volume = "789",
    pages = "366--372",
    year = "2019"
}

@article{Low:1954kd,
    author = "Low, F. E.",
    title = "{Scattering of light of very low frequency by systems of spin $\frac{1}{2}$}",
    doi = "10.1103/PhysRev.96.1428",
    journal = "Phys. Rev.",
    volume = "96",
    pages = "1428--1432",
    year = "1954"
}

@article{Bernabeu:2008wt,
    author = "Bernab{\'e}u, Jose and Prades, Joaquim",
    title = "{The $\sigma \to \gamma \gamma$ Width from Nucleon Electromagnetic Polarizabilities}",
    eprint = "0802.1830",
    archivePrefix = "arXiv",
    primaryClass = "hep-ph",
    reportNumber = "CAFPE-92-08, FTUV-08-0130, UG-FT-222-08",
    doi = "10.1103/PhysRevLett.100.241804",
    journal = "Phys. Rev. Lett.",
    volume = "100",
    pages = "241804",
    year = "2008"
}

@article{Cao:2025ncx,
    author = "Cao, Xiong-Hui and Guo, Feng-Kun and Hanhart, Christoph and Kubis, Bastian",
    title = "{\boldmath{Dispersive analysis of the $J / \psi \rightarrow \pi^0 \gamma^*$ transition form factor with $\rho$--$\omega$ mixing effects}}",
    eprint = "2512.00501",
    archivePrefix = "arXiv",
    primaryClass = "hep-ph",
    doi = "10.1103/gmj6-hd9j",
    journal = "Phys. Rev. D",
    volume = "113",
    number = "7",
    pages = "074030",
    year = "2026"
}

@article{Stoks:1999bz,
    author = "Stoks, V. G. J. and Rijken, T. A.",
    title = "{Soft-core baryon-baryon potentials for the complete baryon octet}",
    eprint = "nucl-th/9901028",
    archivePrefix = "arXiv",
    reportNumber = "ADP-99-3-T348",
    doi = "10.1103/PhysRevC.59.3009",
    journal = "Phys. Rev. C",
    volume = "59",
    pages = "3009--3020",
    year = "1999"
}

@article{Pluemper:1994bs,
    author = "Pl{\"u}emper, D. and Flender, J. and Gari, M. F.",
    title = "{Nucleon-nucleon interaction from meson exchange and nucleonic structure}",
    doi = "10.1103/PhysRevC.49.2370",
    journal = "Phys. Rev. C",
    volume = "49",
    pages = "2370--2378",
    year = "1994"
}

@article{Jade:1998xs,
    author = "J{\"a}de, L.",
    title = "{Consistent description of $NN$ and $\pi N$ interactions using the solitary boson exchange potential}",
    eprint = "nucl-th/9802039",
    archivePrefix = "arXiv",
    reportNumber = "KTH-9801",
    doi = "10.1103/PhysRevC.58.96",
    journal = "Phys. Rev. C",
    volume = "58",
    pages = "96--107",
    year = "1998"
}

@article{Machleidt:1987hj,
    author = "Machleidt, R. and Holinde, K. and Elster, C.",
    title = "{The Bonn Meson-Exchange Model for the Nucleon-Nucleon Interaction}",
    doi = "10.1016/S0370-1573(87)80002-9",
    journal = "Phys. Rept.",
    volume = "149",
    pages = "1--89",
    year = "1987"
}

@misc{Kuang-Dai,
    author = "Kuang, Shi-Qing and Dai, Ling-Yun",
    title = "{Comprehensive amplitude analysis of $\gamma \gamma \to \pi^0 \eta, K \bar{K}$ and $\pi^0 \eta'$ below 1.6~GeV}",
    howpublished = "in preparation"
}

@article{Li:2022vnz,
    author = "Li, X. and others",
    title = "{Proton Compton Scattering from Linearly Polarized Gamma Rays}",
    eprint = "2205.10533",
    archivePrefix = "arXiv",
    primaryClass = "nucl-ex",
    doi = "10.1103/PhysRevLett.128.132502",
    journal = "Phys. Rev. Lett.",
    volume = "128",
    pages = "132502",
    year = "2022"
}

@article{Mornacchi:2022cln,
    author = "Mornacchi, E. and Rodini, S. and Pasquini, B. and Pedroni, P.",
    title = "{First Concurrent Extraction of the Leading-Order Scalar and Spin Proton Polarizabilities}",
    eprint = "2204.13491",
    archivePrefix = "arXiv",
    primaryClass = "hep-ph",
    doi = "10.1103/PhysRevLett.129.102501",
    journal = "Phys. Rev. Lett.",
    volume = "129",
    number = "10",
    pages = "102501",
    year = "2022"
}

@article{A2:2021wmm,
    author = "Mornacchi, E. and others",
    collaboration = "A2",
    title = "{Measurement of Compton Scattering at MAMI for the Extraction of the Electric and Magnetic Polarizabilities of the Proton}",
    eprint = "2110.15691",
    archivePrefix = "arXiv",
    primaryClass = "nucl-ex",
    doi = "10.1103/PhysRevLett.128.132503",
    journal = "Phys. Rev. Lett.",
    volume = "128",
    number = "13",
    pages = "132503",
    year = "2022"
}

@article{Pasquini:2019nnx,
    author = "Pasquini, B. and Pedroni, P. and Sconfietti, S.",
    title = "{Proton scalar dipole polarizabilities from real Compton scattering data, using fixed-$t$ subtracted dispersion relations and the bootstrap method}",
    eprint = "1903.07952",
    archivePrefix = "arXiv",
    primaryClass = "hep-ph",
    doi = "10.1088/1361-6471/ab323a",
    journal = "J. Phys. G",
    volume = "46",
    number = "10",
    pages = "104001",
    year = "2019"
}

@article{Pasquini:2017ehj,
    author = "Pasquini, B. and Pedroni, P. and Sconfietti, S.",
    title = "{First extraction of the scalar proton dynamical polarizabilities from real Compton scattering data}",
    eprint = "1711.07401",
    archivePrefix = "arXiv",
    primaryClass = "hep-ph",
    doi = "10.1103/PhysRevC.98.015204",
    journal = "Phys. Rev. C",
    volume = "98",
    number = "1",
    pages = "015204",
    year = "2018"
}

@article{Krupina:2017pgr,
    author = "Krupina, Nadiia and Lensky, Vadim and Pascalutsa, Vladimir",
    title = "{Partial-wave analysis of proton Compton scattering data below the pion-production threshold}",
    eprint = "1712.05349",
    archivePrefix = "arXiv",
    primaryClass = "nucl-th",
    reportNumber = "MITP-17-100",
    doi = "10.1016/j.physletb.2018.04.066",
    journal = "Phys. Lett. B",
    volume = "782",
    pages = "34--41",
    year = "2018"
}

@article{Detmold:2010ts,
    author = "Detmold, W. and Tiburzi, B. C. and Walker-Loud, A.",
    title = "{Extracting Nucleon Magnetic Moments and Electric Polarizabilities from Lattice QCD in Background Electric Fields}",
    eprint = "1001.1131",
    archivePrefix = "arXiv",
    primaryClass = "hep-lat",
    reportNumber = "JLAB-THY-09-1111, UMD-40762-474",
    doi = "10.1103/PhysRevD.81.054502",
    journal = "Phys. Rev. D",
    volume = "81",
    pages = "054502",
    year = "2010"
}

@article{McGovern:2012ew,
    author = "McGovern, J. A. and Phillips, D. R. and Grie{\ss}hammer, H. W.",
    title = "{Compton scattering from the proton in an effective field theory with explicit Delta degrees of freedom}",
    eprint = "1210.4104",
    archivePrefix = "arXiv",
    primaryClass = "nucl-th",
    reportNumber = "INT-PUB-12-048",
    doi = "10.1140/epja/i2013-13012-1",
    journal = "Eur. Phys. J. A",
    volume = "49",
    pages = "12",
    year = "2013"
}

@article{Lensky:2014efa,
    author = "Lensky, Vadim and McGovern, Judith A.",
    title = "{Proton polarizabilities from Compton data using covariant chiral effective field theory}",
    eprint = "1401.3320",
    archivePrefix = "arXiv",
    primaryClass = "nucl-th",
    doi = "10.1103/PhysRevC.89.032202",
    journal = "Phys. Rev. C",
    volume = "89",
    number = "3",
    pages = "032202",
    year = "2014"
}

@article{Lensky:2015awa,
    author = "Lensky, Vadim and McGovern, Judith and Pascalutsa, Vladimir",
    title = "{Predictions of covariant chiral perturbation theory for nucleon polarisabilities and polarised Compton scattering}",
    eprint = "1510.02794",
    archivePrefix = "arXiv",
    primaryClass = "hep-ph",
    reportNumber = "MITP-15-017",
    doi = "10.1140/epjc/s10052-015-3791-0",
    journal = "Eur. Phys. J. C",
    volume = "75",
    number = "12",
    pages = "604",
    year = "2015"
}

@article{Bernard:1993ry,
    author = "Bernard, V. and Kaiser, Norbert and Mei{\ss}ner, Ulf-G. and Schmidt, A.",
    title = "{Aspects of nucleon Compton scattering}",
    eprint = "hep-ph/9311354",
    archivePrefix = "arXiv",
    reportNumber = "CRN-93-55",
    doi = "10.1007/BF01305891",
    journal = "Z. Phys. A",
    volume = "348",
    pages = "317",
    year = "1994"
}

@article{Bignell:2020xkf,
    author = "Bignell, Ryan and Kamleh, Waseem and Leinweber, Derek",
    title = "{Magnetic polarizability of the nucleon using a Laplacian mode projection}",
    eprint = "2002.07915",
    archivePrefix = "arXiv",
    primaryClass = "hep-lat",
    reportNumber = "ADP-20-6/T1116",
    doi = "10.1103/PhysRevD.101.094502",
    journal = "Phys. Rev. D",
    volume = "101",
    number = "9",
    pages = "094502",
    year = "2020"
}

@article{Kabelitz:2024aye,
    author = "Kabelitz, Thomas and Bignell, Ryan and Kamleh, Waseem and Leinweber, Derek",
    title = "{Magnetic polarizability of octet baryons via lattice QCD}",
    eprint = "2407.01064",
    archivePrefix = "arXiv",
    primaryClass = "hep-lat",
    reportNumber = "ADP-24-13/T1252",
    doi = "10.1103/PhysRevD.110.074514",
    journal = "Phys. Rev. D",
    volume = "110",
    number = "7",
    pages = "074514",
    year = "2024"
}

@article{Kabelitz:2024cig,
    author = "Kabelitz, Thomas and Kamleh, Waseem and Leinweber, Derek",
    title = "{Magnetic polarizability of octet baryons near the physical quark-mass point}",
    eprint = "2412.08960",
    archivePrefix = "arXiv",
    primaryClass = "hep-lat",
    reportNumber = "ADP-24-22/T1261",
    doi = "10.1103/38db-ml51",
    journal = "Phys. Rev. D",
    volume = "112",
    number = "1",
    pages = "014516",
    year = "2025"
}

@article{Engelhardt:2007ub,
    author = "Engelhardt, Michael",
    collaboration = "LHPC",
    title = "{Neutron electric polarizability from unquenched lattice QCD using the background field approach}",
    eprint = "0706.3919",
    archivePrefix = "arXiv",
    primaryClass = "hep-lat",
    doi = "10.1103/PhysRevD.76.114502",
    journal = "Phys. Rev. D",
    volume = "76",
    pages = "114502",
    year = "2007"
}

@article{Lujan:2016ffj,
    author = "Lujan, M. and Alexandru, A. and Freeman, W. and Lee, F. X.",
    title = "{Finite volume effects on the electric polarizability of neutral hadrons in lattice QCD}",
    eprint = "1606.07928",
    archivePrefix = "arXiv",
    primaryClass = "hep-lat",
    doi = "10.1103/PhysRevD.94.074506",
    journal = "Phys. Rev. D",
    volume = "94",
    number = "7",
    pages = "074506",
    year = "2016"
}

@article{COMPTONMAX-lab:2014cve,
    author = "Myers, L. S. and others",
    collaboration = "COMPTON@MAX-lab",
    title = "{Measurement of Compton Scattering from the Deuteron and an Improved Extraction of the Neutron Electromagnetic Polarizabilities}",
    eprint = "1409.3705",
    archivePrefix = "arXiv",
    primaryClass = "nucl-ex",
    doi = "10.1103/PhysRevLett.113.262506",
    journal = "Phys. Rev. Lett.",
    volume = "113",
    number = "26",
    pages = "262506",
    year = "2014"
}

@article{Kossert:2002ws,
    author = "Kossert, K. and others",
    title = "{Quasifree Compton scattering and the polarizabilities of the neutron}",
    eprint = "nucl-ex/0210020",
    archivePrefix = "arXiv",
    reportNumber = "NUCL-EX-0210020",
    doi = "10.1140/epja/i2002-10093-9",
    journal = "Eur. Phys. J. A",
    volume = "16",
    pages = "259--273",
    year = "2003"
}

@article{Griesshammer:2012we,
    author = "Grie{\ss}hammer, H. W. and McGovern, J. A. and Phillips, D. R. and Feldman, G.",
    title = "{Using effective field theory to analyse low-energy Compton scattering data from protons and light nuclei}",
    eprint = "1203.6834",
    archivePrefix = "arXiv",
    primaryClass = "nucl-th",
    reportNumber = "INT-PUB-12-012",
    doi = "10.1016/j.ppnp.2012.04.003",
    journal = "Prog. Part. Nucl. Phys.",
    volume = "67",
    pages = "841--897",
    year = "2012"
}

@article{Hagelstein:2015egb,
    author = "Hagelstein, Franziska and Miskimen, Rory and Pascalutsa, Vladimir",
    title = "{Nucleon Polarizabilities: from Compton Scattering to Hydrogen Atom}",
    eprint = "1512.03765",
    archivePrefix = "arXiv",
    primaryClass = "nucl-th",
    reportNumber = "MITP-15-119",
    doi = "10.1016/j.ppnp.2015.12.001",
    journal = "Prog. Part. Nucl. Phys.",
    volume = "88",
    pages = "29--97",
    year = "2016"
}

@article{Griesshammer:2015ahu,
    author = "Grie{\ss}hammer, Harald W. and McGovern, Judith A. and Phillips, Daniel R.",
    title = "{Nucleon Polarisabilities at and Beyond Physical Pion Masses}",
    eprint = "1511.01952",
    archivePrefix = "arXiv",
    primaryClass = "nucl-th",
    doi = "10.1140/epja/i2016-16139-5",
    journal = "Eur. Phys. J. A",
    volume = "52",
    number = "5",
    pages = "139",
    year = "2016"
}

@article{Guiasu:1976wu,
    author = "Guia{\c{s}}u, I. and Radescu, E. E",
    title = "{Dynamic Electromagnetic Polarizabilities of the Photon}",
    doi = "10.1016/0370-2693(76)90503-7",
    journal = "Phys. Lett. B",
    volume = "62",
    pages = "193--196",
    year = "1976"
}

@article{Guiasu:1976pq,
    author = "Guia{\c{s}}u, I. and Radescu, E. E.",
    title = "{Nucleon Polarizabilities and $\pi^0 \to 2 \gamma, \eta^0 \to 2 \gamma$ Decay Rates from Compton Scattering Amplitudes}",
    reportNumber = "IFA-FT-120-1976",
    doi = "10.1103/PhysRevD.14.1335",
    journal = "Phys. Rev. D",
    volume = "14",
    pages = "1335--1351",
    year = "1976"
}

@article{Guiasu:1978vg,
    author = "Guia{\c{s}}u, I. and Radescu, E. E",
    title = "{Note on Low-Energy Proton Compton Scattering}",
    doi = "10.1103/PhysRevD.18.651",
    journal = "Phys. Rev. D",
    volume = "18",
    pages = "651--659",
    year = "1978"
}

@article{Budnev:1978ke,
    author = "Budnev, V. M. and Karnakov, V. A.",
    title = "{Connection of Proton and Pion Polarizabilities}",
    reportNumber = "TF-100",
    journal = "Sov. J. Nucl. Phys.",
    volume = "30",
    pages = "228",
    year = "1979"
}

@article{Wu:2023uva,
    author = "Wu, Bing and Cao, Xiong-Hui and Dong, Xiang-Kun and Guo, Feng-Kun",
    title = "{{\ensuremath{\sigma}} exchange in the one-boson exchange model involving the ground state octet baryons}",
    eprint = "2312.01013",
    archivePrefix = "arXiv",
    primaryClass = "hep-ph",
    doi = "10.1103/PhysRevD.109.034026",
    journal = "Phys. Rev. D",
    volume = "109",
    number = "3",
    pages = "034026",
    year = "2024"
}

@article{Gryniuk:2015eza,
    author = "Gryniuk, Oleksii and Hagelstein, Franziska and Pascalutsa, Vladimir",
    title = "{Evaluation of the forward Compton scattering off protons: Spin-independent amplitude}",
    eprint = "1508.07952",
    archivePrefix = "arXiv",
    primaryClass = "nucl-th",
    reportNumber = "MITP-15-018",
    doi = "10.1103/PhysRevD.92.074031",
    journal = "Phys. Rev. D",
    volume = "92",
    pages = "074031",
    year = "2015"
}

@article{Levchuk:1999zy,
    author = "Levchuk, M. I. and L'vov, A. I.",
    title = "{Deuteron Compton scattering below pion photoproduction threshold}",
    eprint = "nucl-th/9909066",
    archivePrefix = "arXiv",
    reportNumber = "LPI-HEPD-T-99-6",
    doi = "10.1016/S0375-9474(00)00145-7",
    journal = "Nucl. Phys. A",
    volume = "674",
    pages = "449--492",
    year = "2000"
}

@article{Wang:2023omf,
    author = "Wang, Xuan-He and Zhang, Zhao-Long and Cao, Xiong-Hui and Fan, Cong-Ling and Feng, Xu and Gao, Yu-Sheng and Jin, Lu-Chang and Liu, Chuan",
    title = "{Nucleon Electric Polarizabilities and Nucleon-Pion Scattering at the Physical Pion Mass}",
    eprint = "2310.01168",
    archivePrefix = "arXiv",
    primaryClass = "hep-lat",
    doi = "10.1103/PhysRevLett.133.141901",
    journal = "Phys. Rev. Lett.",
    volume = "133",
    number = "14",
    pages = "141901",
    year = "2024"
}

@article{Gasser:2020mzy,
    author = "Gasser, J. and Leutwyler, H. and Rusetsky, A.",
    title = "{On the mass difference between proton and neutron}",
    eprint = "2003.13612",
    archivePrefix = "arXiv",
    primaryClass = "hep-ph",
    doi = "10.1016/j.physletb.2021.136087",
    journal = "Phys. Lett. B",
    volume = "814",
    pages = "136087",
    year = "2021"
}

@article{Gasser:2020hzn,
    author = "Gasser, J. and Leutwyler, H. and Rusetsky, A.",
    title = "{Sum rule for the Compton amplitude and implications for the proton{\textendash}neutron mass difference}",
    eprint = "2008.05806",
    archivePrefix = "arXiv",
    primaryClass = "hep-ph",
    doi = "10.1140/epjc/s10052-020-08615-2",
    journal = "Eur. Phys. J. C",
    volume = "80",
    number = "12",
    pages = "1121",
    year = "2020"
}

@article{Tomalak:2018dho,
    author = "Tomalak, Oleksandr",
    title = "{Electromagnetic proton{\textendash}neutron mass difference}",
    eprint = "1810.02502",
    archivePrefix = "arXiv",
    primaryClass = "hep-ph",
    reportNumber = "FERMILAB-PUB-19-216-V",
    doi = "10.1140/epjp/s13360-020-00413-9",
    journal = "Eur. Phys. J. Plus",
    volume = "135",
    number = "6",
    pages = "411",
    year = "2020"
}

@article{Gasser:2015dwa,
    author = "Gasser, J. and Hoferichter, M. and Leutwyler, H. and Rusetsky, A.",
    title = "{Cottingham formula and nucleon polarisabilities}",
    eprint = "1506.06747",
    archivePrefix = "arXiv",
    primaryClass = "hep-ph",
    reportNumber = "INT-PUB-15-027",
    doi = "10.1140/epjc/s10052-015-3580-9",
    journal = "Eur. Phys. J. C",
    volume = "75",
    number = "8",
    pages = "375",
    year = "2015",
    note = "[Erratum: Eur.Phys.J.C 80, 353 (2020)]"
}

@article{Thomas:2014dxa,
    author = "Thomas, A. W. and Wang, X. G. and Young, R. D.",
    title = "{Electromagnetic Contribution to the Proton-Neutron Mass Splitting}",
    eprint = "1406.4579",
    archivePrefix = "arXiv",
    primaryClass = "nucl-th",
    reportNumber = "ADP-14-19-T877",
    doi = "10.1103/PhysRevC.91.015209",
    journal = "Phys. Rev. C",
    volume = "91",
    number = "1",
    pages = "015209",
    year = "2015"
}

@article{Walker-Loud:2012ift,
    author = "Walker-Loud, Andre and Carlson, Carl E. and Miller, Gerald A.",
    title = "{The Electromagnetic Self-Energy Contribution to $M_p - M_n$ and the Isovector Nucleon Magnetic Polarizability}",
    eprint = "1203.0254",
    archivePrefix = "arXiv",
    primaryClass = "nucl-th",
    reportNumber = "NT-LBNL-12-002, UCB-NPAT-12-003, NT@UW-12-03",
    doi = "10.1103/PhysRevLett.108.232301",
    journal = "Phys. Rev. Lett.",
    volume = "108",
    pages = "232301",
    year = "2012"
}

@article{Gasser:1974wd,
    author = "Gasser, J. and Leutwyler, H.",
    title = "{Implications of Scaling for the Proton{\textendash}Neutron Mass Difference}",
    reportNumber = "Print-75-0285 (BERN)",
    doi = "10.1016/0550-3213(75)90493-9",
    journal = "Nucl. Phys. B",
    volume = "94",
    pages = "269--310",
    year = "1975"
}

@article{Cottingham:1963zz,
    author = "Cottingham, W. N.",
    title = "{The neutron proton mass difference and electron scattering experiments}",
    doi = "10.1016/0003-4916(63)90023-X",
    journal = "Annals Phys.",
    volume = "25",
    pages = "424--432",
    year = "1963"
}

@article{Erben:2014hza,
    author = "Erben, F. B. and Shanahan, P. E. and Thomas, A. W. and Young, R. D.",
    title = "{Dispersive estimate of the electromagnetic charge symmetry violation in the octet baryon masses}",
    eprint = "1408.6628",
    archivePrefix = "arXiv",
    primaryClass = "nucl-th",
    reportNumber = "ADP-14-22-T880",
    doi = "10.1103/PhysRevC.90.065205",
    journal = "Phys. Rev. C",
    volume = "90",
    number = "6",
    pages = "065205",
    year = "2014"
}

@article{Biloshytskyi:2023fyv,
    author = "Biloshytskyi, Volodymyr and Ciobotaru-Hriscu, Iulian and Hagelstein, Franziska and Lensky, Vadim and Pascalutsa, Vladimir",
    title = "{QED of Bernab{\'e}u-Tarrach sum rule for electric polarizability and its implication for the Lamb shift}",
    eprint = "2305.08814",
    archivePrefix = "arXiv",
    primaryClass = "hep-ph",
    reportNumber = "MITP-23-023; PSI-PR-23-13",
    doi = "10.1103/PhysRevD.109.016026",
    journal = "Phys. Rev. D",
    volume = "109",
    number = "1",
    pages = "016026",
    year = "2024"
}

@article{Bernabeu:1974mb,
    author = "Bernab{\'e}u, J. and Tarrach, R.",
    title = "{Unsubtracted Dispersion Relation for the Longitudinal Compton Amplitude?}",
    reportNumber = "CERN-TH-1937",
    doi = "10.1016/0370-2693(75)90438-4",
    journal = "Phys. Lett. B",
    volume = "55",
    pages = "183--186",
    year = "1975"
}

@article{Pasquini:2014vua,
    author = "Pasquini, B. and Polyakov, M. V. and Vanderhaeghen, M.",
    title = "{Dispersive evaluation of the D-term form factor in deeply virtual Compton scattering}",
    eprint = "1407.5960",
    archivePrefix = "arXiv",
    primaryClass = "hep-ph",
    doi = "10.1016/j.physletb.2014.10.047",
    journal = "Phys. Lett. B",
    volume = "739",
    pages = "133--138",
    year = "2014"
}

@article{Lvov:1998jza,
    author = "L'vov, A. I. and Nathan, A. M.",
    title = "{Sum rule for the backward spin polarizability of the nucleon from a backward dispersion relation}",
    eprint = "nucl-th/9807032",
    archivePrefix = "arXiv",
    doi = "10.1103/PhysRevC.59.1064",
    journal = "Phys. Rev. C",
    volume = "59",
    pages = "1064--1069",
    year = "1999"
}

@article{Pelaez:2015qba,
    author = "Pelaez, J. R.",
    title = "{From controversy to precision on the sigma meson: a review on the status of the non-ordinary $f_0(500)$ resonance}",
    eprint = "1510.00653",
    archivePrefix = "arXiv",
    primaryClass = "hep-ph",
    doi = "10.1016/j.physrep.2016.09.001",
    journal = "Phys. Rept.",
    volume = "658",
    pages = "1",
    year = "2016"
}

@article{Yao:2020bxx,
    author = "Yao, De-Liang and Dai, Ling-Yun and Zheng, Han-Qing and Zhou, Zhi-Yong",
    title = "{A review on partial-wave dynamics with chiral effective field theory and dispersion relation}",
    eprint = "2009.13495",
    archivePrefix = "arXiv",
    primaryClass = "hep-ph",
    doi = "10.1088/1361-6633/abfa6f",
    journal = "Rept. Prog. Phys.",
    volume = "84",
    number = "7",
    pages = "076201",
    year = "2021"
}

@book{Pascalutsa:2024hvy,
    author = "Pascalutsa, Vladimir",
    title = "{Causality Rules (Second Edition)}",
    doi = "10.1088/978-0-7503-3431-0",
    month = "4",
    year = "2024"
}

@article{Gasser:2003hk,
    author = "Gasser, J. and Rusetsky, A. and Scimemi, I.",
    title = "{Electromagnetic corrections in hadronic processes}",
    eprint = "hep-ph/0305260",
    archivePrefix = "arXiv",
    reportNumber = "BUTP-03-06, ECT-03-05",
    doi = "10.1140/epjc/s2003-01383-1",
    journal = "Eur. Phys. J. C",
    volume = "32",
    pages = "97--114",
    year = "2003"
}

@article{Kussner:2024ryb,
    author = {K{\"u}{\ss}ner, Meike},
    collaboration = "BESIII",
    title = "{Recent Highlights on Meson Spectroscopy at BESIII}",
    doi = "10.1051/epjconf/202429101002",
    journal = "EPJ Web Conf.",
    volume = "291",
    pages = "01002",
    year = "2024"
}

@phdthesis{Kussner:2022dft,
    author = {K{\"u}{\ss}ner, Meike},
    title = "{Coupled channel partial wave analysis of two-photon reactions at BESIII}",
    doi = "10.13154/294-8590",
    school = "Ruhr U., Bochum",
    year = "2022"
}

@article{Jing:2025qmi,
    author = "Jing, Hao-Jie and Cao, Xiong-Hui and Guo, Feng-Kun",
    title = "{Discontinuity calculus and applications to two-body coupled-channel scattering}",
    eprint = "2507.06175",
    archivePrefix = "arXiv",
    primaryClass = "hep-ph",
    doi = "10.15302/frontphys.2026.056201",
    journal = "Front. Phys. (Beijing)",
    volume = "21",
    number = "5",
    pages = "056201",
    year = "2026"
}

@article{Moussallam:2013una,
    author = "Moussallam, B.",
    title = "{Unified dispersive approach to real and virtual photon-photon scattering at low energy}",
    eprint = "1305.3143",
    archivePrefix = "arXiv",
    primaryClass = "hep-ph",
    doi = "10.1140/epjc/s10052-013-2539-y",
    journal = "Eur. Phys. J. C",
    volume = "73",
    pages = "2539",
    year = "2013"
}

@article{Chen:2015jgl,
    author = "Chen, Yun-Hua and Daub, Johanna T. and Guo, Feng-Kun and Kubis, Bastian and Mei{\ss}ner, Ulf-G. and Zou, Bing-Song",
    title = "{Effect of $Z_b$ states on $\Upsilon(3S)\to\Upsilon(1S)\pi\pi$ decays}",
    eprint = "1512.03583",
    archivePrefix = "arXiv",
    primaryClass = "hep-ph",
    doi = "10.1103/PhysRevD.93.034030",
    journal = "Phys. Rev. D",
    volume = "93",
    number = "3",
    pages = "034030",
    year = "2016"
}

@article{Wu:2024xwy,
    author = "Wu, Bing and Dong, Xiang-Kun and Du, Meng-Lin and Guo, Feng-Kun and Zou, Bing-Song",
    title = "{Deciphering the mechanism of $J/\psi$-nucleon scattering}",
    eprint = "2410.19526",
    archivePrefix = "arXiv",
    primaryClass = "hep-ph",
    doi = "10.1016/j.fmre.2025.07.005",
    journal = "Fund. Res.",
    volume = "5",
    pages = "2530--2536",
    year = "2025"
}

@article{Belle:2009xpa,
    author = "Uehara, S. and others",
    collaboration = "Belle",
    title = "{High-statistics study of eta pi0 production in two-photon collisions}",
    eprint = "0906.1464",
    archivePrefix = "arXiv",
    primaryClass = "hep-ex",
    doi = "10.1103/PhysRevD.80.032001",
    journal = "Phys. Rev. D",
    volume = "80",
    pages = "032001",
    year = "2009"
}

@article{Ren:2012aj,
    author = "Ren, X. -L. and Geng, L. S. and Martin Camalich, J. and Meng, J. and Toki, H.",
    title = "{Octet baryon masses in next-to-next-to-next-to-leading order covariant baryon chiral perturbation theory}",
    eprint = "1209.3641",
    archivePrefix = "arXiv",
    primaryClass = "nucl-th",
    doi = "10.1007/JHEP12(2012)073",
    journal = "JHEP",
    volume = "12",
    pages = "073",
    year = "2012"
}

@article{Griesshammer:2024twu,
    author = "Grie{\ss}hammer, Harald W. and Liao, Junjie and McGovern, Judith A. and Nogga, Andreas and Phillips, Daniel R.",
    title = "{Compton scattering on $^{4}\hbox {He}$ with nuclear one- and two-body densities}",
    eprint = "2401.16995",
    archivePrefix = "arXiv",
    primaryClass = "nucl-th",
    doi = "10.1140/epja/s10050-024-01339-x",
    journal = "Eur. Phys. J. A",
    volume = "60",
    number = "6",
    pages = "132",
    year = "2024"
}

@article{Li:2019irp,
    author = "Li, X. and others",
    title = "{Compton scattering from $^4$He at the TUNL HI$\gamma$S facility}",
    eprint = "1912.06915",
    archivePrefix = "arXiv",
    primaryClass = "nucl-ex",
    doi = "10.1103/PhysRevC.101.034618",
    journal = "Phys. Rev. C",
    volume = "101",
    number = "3",
    pages = "034618",
    year = "2020"
}

@article{Margaryan:2018opu,
    author = "Margaryan, Arman and Strandberg, Bruno and Grie{\ss}hammer, Harald W. and Mcgovern, Judith A. and Phillips, Daniel R. and Shukla, Deepshikha",
    title = "{Elastic Compton scattering from $^{3}$He and the role of the Delta}",
    eprint = "1804.00956",
    archivePrefix = "arXiv",
    primaryClass = "nucl-th",
    reportNumber = "Eur. Phys. J. A (2018) 54: 125",
    doi = "10.1140/epja/i2018-12554-x",
    journal = "Eur. Phys. J. A",
    volume = "54",
    number = "7",
    pages = "125",
    year = "2018"
}

@article{Kossert:2002jc,
    author = "Kossert, K. and others",
    title = "{Neutron polarizabilities investigated by quasifree Compton scattering from the deuteron}",
    eprint = "nucl-ex/0201015",
    archivePrefix = "arXiv",
    reportNumber = "NUCL-EX-0201015",
    doi = "10.1103/PhysRevLett.88.162301",
    journal = "Phys. Rev. Lett.",
    volume = "88",
    pages = "162301",
    year = "2002"
}

@article{Krause:1990xc,
    author = "Krause, A.",
    title = "{Baryon Matrix Elements of the Vector Current in Chiral Perturbation Theory}",
    doi = "10.5169/seals-116214",
    journal = "Helv. Phys. Acta",
    volume = "63",
    pages = "3--70",
    year = "1990"
}

@article{Oller:2006yh,
    author = "Oller, Jose Antonio and Verbeni, Michela and Prades, Joaquim",
    title = "{Meson-baryon effective chiral lagrangians to $\mathcal{O}(q^3)$}",
    eprint = "hep-ph/0608204",
    archivePrefix = "arXiv",
    reportNumber = "CAFPE-69-06, UGFT-199-06",
    doi = "10.1088/1126-6708/2006/09/079",
    journal = "JHEP",
    volume = "09",
    pages = "079",
    year = "2006"
}

@article{Granados:2017cib,
    author = "Granados, Carlos and Leupold, Stefan and Perotti, Elisabetta",
    title = "{The electromagnetic Sigma-to-Lambda hyperon transition form factors at low energies}",
    eprint = "1701.09130",
    archivePrefix = "arXiv",
    primaryClass = "hep-ph",
    doi = "10.1140/epja/i2017-12324-4",
    journal = "Eur. Phys. J. A",
    volume = "53",
    number = "6",
    pages = "117",
    year = "2017"
}

@article{Hoid:2026atr,
    author = "Hoid, Bai-Long and Danilkin, Igor and Vanderhaeghen, Marc",
    title = "{\boldmath{Dispersive analysis of the $\phi \rightarrow \gamma \pi^0 \pi^0$ process}}",
    eprint = "2602.15100",
    archivePrefix = "arXiv",
    primaryClass = "hep-ph",
    doi = "10.1103/dlcc-jm4d",
    journal = "Phys. Rev. D",
    volume = "113",
    number = "11",
    pages = "114046",
    year = "2026"
}

@article{Mai:2009ce,
    author = "Mai, Maxim and Bruns, Peter C. and Kubis, Bastian and Mei{\ss}ner, Ulf-G.",
    title = "{Aspects of meson-baryon scattering in three and two-flavor chiral perturbation theory}",
    eprint = "0905.2810",
    archivePrefix = "arXiv",
    primaryClass = "hep-ph",
    reportNumber = "HISKP-TH-09-12, FZJ-IKP-TH-2009-10",
    doi = "10.1103/PhysRevD.80.094006",
    journal = "Phys. Rev. D",
    volume = "80",
    pages = "094006",
    year = "2009"
}

@misc{Zhuge:2026dxl,
    author = "Zhuge, Yu and Liu, Zhan-Wei and Leinweber, Derek B. and Thomas, Anthony W.",
    title = "{Lattice QCD constraints on pion electroproduction off a nucleon}",
    eprint = "2603.06055",
    archivePrefix = "arXiv",
    primaryClass = "hep-ph",
    month = "3",
    year = "2026"
}

@article{Brodsky:1971zh,
    author = "Brodsky, Stanley J. and Close, Francis E. and Gunion, J. F.",
    title = "{Compton Scattering and Fixed Poles in Parton Field-Theoretic Models}",
    reportNumber = "SLAC-PUB-0973",
    doi = "10.1103/PhysRevD.5.1384",
    journal = "Phys. Rev. D",
    volume = "5",
    pages = "1384",
    year = "1972"
}

@article{Brodsky:2008qu,
    author = "Brodsky, Stanley J. and Llanes-Estrada, Felipe J. and Szczepaniak, Adam P.",
    title = "{Local Two-Photon Couplings and the $J=0$ Fixed Pole in Real and Virtual Compton Scattering}",
    eprint = "0812.0395",
    archivePrefix = "arXiv",
    primaryClass = "hep-ph",
    reportNumber = "SLAC-PUB-13478",
    doi = "10.1103/PhysRevD.79.033012",
    journal = "Phys. Rev. D",
    volume = "79",
    pages = "033012",
    year = "2009"
}

@article{Danilkin:2026wkl,
    author = "Danilkin, Igor and Pasquini, Barbara and Ronchi, Matteo and Vanderhaeghen, Marc",
    title = "{Subtracted dispersion relations for virtual Compton scattering off the proton}",
    eprint = "2603.09831",
    archivePrefix = "arXiv",
    primaryClass = "hep-ph",
    doi = "10.1103/lv7d-fpls",
    journal = "Phys. Rev. D",
    volume = "114",
    number = "5",
    pages = "054035",
    year = "2026"
}

@article{Brodsky:1972vv,
    author = "Brodsky, Stanley J. and Close, Francis E. and Gunion, J. F.",
    title = "{Phenomenology of Photon Processes, Vector Dominance and Crucial Tests for Parton Models}",
    reportNumber = "SLAC-PUB-1012",
    doi = "10.1103/PhysRevD.6.177",
    journal = "Phys. Rev. D",
    volume = "6",
    pages = "177",
    year = "1972"
}

@article{Geng:2009hh,
    author = "Geng, L. S. and Martin Camalich, J. and Vicente Vacas, M. J.",
    title = "{Leading-order decuplet contributions to the baryon magnetic moments in Chiral Perturbation Theory}",
    eprint = "0903.0779",
    archivePrefix = "arXiv",
    primaryClass = "hep-ph",
    doi = "10.1016/j.physletb.2009.04.061",
    journal = "Phys. Lett. B",
    volume = "676",
    pages = "63--68",
    year = "2009"
}

@article{Cao:2024nxm,
    author = "Cao, Xiong-Hui and Du, Meng-Lin and Guo, Feng-Kun",
    title = "{Photoproduction of the $X(3872)$ beyond vector meson dominance: the open-charm coupled-channel mechanism}",
    eprint = "2401.16112",
    archivePrefix = "arXiv",
    primaryClass = "hep-ph",
    doi = "10.1088/1361-6471/ad6fbc",
    journal = "J. Phys. G",
    volume = "51",
    number = "10",
    pages = "105002",
    year = "2024"
}

@article{Pascalutsa:2000kd,
    author = "Pascalutsa, V.",
    title = "{Correspondence of consistent and inconsistent spin-$3/2$ couplings via the equivalence theorem}",
    eprint = "hep-ph/0008026",
    archivePrefix = "arXiv",
    doi = "10.1016/S0370-2693(01)00140-X",
    journal = "Phys. Lett. B",
    volume = "503",
    pages = "85--90",
    year = "2001"
}

@article{Muller:2015vha,
    author = {M{\"u}ller, Dieter and Semenov-Tian-Shansky, Kirill M.},
    title = "{$J=0$ fixed pole and $D$-term form factor in deeply virtual Compton scattering}",
    eprint = "1507.02164",
    archivePrefix = "arXiv",
    primaryClass = "hep-ph",
    reportNumber = "CPHT-RR023.0715",
    doi = "10.1103/PhysRevD.92.074025",
    journal = "Phys. Rev. D",
    volume = "92",
    number = "7",
    pages = "074025",
    year = "2015"
}

@article{Elitzur:1970yy,
    author = "Elitzur, M. and Harari, H.",
    title = "{Electromagnetic mass differences and inelastic electron scattering}",
    doi = "10.1016/0003-4916(70)90006-0",
    journal = "Annals Phys.",
    volume = "56",
    pages = "81--107",
    year = "1970"
}

@article{Wilcox:2021rtt,
    author = "Wilcox, Walter and Lee, Frank X.",
    title = "{Towards charged hadron polarizabilities from four-point functions in lattice QCD}",
    eprint = "2106.02557",
    archivePrefix = "arXiv",
    primaryClass = "hep-lat",
    reportNumber = "BU-HEPP-21-01",
    doi = "10.1103/PhysRevD.104.034506",
    journal = "Phys. Rev. D",
    volume = "104",
    number = "3",
    pages = "034506",
    year = "2021"
}

@misc{Cao:2026bdc,
    author = "Cao, Xiong-Hui and Dai, Ling-Yun and Guo, Feng-Kun",
    title = "{Precision Sum Rule for Nucleon Isovector Polarizabilities and the Proton--Neutron Mass Difference}",
    eprint = "2609.15154",
    archivePrefix = "arXiv",
    primaryClass = "hep-ph",
    month = "9",
    year = "2026"
}

@article{Bernard:1995dp,
    author = "Bernard, V. and Kaiser, Norbert and Mei{\ss}ner, Ulf-G.",
    title = "{Chiral dynamics in nucleons and nuclei}",
    eprint = "hep-ph/9501384",
    archivePrefix = "arXiv",
    reportNumber = "CRN-95-3, TK-95-1",
    doi = "10.1142/S0218301395000092",
    journal = "Int. J. Mod. Phys. E",
    volume = "4",
    pages = "193--346",
    year = "1995"
}

@article{Anisovich:2016vzt,
    author = "Anisovich, A. V. and others",
    title = "{The impact of new polarization data from Bonn, Mainz and Jefferson Laboratory on $\gamma p \rightarrow \pi N$ multipoles}",
    eprint = "1604.05704",
    archivePrefix = "arXiv",
    primaryClass = "nucl-th",
    doi = "10.1140/epja/i2016-16284-9",
    journal = "Eur. Phys. J. A",
    volume = "52",
    number = "9",
    pages = "284",
    year = "2016"
}

@article{Frink:2006hx,
    author = "Frink, Matthias and Mei{\ss}ner, Ulf-G.",
    title = "{On the chiral effective meson-baryon Lagrangian at third order}",
    eprint = "hep-ph/0609256",
    archivePrefix = "arXiv",
    reportNumber = "HISKP-TH-06-29",
    doi = "10.1140/epja/i2006-10105-x",
    journal = "Eur. Phys. J. A",
    volume = "29",
    pages = "255--260",
    year = "2006"
}

@article{Schneider:2025ndq,
    author = {Schneider, Carolin and R{\"o}nchen, Deborah and Hanhart, Christoph and Mei{\ss}ner, Ulf-G.},
    title = {{Coupled-channel contributions to the GDH sum rule from the J{\"u}lich{\textendash}Bonn approach}},
    eprint = "2508.00598",
    archivePrefix = "arXiv",
    primaryClass = "nucl-th",
    doi = "10.1140/epja/s10050-025-01719-x",
    journal = "Eur. Phys. J. A",
    volume = "61",
    number = "10",
    pages = "241",
    year = "2025"
}

@misc{Kamano:2019gtm,
    author = "Kamano, H. and Lee, T. S. H. and Nakamura, S. X. and Sato, T.",
    title = "{The ANL-Osaka Partial-Wave Amplitudes of $\pi N$ and $\gamma N$ Reactions}",
    eprint = "1909.11935",
    archivePrefix = "arXiv",
    primaryClass = "nucl-th",
    month = "9",
    year = "2019"
}

@misc{ANL-O,
  title = {{ANL-Osaka Partial-Wave Amplitudes}},
  howpublished = {\url{https://www.phy.anl.gov/theory/research/anl-osaka-pwa}}
}

@misc{JuBo,
  title = {{J\"ulich-Bonn-Washington Collaboration}},
  howpublished = {\url{https://collaborations.fz-juelich.de/ikp/meson-baryon/}}
}

@misc{Zhou:2026bgr,
    author = "Zhou, J. and others",
    title = "{First Measurement of Compton Scattering on $^3\text{He}$ at HI$\gamma$S}",
    eprint = "2610.07579",
    archivePrefix = "arXiv",
    primaryClass = "nucl-ex",
    month = "10",
    year = "2026"
}

@article{Jacob:1959at,
    author = "Jacob, M. and Wick, G. C.",
    title = "{On the General Theory of Collisions for Particles with Spin}",
    doi = "10.1006/aphy.2000.6022",
    journal = "Annals Phys.",
    volume = "7",
    pages = "404--428",
    year = "1959"
}

@article{Dong:2020hxe,
    author = {Dong, Xiang-Kun and Guo, Feng-Kun and Zou, Bing-Song},
    title = {{Explaining the Many Threshold Structures in the Heavy-Quark Hadron Spectrum}},
    eprint = {2011.14517},
    archivePrefix = {arXiv},
    primaryClass = {hep-ph},
    doi = {10.1103/PhysRevLett.126.152001},
    journal = {Phys. Rev. Lett.},
    volume = {126},
    number = {15},
    pages = {152001},
    year = {2021}
}

@misc{Cao2026DispDVCS,
    author = "Cao, Xiong-Hui and Guo, Feng-Kun",
    title = "{Unified $t$-channel dispersive analysis of the deeply virtual Compton scattering subtraction constant \mbox{I}: isoscalar contribution}",
    howpublished = "in preparation"
}

@article{Dudek:2016cru,
    author = "Dudek, Jozef J. and Edwards, Robert G. and Wilson, David J.",
    collaboration = "Hadron Spectrum",
    title = "{An $a_0$ resonance in strongly coupled $\pi \eta$, $K\overline{K}$ scattering from lattice QCD}",
    eprint = "1602.05122",
    archivePrefix = "arXiv",
    primaryClass = "hep-ph",
    reportNumber = "JLAB-THY-16-2216, DAMTP-2016-18",
    doi = "10.1103/PhysRevD.93.094506",
    journal = "Phys. Rev. D",
    volume = "93",
    number = "9",
    pages = "094506",
    year = "2016"
}

@article{Briceno:2017qmb,
    author = "Brice{\~n}o, Ra{\'u}l A. and Dudek, Jozef J. and Edwards, Robert G. and Wilson, David J.",
    title = "{Isoscalar $\pi\pi, K\overline{K}, \eta\eta$ scattering and the $\sigma, f_0, f_2$ mesons from QCD}",
    eprint = "1708.06667",
    archivePrefix = "arXiv",
    primaryClass = "hep-lat",
    reportNumber = "JLAB-THY-17-2534",
    doi = "10.1103/PhysRevD.97.054513",
    journal = "Phys. Rev. D",
    volume = "97",
    number = "5",
    pages = "054513",
    year = "2018"
}
\end{document}